\documentclass[prd,nofootinbib,reprint,balancelastpage]{revtex4-2}

\usepackage{amsmath}
\usepackage{amsthm}
\usepackage{amssymb}
\usepackage{bm}      
\usepackage{accents}       
\usepackage[svgnames]{xcolor}

\usepackage{multirow}  
\definecolor{britishracinggreen}{rgb}{0.0, 0.26, 0.15}
\usepackage{hyperref}  
\hypersetup{
    colorlinks=true,
    linkcolor=DarkBlue,
    citecolor=britishracinggreen,
    urlcolor=DarkBlue,
    }

\numberwithin{equation}{section}
\renewcommand{\theequation}{\arabic{section}.\arabic{equation}}

\newcommand{\beq}[1]{\begin{equation} \label{#1}}
\newcommand{\eeq}{\end{equation}}

\newcommand{\bea}[1]{\begin{eqnarray} \label{#1}}
\newcommand{\eea}{\end{eqnarray}}

\newcommand{\HIDE}[1]{}

\DeclareMathAlphabet{\mathbbmsl}{U}{bbm}{m}{sl}
\newcommand{\fatm}[1]{\hspace{-0.8pt}\mathbbmsl{#1}\hspace{-0.8pt}}

\definecolor{bulgarianrose}{rgb}{0.28, 0.02, 0.03} 
\newcommand{\ggColor}[1]{{\color{bulgarianrose}{#1}}} 

\newcommand{\za}{{a}}   

\newcommand{\zi}{{i}}   
\newcommand{\zj}{{j}}   
\newcommand{\zk}{{k}}   

\newcommand{\Zm}{{\mu}}      
\newcommand{\Zn}{{\nu}}      

\newcommand{\zA}{{\mathchoice{A}{A}{\scriptscriptstyle{A}}{A}}}   
\newcommand{\zB}{{\mathchoice{B}{B}{\scriptscriptstyle{B}}{B}}}   
\newcommand{\zC}{{\mathchoice{C}{C}{\scriptscriptstyle{C}}{C}}}   

\newcommand{\zI}{{\mathchoice{I}{I}{\scriptscriptstyle{I}}{I}}}   
\newcommand{\zJ}{{\mathchoice{J}{J}{\scriptscriptstyle{J}}{J}}}   
\newcommand{\zK}{{\mathchoice{K}{K}{\scriptscriptstyle{K}}{K}}}   
\newcommand{\zL}{{\mathchoice{L}{L}{\scriptscriptstyle{L}}{L}}}   

\newcommand{\ZA}{{\mathchoice{\mathcal{A}}{\mathcal{A}}{\scriptstyle{\mathcal{A}}}{\mathcal{A}}}}   
\newcommand{\ZB}{{\mathchoice{\mathcal{B}}{\mathcal{B}}{\scriptstyle{\mathcal{B}}}{\mathcal{B}}}}   
\newcommand{\ZC}{{\mathchoice{\mathcal{C}}{\mathcal{C}}{\scriptstyle{\mathcal{C}}}{\mathcal{C}}}}   
\newcommand{\ZD}{{\mathchoice{\mathcal{D}}{\mathcal{D}}{\scriptstyle{\mathcal{D}}}{\mathcal{D}}}}   

\newcommand{\ZI}{{\mathchoice{\mathcal{I}}{\mathcal{I}}{\scriptstyle{\mathcal{I}}}{\mathcal{I}}}}   
\newcommand{\ZJ}{{\mathchoice{\mathcal{J}}{\mathcal{J}}{\scriptstyle{\mathcal{J}}}{\mathcal{J}}}}   

\newcommand{\cza}{\alpha}
\newcommand{\czb}{\beta}
\newcommand{\czc}{\gamma}
\newcommand{\czd}{{\,\delta}}
\newcommand{\cze}{\epsilon}
\newcommand{\czz}{\zeta}

\newcommand{\ccza}{{\ggColor{a}}}   
\newcommand{\cczb}{{\ggColor{b}}}   
\newcommand{\cczc}{{\ggColor{c}}}   
\newcommand{\cczd}{{\ggColor{d}}}   

\newcommand{\gC}{\mathcal{C}}          
\newcommand{\agC}{\bar{\mathcal{C}}}   
  \newcommand{\agB}{\pi}           
\newcommand{\egC}{\mathcal{C}'}        
  \newcommand{\egB}{\pi'}                

\newcommand{\GG}{\textsl{g}}  

\newcommand{\AB}[2]{{\mathchoice {\big(#1 {\hspace{.4pt},\hspace{.4pt}} #2\big)} {\big(#1{\hspace{.4pt},\hspace{.4pt}}#2\big)} {(#1\,,\,#2)} {(#1,#2)} }}    

\newcommand{\Z}{\mathbb Z}    

\newcommand{\var}{\delta} 

\newcommand{\Tr}{\mathop{\mathrm{Tr}}}

\DeclareMathOperator{\rank}{\,\mathrm{rank}\,}
\DeclareMathOperator{\corank}{\,\mathrm{corank}\,}

\newcommand{\gh}[1]{\mathop{gh}{(#1)}}  

\newcommand{\vlv}{\vphantom{|}}  

\newcommand{\auxtitle}[1]{\vspace{2mm}\begin{center} \emph{#1} \end{center} \vspace{2mm}}

\begin{document}

\title{Restricted Gauge Theory Formalism and Unimodular Gravity}

\author{A.\,O. Barvinsky}
\email{barvin@td.lpi.ru}

\author{D.\,V. Nesterov}
\email{nesterov@lpi.ru}

\affiliation{I.E.Tamm Theory Department, P.N. Lebedev Physical Institute of the Russian Academy of Sciences,  Leninsky Prospekt 53, Moscow 119991, Russia}

\begin{abstract}
 \begin{center}
  (Received 29 December 2022; revised 25 May 2023; accepted 25 May 2023; published 7 September 2023)\\
 \end{center}

We develop a Lagrangian quantization formalism for a class of theories obtained by the restriction of the configuration space of gauge fields from a wider (parental) gauge theory. This formalism is based on application of the Batalin-Vilkovisky technique for quantization of theories with linearly dependent generators, their linear dependence originating from a special type of projection from the originally irreducible gauge generators of the parental theory. The algebra of these projected generators is shown to be closed for parental gauge algebras closed off shell. We demonstrate that new physics of the restricted theory, as compared to its parental theory, is associated with the rank deficiency of a special gauge-restriction operator reflecting the gauge transformations of the restriction constraints functions --- this distinguishes the restricted theory from its partial gauge fixing. As a byproduct of this technique a workable algorithm for the one-loop effective action in generic first-stage reducible theory was constructed, along with the explicit set of tree-level Ward identities for gauge field, ghost, and ghosts-for-ghosts propagators. The formalism is applied to unimodular gravity theory, and its one-loop effective action is obtained in terms of functional determinants of minimal second-order operators, calculable on generic backgrounds by Schwinger-DeWitt technique of local curvature expansion. The result is shown to be equivalent to Einstein gravity theory with a cosmological term up to a special contribution of the global degree of freedom associated with the variable value of the cosmological constant. The role of this degree of freedom in a special duality relation between Einstein theory and unimodular gravity is briefly discussed.\\

\hspace{-4.5mm} DOI:\href{https://doi.org/10.1103/PhysRevD.108.065004}{10.1103/PhysRevD.108.065004}
\vspace{10mm}
\end{abstract}

\maketitle

\if{
I. Introduction 2
II. Effective action for theories with first-stage
reducible gauge generators 3
A. Batalin-Vilkovisky formalism for reducible
gauge theories 3
B. Stationary point of the gauge-fixed master
action 6
C. One-loop contribution to the generating
functional 8
D. Ward identities and gauge independence of
the effective action 8
III. Reducible gauge structure of restricted gauge
theories 9
A. Two representations of a restricted theory 9
B. Gauge symmetry and reducibility 11
C. Gauge algebra of a restricted theory 12
IV. One-loop effective action of restricted gauge
theory 13
A. Two representations of the one-loop effective
action 13
B. Independence of projector parameter 15
C. Canonical normalization of generators 15
D. The difference between parental and restricted
theories in the one-loop approximation 18
V. Unimodular gravity theory 19
A. Gauge fixing and propagators 20
B. Reduction of functional determinants 22
VI. Conclusion 24
Acknowledgements 26
A. Moore-Penrose inverse and variation of
projectors 26
B. Gauge algebra of the restricted theory from a
parental theory 27
C. Determinant relation 28
References 29
}\fi

\section*{CONTENTS}

\newcommand{\tocBSectVSPACE}{{\vspace{3mm}}} 
\newcommand{\tocASectVSPACE}{{\vspace{2mm}}} 
\newcommand{\tocBSSectVSPACE}{{\vspace{1mm}}} 
\newcommand{\tocPNumHSPACE}{{\hspace{4mm}}}
\newcommand{\tocSSectHSPACE}{{\phantom{A.}\hspace{1.1mm}}}
\newcommand{\phl}{\phantom{1}}
\newcommand{\PGRef}[1]{{\hypersetup{hidelinks}\pageref{#1}}}
\newcommand{\opt}[1]{}
\newcommand{\msp}{\hspace{-1pt} }
\newcommand{\hsp}{\hspace{0.5pt} }
\newcommand{\lsp}{\hspace{1pt} }
\newcommand{\lhsp}{\hspace{1.5pt} }
\newcommand{\llsp}{\hspace{2pt} }
\newcommand{\lllsp}{\hspace{3pt} }
{
\hspace{-5mm}
\begin{tabular}
{r l r c}
 
  
  \ref{Sect:Introduction}. 
  & \hyperref[Sect:Introduction]{Introduction\hfill}
  \opt{&}
  & {\phl}\PGRef{Sect:Introduction}
  & \tocBSectVSPACE
  \\

  \ref{Sect:BV_EA_1stReducible}. 
  & \hyperref[Sect:BV_EA_1stReducible]{Effective action for theories with first-stage re-\HIDE{ducible gauge generators\hfill}}
  \opt{&}
  & 
      & 
  \\
  {} 
  & \hyperref[Sect:BV_EA_1stReducible]{\HIDE{Effective action for theories with first-stage re-}ducible gauge generators\hfill}
  \opt{&}
  & \HIDE{\tocPNumHSPACE} {\phl}\PGRef{Sect:BV_EA_1stReducible} 
      & \tocASectVSPACE
  \\

      & \hyperref[SSect:Reducible_BV_Setup]{A.\hspace{1.2mm}Batalin-Vilkovisky\lsp formalism\lsp for\lsp reducible\HIDE{gauge theories}\hfill}
  \opt{&}
      & 
      & 
      \\
      & \hyperref[SSect:Reducible_BV_Setup]{\HIDE{A.\hspace{1.2mm}Batalin-Vilkovisky formalism for reducible}{\tocSSectHSPACE}gauge theories\hfill}
  \opt{&}
      & \HIDE{\tocPNumHSPACE} {\phl}\PGRef{SSect:Reducible_BV_Setup} 
      & \tocBSSectVSPACE
      \\

      & \hyperref[SSect:StatSurface_and_RegularityCond]{B.\hspace{1.2mm}Stationary point\lsp of the\lsp gauge-fixed master\HIDE{action}\hfill}
  \opt{&}
      & 
      & 
      \\
      & \hyperref[SSect:StatSurface_and_RegularityCond]{\HIDE{B.\hspace{1.2mm}Stationary point of the gauge-fixed master}{\tocSSectHSPACE}action\hfill}
  \opt{&}
      & \HIDE{\tocPNumHSPACE} {\phl}\PGRef{SSect:StatSurface_and_RegularityCond} 
      & \tocBSSectVSPACE
      \\

      & \hyperref[SSect:BV_EA]{C.\hspace{1.2mm}One-loop\lllsp contribution\lllsp to\lllsp the\lllsp generating\HIDE{functional}\hfill}
  \opt{&}
      & 
      & 
      \\
      & \hyperref[SSect:BV_EA]{\HIDE{C.\hspace{1.2mm}One-loop contribution to the generating}{\tocSSectHSPACE}functional\hfill}
  \opt{&}
      & \HIDE{\tocPNumHSPACE} {\phl}\PGRef{SSect:BV_EA} 
      & \tocBSSectVSPACE
      \\

      & \hyperref[SSect:BV_WardIdentities]{D.\hspace{1.2mm}Ward identities and gauge independence\msp of\HIDE{the effective action}\hfill}
  \opt{&}
      & 
  & 
  \\
      & \hyperref[SSect:BV_WardIdentities]{\HIDE{D.\hspace{1.2mm}Ward identities and gauge independence of}{\tocSSectHSPACE}the effective action\hfill}
  \opt{&}
      & \HIDE{\tocPNumHSPACE} {\phl}\PGRef{SSect:BV_WardIdentities} 
  & \tocBSectVSPACE
  \\

  \ref{Sect:Restricting_and_Reducibility}. 
  & \hyperref[Sect:Restricting_and_Reducibility]{Reducible\hsp gauge\hsp structure\lsp of\lsp restricted\hsp gauge\HIDE{theories}\hfill}
  \opt{&}
  & 
      & 
  \\
  & \hyperref[Sect:Restricting_and_Reducibility]{\HIDE{Reducible gauge structure of restricted gauge}theories\hfill}
  \opt{&}
  & \HIDE{\tocPNumHSPACE} {\phl}\PGRef{Sect:Restricting_and_Reducibility} 
      & \tocASectVSPACE
  \\

      & \hyperref[SSect:Restricted_and_Reduced_Reps]{A.\hspace{1.2mm}Two representations of a restricted theory\hfill}
  \opt{&}
      & {\phl}\PGRef{SSect:Restricted_and_Reduced_Reps} 
      & \tocBSSectVSPACE
      \\

      & \hyperref[SSect:Restricted_Gauge_Reducibility]{B.\hspace{1.2mm}Gauge symmetry and reducibility\hfill}
  \opt{&}
      & \hspace{0.5pt}\PGRef{SSect:Restricted_Gauge_Reducibility}\hspace{-0.5pt} 
      & \tocBSSectVSPACE
      \\

      & \hyperref[SSect:Restricted_Gauge_Algebra]{C.\hspace{1.2mm}Gauge algebra of a restricted theory\hfill}
  \opt{&}
      & \PGRef{SSect:Restricted_Gauge_Algebra} 
  & \tocBSectVSPACE
  \\

  \ref{Sect:1loop_EA_Restricted}. 
  & \hyperref[Sect:1loop_EA_Restricted]{One-loop\lhsp effective\lhsp action\llsp of\llsp restricted\lhsp gauge\HIDE{theory}\hfill}
  \opt{&}
  & 
  & 
  \\
  & \hyperref[Sect:1loop_EA_Restricted]{\HIDE{One-loop effective action of restricted gauge}theory\hfill}
  \opt{&}
  & \HIDE{\tocPNumHSPACE} \PGRef{Sect:1loop_EA_Restricted} 
  & \tocASectVSPACE
  \\
\end{tabular}

\vspace{2cm}
\phantom{.}
\vspace{5.4mm}

\hspace{-5mm}
\begin{tabular}
{r l c r c}

      & \hyperref[SSect:1loop_EA_Restricted]{\hspace{-1pt}A.\hspace{1.2mm}Two representations of the one-loop effective\HIDE{action}\hfill}
      && 
      & 
      \\
      & \hyperref[SSect:1loop_EA_Restricted]{\HIDE{A.\hspace{1.2mm}Two representations of the one-loop effective}{\tocSSectHSPACE}action\hfill}
      && \HIDE{\tocPNumHSPACE} \PGRef{SSect:1loop_EA_Restricted} 
      & \tocBSSectVSPACE
      \\

      & \hyperref[SSect:Projector_Param_Independence]{\hspace{-1pt}B.\hspace{1.2mm}Independence of projector parameter\hfill}
      && \PGRef{SSect:Projector_Param_Independence} 
      & \tocBSSectVSPACE
      \\

      & \hyperref[SSect:Canon_Measure_Normalization]{\hspace{-1pt}C.\hspace{1.2mm}Canonical normalization of generators\hfill}
      && \PGRef{SSect:Canon_Measure_Normalization} 
      & \tocBSSectVSPACE
      \\

      & \hyperref[SSect:1loop_EA_Difference]{\hspace{-1pt}D.\hspace{1.2mm}The difference between parental and restrict-\HIDE{ed theories in the one-loop approximation}\HIDE{\hfill}}
      && 
  & 
  \\
      & \hyperref[SSect:1loop_EA_Difference]{\HIDE{D.\hspace{1.2mm}The difference between parental and restrict-}{\tocSSectHSPACE}ed theories in the one-loop approximation\hfill}
      && \HIDE{\tocPNumHSPACE} \PGRef{SSect:1loop_EA_Difference} 
  & \tocBSectVSPACE
  \\

  \ref{Sect:UMG_EA}. 
  & \hyperref[Sect:UMG_EA]{Unimodular gravity theory\hfill}
  && \PGRef{Sect:UMG_EA} 
      & \tocASectVSPACE
  \\

      & \hyperref[SSect:UMG_gf_Propagators]{A.\hspace{1.2mm}Gauge fixing and propagators\hfill}
      && \PGRef{SSect:UMG_gf_Propagators} 
      & \tocBSSectVSPACE
      \\

      & \hyperref[SSect:UMG_1loop_Det]{B.\hspace{1.2mm}Reduction of functional determinants\hfill}
      && \PGRef{SSect:UMG_1loop_Det} 
  & \tocBSectVSPACE
  \\

  \ref{Sect:Conclusion}. 
  & \hyperref[Sect:Conclusion]{Conclusion\hfill}
  && \PGRef{Sect:Conclusion}
  & \tocASectVSPACE
  \\

  {} 
  & \hyperref[Sect:Acknowledgements]{Acknowledgements\hfill}
  && \PGRef{Sect:Acknowledgements}
  & \tocASectVSPACE
  \\


  \ref{ASect:Projectors_Variation}. 
  & \hyperref[ASect:Projectors_Variation]{Moore-Penrose inverse\lsp and\lsp variation of projec-\HIDE{tors}\hfill}
  && 
  & 
  \\
  & \hyperref[ASect:Projectors_Variation]{\HIDE{Moore-Penrose inverse and variation of projec}tors\hfill}
  && \HIDE{\tocPNumHSPACE} \PGRef{ASect:Projectors_Variation}
  & \tocASectVSPACE
  \\

  \ref{ASect:Gauge_Struct_Restriction}. 
  & \hyperref[ASect:Gauge_Struct_Restriction]{Gauge\lsp algebra\lsp of\lsp the\lsp restricted\lsp theory\lsp from\lsp a \HIDE{parental theory}\hfill}
  && 
  & 
  \\
  & \hyperref[ASect:Gauge_Struct_Restriction]{\HIDE{Gauge algebra of the restricted theory from a}
parental theory\hfill}
  && \HIDE{\tocPNumHSPACE} \PGRef{ASect:Gauge_Struct_Restriction}
  & \tocASectVSPACE
  \\

  \ref{ASect:Determinant_Relation}. 
  & \hyperref[ASect:Determinant_Relation]{Determinant relation\hfill}
  && \PGRef{ASect:Determinant_Relation}
  & \tocASectVSPACE
  \\

 {}
 & \hyperref[Sect:References]{References\hfill}
 && \PGRef{Sect:References}
 \\

\end{tabular}
}

\if{
\begin{tabular*}{80mm}
{r p{72mm} c p{4mm}c}
  \ref{Sect:Introduction}. 
  & \hyperref[Sect:Introduction]{Introduction\hfill}
  && {\phl}\PGRef{Sect:Introduction}
  & \tocBSectVSPACE
  \\

  \ref{Sect:BV_EA_1stReducible}. 
  & \hyperref[Sect:BV_EA_1stReducible]{Effective action for theories with first-stage reducible gauge generators\hfill}
  && \tocPNumHSPACE {\phl}\PGRef{Sect:BV_EA_1stReducible} 
      & \tocASectVSPACE
  \\

      & \hyperref[SSect:Reducible_BV_Setup]{A.\hspace{1.2mm}Batalin-Vilkovisky formalism for reducible gauge theories\hfill}
      && \tocPNumHSPACE {\phl}\PGRef{SSect:Reducible_BV_Setup} 
      & \tocBSSectVSPACE
      \\

      & \hyperref[SSect:StatSurface_and_RegularityCond]{B.\hspace{1.2mm}Stationary point of the gauge-fixed master action\hfill}
      && \tocPNumHSPACE {\phl}\PGRef{SSect:StatSurface_and_RegularityCond} 
      & \tocBSSectVSPACE
      \\

      & \hyperref[SSect:BV_EA]{C.\hspace{1.2mm}One-loop contribution to the generating functional\hfill}
      && \tocPNumHSPACE {\phl}\PGRef{SSect:BV_EA} 
      & \tocBSSectVSPACE
      \\

      & \hyperref[SSect:BV_WardIdentities]{D.\hspace{1.2mm}Ward identities and gauge independence of the effective action\hfill}
      && \tocPNumHSPACE {\phl}\PGRef{SSect:BV_WardIdentities} 
  & \tocBSectVSPACE
  \\

  \ref{Sect:Restricting_and_Reducibility}. 
  & \hyperref[Sect:Restricting_and_Reducibility]{Reducible gauge structure of restricted gauge
theories\hfill}
  && \tocPNumHSPACE {\phl}\PGRef{Sect:Restricting_and_Reducibility} 
      & \tocASectVSPACE
  \\

      & \hyperref[SSect:Restricted_and_Reduced_Reps]{A.\hspace{1.2mm}Two representations of a restricted theory\hfill}
      && {\phl}\PGRef{SSect:Restricted_and_Reduced_Reps} 
      & \tocBSSectVSPACE
      \\

      & \hyperref[SSect:Restricted_Gauge_Reducibility]{B.\hspace{1.2mm}Gauge symmetry and reducibility\hfill}
      && \hspace{0.5pt}\PGRef{SSect:Restricted_Gauge_Reducibility} 
      & \tocBSSectVSPACE
      \\

      & \hyperref[SSect:Restricted_Gauge_Algebra]{C.\hspace{1.2mm}Gauge algebra of a restricted theory\hfill}
      && \PGRef{SSect:Restricted_Gauge_Algebra} 
  &\tocBSectVSPACE
  \\

  \ref{Sect:1loop_EA_Restricted}. 
  & \hyperref[Sect:1loop_EA_Restricted]{One-loop effective action of restricted gauge theory\hfill}
  && \tocPNumHSPACE \PGRef{Sect:1loop_EA_Restricted} 
  & \tocASectVSPACE
  \\

\end{tabular*}

\vspace{2cm}
\phantom{.}
\vspace{5.3mm}

\begin{tabular*}{80 mm} 
{r p{72mm} c p{4mm} c}
      & \hyperref[SSect:1loop_EA_Restricted]{A.\hspace{1.2mm}Two representations of the one-loop effective action\hfill}
      && \tocPNumHSPACE \PGRef{SSect:1loop_EA_Restricted} 
      & \tocBSSectVSPACE
      \\

      & \hyperref[SSect:Projector_Param_Independence]{B.\hspace{1.2mm}Independence of projector parameter\hfill}
      && \PGRef{SSect:Projector_Param_Independence} 
      & \tocBSSectVSPACE
      \\

      & \hyperref[SSect:Canon_Measure_Normalization]{C.\hspace{1.2mm}Canonical normalization of generators\hfill}
      && \PGRef{SSect:Canon_Measure_Normalization} 
      & \tocBSSectVSPACE
      \\

      & \hyperref[SSect:1loop_EA_Difference]{D.\hspace{1.2mm}The difference between parental and restricted theories in the one-loop approximation\HIDE{\hfill}}
      && \tocPNumHSPACE \PGRef{SSect:1loop_EA_Difference} 
  & \tocBSectVSPACE
  \\

  \ref{Sect:UMG_EA}. 
  & \hyperref[Sect:UMG_EA]{Unimodular gravity theory\hfill}
  && \PGRef{Sect:UMG_EA} 
      & \tocASectVSPACE
  \\

      & \hyperref[SSect:UMG_gf_Propagators]{A.\hspace{1.2mm}Gauge fixing and propagators\hfill}
      && \PGRef{SSect:UMG_gf_Propagators} 
      & \tocBSSectVSPACE
      \\

      & \hyperref[SSect:UMG_1loop_Det]{B.\hspace{1.2mm}Reduction of functional determinants\hfill}
      && \PGRef{SSect:UMG_1loop_Det} 
  & \tocBSectVSPACE
  \\

  \ref{Sect:Conclusion}. 
  & \hyperref[Sect:Conclusion]{Conclusion\hfill}
  && \PGRef{Sect:Conclusion}
  & \tocASectVSPACE
  \\

  {} 
  & \hyperref[Sect:Acknowledgements]{Acknowledgements\hfill}
  && \PGRef{Sect:Acknowledgements}
  & \tocASectVSPACE
  \\


  \ref{ASect:Projectors_Variation}. 
  & \hyperref[ASect:Projectors_Variation]{Moore-Penrose inverse and variation of
projectors\hfill}
  && \tocPNumHSPACE \PGRef{ASect:Projectors_Variation}
  & \tocASectVSPACE
  \\

  \ref{ASect:Gauge_Struct_Restriction}. 
  & \hyperref[ASect:Gauge_Struct_Restriction]{Gauge algebra of the restricted theory from a
parental theory\hfill}
  && \tocPNumHSPACE \PGRef{ASect:Gauge_Struct_Restriction}
  & \tocASectVSPACE
  \\

  \ref{ASect:Determinant_Relation}. 
  & \hyperref[ASect:Determinant_Relation]{Determinant relation\hfill}
  && \PGRef{ASect:Determinant_Relation}
  & \tocASectVSPACE
  \\

 {}
 & \hyperref[Sect:References]{References\hfill}
 && \PGRef{Sect:References}
 \\

\end{tabular*}
}\fi

\if{
{
    \phantom{.}
    \newpage
    \ref{Sect:Introduction}. 
    \hyperref[Sect:Introduction]{Introduction} \hfill
    2
    \\

    \ref{Sect:BV_EA_1stReducible}. 
    \hyperref[Sect:BV_EA_1stReducible]{Effective action for theories with first-stage reducible gauge generators} \hfill
    3

    \leftskip 10pt
    \hyperref[SSect:Reducible_BV_Setup]{A}.
    \hyperref[SSect:Reducible_BV_Setup]{Batalin-Vilkovisky formalism for reducible gauge theories} \hfill
    3

    \hyperref[SSect:StatSurface_and_RegularityCond]{B}.
    \hyperref[SSect:StatSurface_and_RegularityCond]{Stationary point of the gauge-fixed master
    action} \hfill
    6

    \hyperref[SSect:BV_EA]{C}.
    \hyperref[SSect:BV_EA]{One-loop contribution to the generating functional} \hfill
    8

    \hyperref[SSect:BV_WardIdentities]{D}.
    \hyperref[SSect:BV_WardIdentities]{Ward identities and gauge independence of the effective action} \hfill
    8
    \\

    \leftskip 0pt

    III. Reducible gauge structure of restricted gauge
    theories \hfill 9

    \leftskip 10pt

    A. Two representations of a restricted theory \hfill 9

    B. Gauge symmetry and reducibility \hfill 11

    C. Gauge algebra of a restricted theory \hfill 12
    \\

    IV. One-loop effective action of restricted gauge
    theory \hfill 13

    A. Two representations of the one-loop effective
    action \hfill 13

    B. Independence of projector parameter \hfill 15

    C. Canonical normalization of generators \hfill 15

    D. The difference between parental and restricted
    theories in the one-loop approximation \hfill 18
    \\

    V. Unimodular gravity theory \hfill 19

    A. Gauge fixing and propagators \hfill 20

    B. Reduction of functional determinants \hfill 22
    \\

    VI. Conclusion \hfill 24
    \\

    Acknowledgements \hfill 26
    \\

    A. Moore-Penrose inverse and variation of
    projectors \hfill 26
    \\

    B. Gauge algebra of the restricted theory from a
    parental theory \hfill 27
    \\

    C. Determinant relation \hfill 28
    \\

    References \hfill 29
}
}\fi

\if{
\newpage
\setcounter{tocdepth}{2}
\tableofcontents
}\fi



\newpage
 \section{Introduction}
 \label{Sect:Introduction}
The purpose of this paper is to discuss an interesting property that arises when one restricts a configuration space of gauge theory. This is the origin of a reducible gauge structure characterized by linear dependence of the generators of gauge transformations. Both mechanisms --- reduction of the configuration space and reducibility of gauge symmetry --- are widely known and applied in various areas of field theory. In the gravity theory context one of the most important examples of the field space restriction is unimodular gravity (UMG) theory which was suggested by Einstein soon after the invention of general relativity \cite{Einstein_UMG}. Much later it was considered in the context of particle physics \cite{Bij-vanDam-Ng}, in the context of a spacetime covariant formulation \cite{Henneaux_T}, as a problem of time and the cosmological constant problem \cite{Unruh}, and then applied within the dark energy paradigm \cite{Shaposhnikov} with the emphasis on purely technical issues of perturbation theory, etc. Extension beyond the unimodular constraint on the metric of spacetime --- the so-called generalized unimodular gravity\HIDE{ (GUMG)} \cite{Barvinsky:2017pmm} --- also turned out to be rather fruitful from the viewpoint of the generation of viable dark energy and inflation scenarios \cite{Barvinsky:2019agh,Barvinsky:2019qzx}.

From a field-theoretical point of view UMG is interesting because it is generally observationally equivalent to general relativity (see recent review \cite{Carballo-Rubio:2022ofy}), but even semiclassically raises the issues of such an equivalence \cite{Alvarez:2005iy} depending on the subtleties of local physics vs global behavior encoded in boundary conditions, finiteness of the spacetime volume, thermodynamical setup in gravity theory \cite{Fiol:2008vk}, etc. Unimodular gravity is an interesting object of group theoretical analysis \cite{Alvarez:2006uu} and quantization \cite{deLeonArdon:2017qzg,Percacci:2017fkn,Percacci:2017fsy,Kugo:2022iob, Kugo:2022dui}, especially regarding its relation to quantization of Einstein theory \cite{Bufalo:2015wda} as a sort of a parental theory whose field space reduction leads to the unimodular gravity model.

On the other hand, it has been conjectured \cite{Barvinsky:2017pmm} that the formalism of quantized unimodular and generalized unimodular gravity theories can be developed along the lines of gauge theory quantization in models with linearly dependent (or reducible) generators \cite{Batalin:1983ggl,Batalin:1983pz}. This observation follows from a simple fact that the restriction of the configuration space of metrics to the subspace of metrics with a unit determinant results in the relevant reduction in the space of diffeomorphism invariance parameters, which can be attained by a projection procedure. Then, if one wants (for the sake of retaining manifest covariance) to describe the restricted theory in terms of the parental one, this projection applied to the original gauge generators makes them linearly dependent.

To the best of our knowledge this relation between the reducible nature of gauge generators and the reduction of the field space has not yet been fully exploited even in the unimodular gravity context, as well as its extension to more or less general gauge theories. The studies of theories with restrictions on gauge transformations, the theories with the so-called unfree gauge parameters, were recently intensively conducted in a series of publications
\cite{Kaparulin:2019quz,Lyakhovich:unfree-hamiltonian}
with the purpose of constructing the BFV (Batalin-Fradkin-Vilkovisky) formalism transcending these restrictions to the ghost sector of the model and starting from the first principles of Hamiltonian formalism and canonical quantization. Here we choose a somewhat different approach (closer to the Lagrangian formalism) and try to develop the idea of \cite{Barvinsky:2017pmm} for a generic parental gauge field theory with the closed algebra of its irreducible generators becoming linearly dependent in the course of a projection induced by the restriction of the original configuration space.

Our method is based on the explicit construction of a projector operator with which the generators of the original parental theory are projected onto the generators of the residual symmetries of the restricted theory. Linear dependence of the latter is then treated by the quantization method of Batalin and Vilkovisky (BV) for reducible gauge theories \cite{Batalin:1983ggl,Batalin:1983pz} --- a brief review of this formalism with all necessary notations and definitions is presented in Sec.\,\ref{SSect:Reducible_BV_Setup}. 
 As a by-product of using BV formalism we develop in much detail the gauge-fixing procedure for gauge theory of the first-stage reducibility. Quite interestingly, despite numerous applications of this method, current literature does not present a workable algorithm even for the one-loop approximation in generic reducible gauge theories. Implicitly this algorithm is, of course, contained in the pioneering works \cite{Batalin:1983ggl,Batalin:1983pz}, but its concrete realization with all the details of Ward identities providing gauge independence of the on-shell effective action is still missing. We close this omission and suggest the recipe for all the elements of Feynman diagrammatic technique and the relevant algorithm of the one-loop effective action in the above class of theories at the level of presentation characteristic of a folklore use of the well-known Faddeev-Popov prescription.

The gauge-fixing algorithm here turns out to be more involved than in a conventional Faddeev-Popov formalism for irreducible theories, because it includes gauge fixing for original gauge fields, gauge fixing for ghosts and the corresponding ghosts for ghosts. Even for first-stage reducible theories\HIDE{, for which the hierarchy of structure functions of the gauge algebra terminates at the first step} (when there is no higher-order zero vectors for reducibility generators --- zero vectors of original gauge generators)\HIDE{,} this gauge-fixing procedure involves, apart from the usual gauge conditions, at least three extra gauge-fixing elements and the construction of a special projector in the gauge-breaking term of the theory. The main result for the one-loop effective action of gauge theories of the first-stage reducibility is assembled in the sequence of Eqs.\,(\ref{Z_1loop_1st})--(\ref{Xalphai}) of Sec.\,\ref{SSect:BV_EA}.

Then in Sec.\,\ref{Sect:Restricting_and_Reducibility} we explain how a reducible gauge structure with linearly-dependent generators arises in theories with restricted configuration space of gauge fields. In particular, we raise the question why and when constraining this space can be regarded merely as a partial gauge fixing, or on the contrary, forming a new physical theory inequivalent to the parental one. It turns out that the answer is based on the rank of a special \emph{gauge-restriction operator}. The rank deficiency of this operator, or the presence of its nontrivial zero modes, signifies new physics of a restricted theory as compared to the parental model whose field space is subject to restriction. This gauge-restriction operator allows one to construct the projector on the space of residual gauge symmetries. The projected gauge generators form the algebra which is closed as long as the original parental theory has a closed algebra both on and off shell.

The zero vectors of this projector become first-stage reducibility generators of the BV method of treatment of the restricted theory, and they represent a free element of the formalism, whose choice is limited only by a certain rank restriction condition.  It turns out, however, that their special normalization fully provides on-shell independence of the one-loop effective action of this choice, whereas the normalization of the gauge generators of the parental theory can be fixed by matching it with the canonical formalism and canonical quantization of parental theory.

The one-loop effective action for generic restricted theory is discussed in Sec.\,\ref{SSect:1loop_EA_Restricted}. Discussion of correct normalization of the gauge generators in a covariant formalism is the subject of Secs.\,\ref{SSect:Projector_Param_Independence} and
\ref{SSect:Canon_Measure_Normalization}. In Sec.\,\ref{SSect:1loop_EA_Difference} we discuss the factor in one-loop effective action which distinguishes restricted and parental theories at the quantum level.

In Sec.\,\ref{Sect:UMG_EA} we apply the above formalism to {the} unimodular gravity theory treated as descending from its parental version --- Einstein general relativity. We construct all the elements of its gauge-fixing procedure in the background covariant gauge \cite{DeWitt} with the background covariance property extended to all the objects of the Feynman diagrammatic technique. Then, according to the general algorithm, we build on shell the one-loop effective action of the UMG theory. This confirms conclusions of \cite{Percacci:2017fsy} which were attained, however, with certain assumptions on the treatment of the group of diffeomorphisms volume factored out of the partition function. No such assumptions are needed in our approach which is fully determined by the requirement of gauge independence of the on-shell physical results.

Moreover, our approach allows us to disentangle the one-loop contribution of a global zero mode responsible for the ``new'' physics contained in the UMG theory as compared to general relativity. The UMG one-loop effective action coincides with that of the Einstein theory having a cosmological constant modulo a special contribution of the global degree of freedom associated with the variable value of this constant. The resulting expression for the effective action, which is usually presented in terms of functional determinants on irreducible (transverse and transverse-traceless) subspaces of the full tensor and vector field space, is built in terms of the functional determinants of the minimal operators. These determinants admit for their calculation the application of the Schwinger-DeWitt curvature expansion \cite{Schwinger-DeWitt}. This property is important for calculations on generic backgrounds not exhausted by homogeneous spaces.

In the Conclusion we summarize the results and briefly discuss the role of a global extra degree of freedom in the UMG theory as a cosmological constant. In particular, we show how this reveals the duality of UMG and Einstein theories analogous to the transition between statistical ensembles with fixed observables related by the Laplace duality transformations. In the appendixes we discuss the Moore-Penrose construction of inverse operators for degenerate operators \cite{MoorePenrose}, as well as the derivation of the gauge algebra of the projected generators and some technical details of determinant relations needed for the construction of the generator basis.

 \section{Effective action for theories with first-stage reducible gauge generators}
  \label{Sect:BV_EA_1stReducible}
In this section we briefly review Batalin-Vilkovisky (BV) formalism for the first-stage reducible gauge theories. Throughout this section (and in the most of other sections unless otherwise indicated) we use DeWitt condensed notations. All indices are of condensed nature and combine discrete bundle and continuous spacetime{-}point labels. Two-index quantities are two-point kernels and summation over contracted indices implies the corresponding spacetime integration. Ranges of condensed indices \HIDE{to be summed and integrated over formally{,} }reflect the continuum of spacetime points and the discrete spin-tensor labels of fields. The ranks of two-point kernels of linear operators and forms thus refer to their functional space dimensionality.

   \subsection{Batalin-Vilkovisky formalism for reducible gauge theories}
    \label{SSect:Reducible_BV_Setup}
Consider a generic gauge theory of fields $\phi^\zi$ (with the range of field indices $\zi$ formally denoted in what follows by $n$). Let it be described by the action $S[\,\phi\,]$ which is invariant under the local gauge transformations with generators $\mathcal{R}^\zi_\cza=\mathcal{R}^\zi_\cza(\phi)$ and infinitesimal gauge parameters $\xi^\cza$ (the range of gauge indices $\cza$ being denoted by  $m_0<n$)
 \bea{}
  &&\var_\xi \phi^\zi
  = \mathcal{R}^\zi_\cza \xi^\cza,
  \label{GaugeTransfsLagrange}\\
  &&S_{,\zi} \mathcal{R}^\zi_{\cza}
  =0,
  \;\quad\;
  S_{,\zi}\equiv\frac{\var S}{\var \phi^\zi}. \label{NoetherId}
 \eea
Here, and in what follows, comma denotes a functional derivative with respect to the relevant field variable.

The algebra of gauge generators begins with the relations --- corollaries of the Jacobi identity for a double commutator of gauge transformations
 \beq{Jacobi_Id_Lagrange}
   \mathcal{R}^\zi_{\cza,\zj} \mathcal{R}^\zj_{\czb} -  \mathcal{R}^\zi_{\czb,\zj} \mathcal{R}^\zj_{\cza}
  \,=\,  \mathcal{R}^\zi_{\czc} C^{\czc}_{\cza\czb} + E^{\zi\zj}_{\cza\czb}S_{,\zj}
 \eeq
with the structure functions $ C^{\czc}_{\cza\czb}(\phi) = - C^{\czc}_{\czb\cza}(\phi)$, $E^{\zi\zj}_{\cza\czb} (\phi) = - E^{\zi\zj}_{\czb\cza} (\phi) = - E^{\zj\zi}_{\cza\czb} (\phi)$. For closed algebras (in contrast to so-called open algebras) $E^{\zi\zj}_{\cza\czb} (\phi)$ and higher-order structure functions of the gauge algebra vanish \cite{Henneaux:1992ig,Gomis:1994he}.

The theory is \emph{reducible} when its gauge generators are linearly dependent on shell, that is on solutions of classical equations $S_{,\zi}=0$, so that there exist $m_1 $ reducibility generators $\mathcal{Z}^\cza_{\,\ccza}=\mathcal{Z}^\cza_{\,\ccza}(\phi)$ --- right zero-eigenvalue eigenvectors of the original generators $\mathcal{R}^\zi_\cza$ {parametrized} by indices $\ccza$ of range $m_1$. This implies relations
 \beq{ReducibilityId}
  \mathcal{R}^\zi_\cza \mathcal{Z}^\cza_{\,\ccza}
  \,=\, - B^{\zi\zj}_{\,\ccza} S_{,\zj}
 \eeq
with some coefficient functions $B^{\zi\zj}_{\,\ccza}(\phi)=-B^{\zj\zi}_{\,\ccza}(\phi)$. For the {\em first-stage} reducible theories, which we consider in this section,  the  generators $\mathcal{Z}^\cza_{\,\ccza}$ form on shell a {complete} independent set, whereas for higher-order stages of reducibility they also become linearly dependent with higher-order reducibility generators and so on.

    \auxtitle{Master action}

\newcommand{\PhiBV}{\varPhi\vphantom{\big|}_{\scriptscriptstyle{\!BV}}} 
Batalin-Vilkovisky quantization of gauge theories starts with the construction of \emph{master action}
$S^{\scriptscriptstyle{BV}}[\,\varPhi, \varPhi^*]$ which is a certain extension of the original action classical $S[\,\phi\,]$ onto the configuration space of fields and antifields $(\varPhi^\ZI, \varPhi^*_\ZI)$. This even-dimensional space is $\Z$-graded with respect to the so-called ghost number $\gh{\varPhi^*_\ZI}=-\gh{\varPhi^\ZI}-1$ and $\Z_2$-graded with respect to their Grassmann parity $\epsilon$, \,$\epsilon(\varPhi^*_\ZI)=\epsilon(\varPhi^\ZI)+1$. Original fields $\phi^\zi$ become a part of extended fields set $\varPhi^\ZI$ and carry zero ghost number. To avoid messy sign factors we assume that all $\phi^\zi$ are bosonic (even) fields, $\epsilon(\phi^\zi)=0$, and Grassmann parity of ghost fields and antifields coincides with $\Z_2$ parity of their ghost number (those with even ghost numbers are commuting fields, those with odd ghosts number --- anticommuting).

Extension from $S[\,\phi\,]$ to $S^{\scriptscriptstyle{BV}}[\,\varPhi, \varPhi^*]$ is governed by the \emph{master equation} on $S^{\scriptscriptstyle{BV}}[\,\varPhi, \varPhi^*]$
 \beq{masterequation} 
  \AB{S^{\scriptscriptstyle{BV}}}{S^{\scriptscriptstyle{BV}}}
  \,\equiv\,
   \frac{{\var^{(r)}} S^{\scriptscriptstyle{BV}}} {\var {\varPhi^\ZI}}
      \frac{{\var^{(l)}} S^{\scriptscriptstyle{BV}}} {\var {\varPhi^*_\ZI}}
     -
      \frac{{\var^{(r)}} S^{\scriptscriptstyle{BV}}} {\var {\varPhi^*_\ZI}}
      \frac{{\var^{(l)}} S^{\scriptscriptstyle{BV}}} {\var {\varPhi^\ZI}}
      \,=\, 0
 \eeq
in terms of the {antibracket} $(G,H)$ \cite{Batalin:1981jr} which is defined for any two functionals $G$ and $H$ of fields and antifields. Here $ (r) $ and $ (l)$ label, respectively, right and left derivatives.

The solution $S^{\scriptscriptstyle{BV}}[\,\varPhi, \varPhi^*]$ is obtained as a perturbative expansion in powers of ghost fields and antifields. To make this solution proper and allow one to perform a further gauge-fixing procedure one should specify a BV-extended configuration space and impose certain initial conditions to (\ref{masterequation}) which should explicitly encode the gauge structure of the original action $S[\,\phi\,]$.

Any proper solution of master equation is inevitably gauge invariant
\cite{Batalin:1983ggl,Henneaux:1992ig,Batalin:1981jr,Henneaux:1989jq,Gomis:1994he,BV3}. In terms of the total set of BV fields and antifields $\PhiBV^\ZA = (\varPhi^\ZI,\varPhi^*_\ZI)$, $\ZA=1,..,2N$, $\ZI=1,..,N$, the master equation reads
 \beq{masterequation2}
  \frac{{\var^{r}} S^{\scriptscriptstyle{BV}} } {\var \PhiBV^\ZA}
    \zeta^{\ZA\ZB}
  \frac{{\var^{l}} S^{\scriptscriptstyle{BV}}} {\var \PhiBV^\ZB}
  = 0,\;\quad\;
  \zeta^{\ZA\ZB} =\left[
                   \begin{array}{cc}
                     0 & \;\;\;1 \;\\
                     -1 & \;\;\; 0 \;\\
                   \end{array}
                 \right].
 \eeq
By differentiating (\ref{masterequation2}) one gets the Noether identities
 \beq{NoetherIdmaS}
  \frac{{\var^{r}} S^{\scriptscriptstyle{BV}}} {\var \PhiBV^\ZA}
   \mathcal{R}^\ZA_{\,\ZC}
  = 0,
  \;\quad\;
  \mathcal{R}^\ZA_{\,\ZC}
   =
   \zeta^{\ZA\ZB}
  \frac{\var^{l} \var^{r} S^{\scriptscriptstyle{BV}}} {\var \PhiBV^\ZB  \var \PhiBV^\ZC}
  ,
 \eeq
featuring $2N$ gauge generators $\mathcal{R}^\ZA_{\,\ZC}$ of master action gauge symmetry.

These generators, however, form a reducible set because
the second variational derivative of (\ref{masterequation2})
shows their $N$-reducibility on shell,
 $
  \zeta^{-1}_{\ZD\ZB}
  \mathcal{R}^\ZB_{\,\ZA} \mathcal{R}^\ZA_{\,\ZC}\propto\var S^{\scriptscriptstyle{BV}}/\var \PhiBV^\ZA =0
 $.
Thus, the master action $S^{\scriptscriptstyle{BV}}$
possess $N$ on-shell independent symmetries --- half the dimension of the BV configuration space $(\varPhi^\ZI,\varPhi^*_\ZI)$.
\footnote{
 Nilpotence of $2N \times 2N$ matrix implies that its rank is at most $N$. Considering a \emph{proper} solution implies that rank of the Hessian $ \var^{l} \var^{r} S^{\scriptscriptstyle{BV}}/\var \PhiBV^\ZA \var \PhiBV^\ZB$ equals $N$. Otherwise the set of Noether identities (\ref{NoetherIdmaS}) is incomplete and there are more then $N$ gauge generators since there are more then $N$ zero modes of the Hessian.
}

Convenient gauge fixing of the master action gauge symmetry (\ref{NoetherIdmaS})
consists in expressing antifields in terms of fields via $N$ gauge conditions
\cite{Batalin:1983ggl,Henneaux:1992ig,Batalin:1981jr,Henneaux:1989jq,Gomis:1994he,BV3}
 \beq{GaugeFixingConditions} 
  \varPhi^*_\ZI - \frac{\var \varPsi[\,\varPhi\,]}{\var \varPhi^\ZI}
  = 0 .
 \eeq
It is {parametrized} by a single functional of fields --- gauge fermion $\varPsi[\,\varPhi\,]$ of ghost number $-1$ and leads to gauge-fixed action
 \beq{GaugeFixedAction}
  S_\varPsi[\,\varPhi\,]
  = S^{\scriptscriptstyle{BV}}\big[\,\varPhi\hspace{1pt}, \var \varPsi[\hspace{1pt}\varPhi\hspace{1pt}]/\var \varPhi\,\big] ,
 \eeq
which is functional of only BV fields $\varPhi^\ZI$. For the correctly chosen gauge fermion the  gauge-fixed action (\ref{GaugeFixedAction}) appears to be a nondegenerate
 $
  \mathrm{sdet}\,[\,\var^{(l)}\var^{(r)} S_\varPsi/\var \varPhi^\ZI\,\var \varPhi^\ZJ \,]_{\,\var S_\varPsi/\var \varPhi=0}
  \neq 0
 $
and yields a perturbatively consistent path integral for the generating functional which modulo the contribution of local measure reads
 \beq{GeneratingFunctional}
  Z = \int D\varPhi\; e^{\, i\, S_{\varPsi}[\,\varPhi\,]} .
 \eeq

The master equation for $S^{\scriptscriptstyle{BV}}[\,\varPhi, \varPhi^*]$ (\ref{masterequation}) provides the corner stone of gauge theory quantization --- independence of the on-shell generating functional of the choice of gauge fermion. Under the gauge fermion change $\varPsi\to\varPsi+\Delta\varPsi$, in view of integration by parts the change in $Z$ reads
 \bea{}
  \Delta Z
  &=&
  -\int D\varPhi\,\Delta\varPsi\,\left(\,i\,
  \frac{\var^{(r)}S^{\scriptscriptstyle{BV}}}{\var\varPhi^\ZI} \frac{\var^{(l)}S^{\scriptscriptstyle{BV}}}{\var\varPhi^*_\ZI}
  \right.\nonumber\\
  &&\quad\left.
  \,+\,\frac{\var^{(r)}}{\var\varPhi^\ZI}
  \frac{\var^{(l)}S^{\scriptscriptstyle{BV}}}{\var\varPhi^*_\ZI}\right) \Big|_{\,\varPhi^* = \frac{\var\varPsi}{\var\varPhi}}
  \,e^{\,i\,S_\varPsi},
\eea
where extra terms vanish in view of Grassmann symmetry properties. This expression equals zero because the first term vanishes in view of the master equation $\var^{(r)}\! S^{\scriptscriptstyle{BV}}\!\!/\var\varPhi^*_\ZI \:  \var^{(l)}\! S^{\scriptscriptstyle{BV}}\!\!/\var\varPhi^\ZI =\tfrac12(S^{\scriptscriptstyle{BV}},S^{\scriptscriptstyle{BV}})$, while the second term in local theories is proportional to $\partial...\partial\,\delta(0)$ and is supposed to be either canceled by a local measure or killed by dimensional regularization. The local measure is not rigorously available within the BV formalism whose incompleteness can be disregarded for theories with local gauge algebra and within the class of local gauge fermions. This measure can be attained within the canonical BFV quantization formalism, which will be used below to justify the application of the BV method for {\em restricted} gauge theories in which locality of their generators will be spoiled by generically nonlocal projectors.

    \auxtitle{Minimal and nonminimal proper solutions}

Construction of BV master action begins with finding a proper solution of master equation (\ref{masterequation}) on the \emph{minimal} sector of fields and antifields. For the first-stage reducible theories  (\ref{NoetherId}),(\ref{ReducibilityId}) this requires introduction of ghosts $\gC^\cza$, $\gh{\gC^\cza}=1$, the ghosts for ghosts $\gC^\ccza$, $\gh{\gC^\ccza}=2$, and antifields to all fields and ghosts \cite{Batalin:1983ggl,Henneaux:1992ig,Gomis:1994he},
 \beq{ConfigSpaceMin}
  \varPhi_{\rm min}=(\phi^\zi,\gC^\cza,\gC^\ccza ),
  \quad
  \varPhi^*_{\rm min}=(\phi^*_\zi,\gC^*_\cza,\gC^*_\ccza)
 \eeq

The \emph{minimal} proper solution $S^{\rm min}[\,\varPhi_{\rm min}, \varPhi^*_{\rm min}\,]$ of the master equation can be represented as the series in powers of antifields
 \beq{maSminExpr}
  S^{\rm min} \,=\, S + \phi^*_\zi \mathcal{R}^\zi_\cza \gC^\cza +  \gC^*_\cza \mathcal{Z}^\cza_{\,\ccza} \gC^\ccza + ...
 \eeq
where $S=S[\,\phi\,]$ is the action of the initial gauge theory, two terms bilinear in ghost fields and antifields serve as initial conditions for the master equation which guaranty that such a solution of master equation is proper, $\mathcal{R}^\zi_\cza $ and $\mathcal{Z}^\cza_{\,\ccza}$ are gauge and reducibility generators (\ref{NoetherId}),(\ref{ReducibilityId}), dots stand for higher-order terms in powers of ghost fields and antifields. All higher-order terms can be found within the iterative procedure of solving the master equation which leads to \cite{Henneaux:1992ig,Gomis:1994he}
 \bea{maSminExpr_ext}
   S^{min} 
  &
   \,=\,
  &
    S 
   \,+
    \phi^*_\zi \mathcal{R}^\zi_\cza \gC^\cza
   +
    \gC^*_\cza
     \left( \mathcal{Z}^\cza_{\,\ccza} \gC^\ccza
        +
        \tfrac12 C_{\czb\czc }^\cza
        \gC^\czc\gC^\czb
     \right)\nonumber\\
     &&
   +
    \gC^*_\ccza
     \left( A_{\cczb\cza}^\ccza \gC^\cczb \gC^\cza
       +
        \tfrac12 F_{\cza \czb \czc }^\ccza
        \gC^\czc \gC^\czb \gC^\cza
     \right)
  \nonumber\\
  &
  &
   +\phi^*_\zi\phi^*_\zj
    \left( {\tfrac12  B_\ccza^{\zj\zi}\gC^\ccza
       +
       \tfrac14 E_{\cza \czb }^{\zj\zi}
       \gC^\czb \gC^\cza }
    \right)
  \nonumber
   + \ldots\,,
 \eea
where the coefficient functions of higher-order terms originate from higher-order structure functions of gauge algebra and the dots denote terms with antifields of the total ghost number $-3$ and lower.

Meaningful gauge fixing of the BV master action requires extension of the proper solution to nonminimal configuration space. Motivation for this is that gauge fixing (\ref{GaugeFixingConditions})
is performed through properly constructed \emph{gauge fermion} $\varPsi[\,\varPhi\,]$ --- the functional of only the BV fields $\varPhi^\ZI$. But the gauge fermion has a negative ghost number $-1$  and thus cannot be constructed as a functional of the minimal fields  $\phi^\zi,\gC^\cza,\gC^\ccza$ with nonnegative ghost numbers. Therefore, the set of fields should be further extended to contain auxiliary ghosts with negative ghost numbers.

Standard scheme for first-stage reducible theories presumes introducing auxiliary ghosts $\agC_\cza, \agC_\ccza, \egC^\ccza$, their partners $\agB_\cza, \agB_\ccza, \egB^\ccza$ with the higher ghost number and the corresponding set of antifields \cite{Henneaux:1992ig, Gomis:1994he}, so that nonminimal BV configuration space reads
 \bea{ConfigSpaceNonmin}
  \hspace{-5mm}\varPhi &=& ( 
           \underbrace{\phi^\zi,\gC^\cza,\gC^\ccza}_{\varPhi_{\rm min}},
          \agC_\cza, \agC_\ccza, \egC^\ccza
          ,\agB_\cza, \agB_\ccza, \egB^\ccza ),
 \nonumber\\
  \hspace{-5mm}\varPhi^* &=&
  (\underbrace{\phi^*_\zi,\gC^*_\cza,\gC^*_\ccza}_{\varPhi^*_{\rm min}},
           {\agC^*}^\cza, {\agC^*}^\ccza, \egC^*_{\,\ccza},
           {\agB^*}^\cza\!, {\agB^*}^\ccza\!, {\egB^*}_{\!\!\!\ccza}).
 \eea
Here the fields $\agC_\cza, \agC_\ccza$ are often referred to as antighosts, $\egC^\ccza$ --- extraghosts, $\agB_\cza, \agB_\ccza, \egB^\ccza$ --- Lagrange multiplier or Nakanishi-Lautrup fields.

Ghost numbers of the nonminimal configuration space of the first-stage reducible gauge theory are listed in Table \ref{Table:BVFieldContent}, where antifields ${\agB^*}^\cza, {\agB^*}^\ccza, {\egB^*}_{\!\!\!\ccza}$ are omitted since they do not appear in the BV procedure.
\def \TblGhNum
{
  \begin{tabular}{c||c|c c||c|c c||}
    \cline{2-7}
    & \quad$\varPhi$\quad \vlv & $\gh{\varPhi}$ && \quad$\varPhi^*$\!\quad \vlv & $\gh{\varPhi^*}$ &  \\
    \cline{2-7}
    \cline{2-7}
    \multirow{10}{0cm}{} 
    & \multicolumn{6}{|c||}{\textit{minimal sector}} \\
    \cline{2-7}
    \cline{2-7}
    \vphantom{{$I^{^I}$}}
    &  $\phi^\zi$   & $0$  &
    & $\phi^*_\zi$  & $-1$ &
    \\
    &  $\gC^\cza$    & $+1$  &
    & $\gC^*_\cza$   & $-2$  &
    \\
    &   $\gC^\ccza$   & $+2$  &
    & $\gC^*_\ccza$  & $-3$  &
    \\[0.5ex]
    \cline{2-7}
    & \multicolumn{6}{|c||}{\textit{auxiliary sector}} \\
    \cline{2-7}
    \cline{2-7}
    \vphantom{{$I^{^I}$}}
    &  $\agC_\cza$    & $-1$  &
    &  ${\agC^*}^\cza$   & $0$   &
    \\
    &  $\qquad\agB_\cza$    & $\qquad 0$   &
    &    &   &
    \\
    &  $\agC_\ccza$   & $-2$  &
    & ${\agC^*}^\ccza$  & $+1$  &
    \\
    &  $\qquad\agB_\ccza$    & $\qquad -1$   &
    &    &   &
    \\
    &   $\egC^\ccza$   & $0$   &
    & $\egC^*_\ccza$  & $-1$  &
    \\
    &  $\qquad\egB^\ccza$    & $\qquad +1$   &
    &    &   &
    \\[0.5ex]
    \cline{2-7}
  \end{tabular}
}
\begin{table}[h!]
 \centering
  \TblGhNum
 \caption{BV configuration space of the first-stage reducible gauge theory.}
 \label{Table:BVFieldContent}
\end{table}

The \emph{nonminimal} proper solution $S^{\scriptscriptstyle{BV}}[\,\varPhi,\varPhi^*]$ of the master equation on such space is just a sum of the minimal master action $S^{\rm min}[\,\varPhi_{\rm min},\varPhi^*_{\rm min}]$ (\ref{maSminExpr}) and the contribution of auxiliary fields $S^{\rm aux}[\agB_\cza, \agB_\ccza, \egB^\ccza, {\agC^*}^\cza, {\agC^*}^\ccza, \egC^*_\ccza]$
 \bea{}
  S^{\scriptscriptstyle{BV}}
   &=&  S^{\rm min} + S^{\rm aux}, \label{maSnonminExpr}\\
 S^{\rm aux}
   &=& \,\agB_{\cza} {\agC^*}^{\cza}
     + \,\agB_{\ccza} {\agC^*}^{\ccza}
     + \,\egC^*_{\ccza} \egB^{\ccza}. \label{maSaux1}
 \eea
The master action (\ref{maSnonminExpr}) obviously satisfies the master equation (\ref{masterequation}) on the nonminimal configuration space since auxiliary and minimal sectors are so far decoupled (these sectors will be mixed after gauge fixing) so that $(S^{\rm min} + S^{\rm aux},S^{\rm min} + S^{\rm aux})=(S^{\rm min},S^{\rm min})+(S^{\rm aux},S^{\rm aux})$, and by construction $S^{\rm min}$ and $S^{\rm aux}$ separately nilpotent in antibracket.

    \auxtitle{Choice of a gauge fermion for the Gaussian gauge fixing}

Grading restriction $\gh{\varPsi}=-1$ on gauge fermion $\varPsi[\,\varPhi\,]$ admits the following form at most quadratic in ghosts and auxiliary fields
 \bea{GaugeFermion_1st}
  \varPsi[\,\varPhi\,]
   \;&=&\;  \agC_\cza \big(\chi^\cza(\phi) + \sigma^{\cza}_{\,\ccza}(\phi)\, \egC^\ccza\big)
    + \agC_\ccza\, \omega^{\ccza}_{\,\cza}(\phi)\, \gC^\cza
  \nonumber\\
  &&
   + \tfrac12\, \big(
     \agC_\cza \varkappa^{\cza\czb} \agB_{\czb}
    + \agC_\ccza \rho^{\ccza}_{\,\cczb}  \egB^{\cczb}
    {+\,}
     \agB_{\ccza} \rho^{\ccza}_{\,\cczb}  \egC^{\cczb}
       \big).
 \;\;\;\;
 \eea 
The latter group of terms linear in $\agB_\cza, \agB_\ccza, \egB^\ccza$ allows us to  implement \emph{Gaussian} (Faddeev-Popov) type of gauge fixing when $\varkappa^{\cza\czb}$, $\rho^{\ccza}_{\,\cczb}$ are \emph{nondegenerate}
 \beq{}
  \det{\varkappa^{\cza\czb}} \neq0,    \qquad
  \det{\rho^{\ccza}_{\,\cczb}} \neq0.
 \eeq
In generic theories it is convenient to choose $\phi$-independent coefficient functions $\varkappa^{\cza\czb}$, $\rho^{\ccza}_{\,\cczb}$. Such a choice guarantees the possibility to integrate out auxiliary fields  $\agB_\cza, \agB_\ccza, \egB^\ccza$ to obtain Faddeev-Popov representation for the generating functional.

Here we take into account that for $\phi$-dependent functions $\varkappa^{\cza\czb}(\phi), \rho^{\ccza}_{\,\cczb}(\phi)$ terms linear in companion fields $\agB_\cza, \agB_\ccza, \egB^\ccza$ will arise in gauge-fixed minimal master action $S_\varPsi^{\rm min}$ instead of each antifield, $\phi^*_\zi=\var\varPsi/\var\phi^\zi=\tfrac12{\mathcal C}_\cza(\var\varkappa^{\cza\czb}/\var\phi^\zi)\,\agB_\czb+...\,$. Therefore, for open algebras and algebras with higher-structure functions, whose $S^{\rm min}$ contains second and higher powers of antifields, $\agB_\cza, \agB_\ccza, \egB^\ccza$ dependence of $S_\varPsi^{\rm min}$ becomes more than quadratic, which does not allow us to integrate them out via Gaussian integration.

After the gauge fixing (\ref{GaugeFixedAction}), $\varPhi^*\to\var\varPsi/\var\varPhi$, with the gauge fermion (\ref{GaugeFermion_1st}) one gets the nonminimal gauge-fixed action (\ref{GaugeFixedAction}) which according to (\ref{maSnonminExpr}) is $S_\varPsi^{} =  S_\varPsi^{\rm min}+S_\varPsi^{\rm aux}$ with
  \bea{}
   \hspace{-1cm}
    S^{\rm min}_{\varPsi} 
   &=&\;
    S +
    \agC_\cza \big( X^ {\cza}_{,\zi} \mathcal{R}^\zi_\czb\big) \gC^\czb
    \!+\!
    \agC_\ccza \big(\omega^{\ccza}_{\,\cza} \mathcal{Z}^\cza_{\,\cczb}\big) \gC^\cczb
    \!+ ...,
   \label{maSmin1GF_}\\
   \hspace{-1cm}
    S^{\rm aux}_{\varPsi} 
   &=&\;
    \tfrac12 \agB_{\cza}\varkappa^{\cza\czb} \agB_{\czb}
    + \agB_{\cza}  X^ \cza 
    +  \HIDE{1}\agB_{\ccza} \rho^{\ccza}_{\,\cczb} \egB^{\cczb}
   \nonumber\\
   &&+ \agB_{\ccza} \omega^{\ccza}_{\,\cza} \gC^\cza
    + \agC_\cza \sigma^{\cza}_{\,\cczb} \egB^{\cczb},
   \label{maSaux1GF_}
  \eea
where $S=S[\,\phi\,]$ is the action of the initial gauge theory and dots hide terms more then quadratic in ghost fields with nonzero ghost numbers.\footnote{
 Terms explicated in (\ref{maSmin1GF_}) and (\ref{maSaux1GF_}) are sufficient in the one-loop approximation. The gauge-fixed action with several higher-order terms reads \;
 \if{ \bea{maSmin1GF__} 
   &&\quad S^{\rm min}_{\varPsi} [\varPhi] =
    S [\phi]
   \!+\!
    \agC_\cza \big( X^ {\cza}_{,\zi} \mathcal{R}^\zi_\czb\big) \gC^\czb
   \!+\!
    \agC_\ccza \big(\omega^{\ccza}_{\,\cza} \mathcal{Z}^\cza_{\,\cczb}\big) \gC^\cczb \!\!+
      \agC_\ccza
     \big(\omega^{\ccza}_{\cza,\zi} \mathcal{R}^\zi_\czb
 \nonumber\\
 &&\qquad\qquad
    + \tfrac12\omega^{\ccza}_{\,\cza}
         C_{\czb\czc}^\cza \big)
        \gC^\czc\gC^\czb
   +  \agC_\cza  \agC_\czb \big(\tfrac12 X^ \cza_{,\zi}  X^ \cza_{,\zj}
      B_\ccza^{\zj\zi} \big) \gC^\ccza
   + \ldots\,\,.
 \nonumber
 \eea
 }\fi
  \( 
    S^{\rm min}_{\varPsi} [\,\varPhi\,]
    \,=\, 
    S [\,\phi\,]
    \,+\,
    \agC_\cza \big( X^ {\cza}_{,\zi} \mathcal{R}^\zi_\czb\big) \gC^\czb
    +
    \agC_\ccza \big(\omega^{\ccza}_{\,\cza} \mathcal{Z}^\cza_{\,\cczb}\big) \gC^\cczb
    +\,
     \agC_\ccza
     \big(\omega^{\ccza}_{\cza,\zi} \mathcal{R}^\zi_\czb + \tfrac12\omega^{\ccza}_{\,\cza}
         C_{\czb\czc}^\cza \big)
        \gC^\czc\gC^\czb
    + \agC_\cza  \agC_\czb \big(\tfrac12 X^ \cza_{,\zi}  X^ \cza_{,\zj}
      B_\ccza^{\zj\zi} \big) \gC^\ccza
    + \ldots\,\,.
  \) 
}
We introduce the combination 
  \beq{defX}
    X^ {\cza}
    \,\equiv\,
    \chi^\cza(\phi)  +  \sigma^{\cza}_{\,\cczc}(\phi)\, \egC^\cczc
  \eeq
which depends on fields $\phi^\zi,\egC^\ccza$ with zero ghost number.
All the dependence on $\egC$ in the gauge-fixed action $S_\varPsi^{}$ is now hidden within $ X^ \cza$ and its variational derivative
  \bea{X=chi}
    X^{\cza}_{,\zi}
    \,\equiv\, \var X^{\cza}/\var \phi^\zi
    \,=\, \chi^\cza_{\,,\zi} + \sigma^{\cza}_{\,\cczc,\zi} \egC^\cczc.
  \eea

    \auxtitle{Reduction of Lagrange multiplier fields}

For $\phi$-independent matrices $\varkappa^{\cza\czb}$ and $\rho^{\ccza}_{\,\cczb}$ --- the case which we consider in what follows --- Gaussian integration over the fields  $\agB_\cza, \agB_\ccza, \egB^\ccza$
in the generating functional (\ref{GeneratingFunctional}) gives
 \beq{redGeneratingFunctional}
  Z
  = \int
   D \varPhi_{\rm red}
  \; \big(\!\det \varkappa^{\cza\czb}\big)^{-1/2} \; {\det}\, \rho^{\ccza}_{\,\cczb}
  \; e^{\, i\,S^{\scriptscriptstyle{FP}}[\,\varPhi_{\rm red}] },
 \eeq
where $\varPhi_{\rm red}$ is the reduced nonminimal set of fields $\varPhi_{\rm red} = (\phi^\zi,\egC^\ccza,\gC^\cza,\agC_\cza,\gC^\ccza,\agC_\ccza)$, cf. Eq.\,(\ref{ConfigSpaceNonmin}), and $S^{\scriptscriptstyle{FP}}[\,\varPhi_{\rm red}]$ is the corresponding Faddeev-Popov reduced gauge-fixed action
 \bea{maSredGF}
  S^{\scriptscriptstyle{FP}}[\,\varPhi_{\rm red}]
  \,&=&\,
   S 
    \,- \tfrac12  X^\cza {\varkappa}_{\cza\czb} X^\czb
 \nonumber\\
 &&
   \,+\, \agC_\cza
         \big(  X^ {\cza}_{,\zi}
             \mathcal{R}^\zi_\czb  -  \sigma^{\cza}_{\,\ccza}\, \rho^{-1\,\ccza}_{\;\;\;\;\;\,\cczb}\, \omega^{\cczb}_{\,\czb}\big)\gC^\czb
 \nonumber\\
 &&
   +\, \agC_\ccza \big( \omega^{\ccza}_{\,\cza} \mathcal{Z}^\cza_{\,\cczb} \big) \gC^\cczb+ ...\,.
   \;\;\;\;\;
 \eea
Here we have introduced the matrices ${\varkappa}_{\cza\czb}$, ${\rho^{-1}}^{\ccza}_{\,\cczb}$ inverse respectively to $\varkappa^{\czb\cza}$, $\rho^{\cczb}_{\,\ccza}$. Dots denote terms which are more than quadratic in ghost fields $\gC^\cza,\agC_\cza,\gC^\ccza,\agC_\ccza$. Such terms are irrelevant in the one-loop approximation.

The obtained reduced action is nondegenerate in the sense that its Hessian at the stationary points of the classical BV action represents the nondegenerate operator
 \beq{redNondegeneracy} 
  \mathrm{sdet}\, {
   \frac{\var^{(l)}\var^{(r)} S^{\scriptscriptstyle{FP}}}{\var \varPhi_{\rm red} \, \var \varPhi_{\rm red}}
  } \Bigg|_{\,\var S^{\scriptscriptstyle{FP}}/\var \varPhi_{\rm red}=0}
  \!\!\!\!\!\!\!\!\!
   \neq 0
 \eeq
for the relevant set of boundary conditions. This property is inherited from the nondegeneracy of the unreduced gauge-fixed action $S_\varPsi[\,\varPhi\,]$ (with appropriately chosen gauge fermion){,} in which auxiliary nondynamical variables were integrated out (being expressed from their own equations of motion). For nondegenerate gauge-fixing matrices $\varkappa^{\czb\cza}$, $\rho^{\cczb}_{\,\ccza}$ \HIDE{companion }fields $\agB_\cza, \agB_\ccza, \egB^\ccza$ form the set of such auxiliary variables.

Now we will analyze nondegeneracy in various sectors of the reduced configuration space.

    \subsection{Stationary point of the gauge-fixed master action}
      \label{SSect:StatSurface_and_RegularityCond}

The simplest is the ghost sector consisting of the fields $\gC^\cza,\agC_\cza,\gC^\ccza,\agC_\ccza$ (the field $\egC^\ccza$ with a zero ghost number does not belong to this sector even though it is usually called the extraghost \cite{Batalin:1983ggl} --- rather it belongs to zero ghost number sector where it plays a special role in gauge-fixing procedure for $\phi^\zi$). The variational equations for the fields $\gC^\cza,\agC_\cza,\gC^\ccza, \agC_\ccza$ obviously make them vanishing under zero boundary conditions
  \bea{ghostSectorSolutions}
    \gC^\cza=  \agC_\cza =  \gC^\ccza  =  \agC_\ccza  =  0,
  \eea
provided the kernels of their bilinear forms in gauge-fixed action (\ref{maSredGF}) represent invertible operators having well defined Green's functions
  \bea{}
   && F^\cza_{\;\czb}=
            X^ {\cza}_{,\zi} \mathcal{R}^\zi_\czb  -  \sigma^{\cza}_{\,\ccza} {\rho^{-1}}^{\ccza}_{\,\cczb}\, \omega^{\cczb}_{\,\czb},\quad
            \det F^\cza_{\;\czb} \neq 0,\quad
   \label{gRegCond}
   \\
   && F^\ccza_{\;\cczb}
    =\omega^{\ccza}_{\,\cza} \mathcal{Z}^\cza_{\,\cczb},\quad
    \det F^\ccza_{\;\cczb}  \neq 0.
   \label{ggRegCond}
  \eea
The interpretation of these operators and their properties is obvious. The first term of the ghost operator $F^\cza_{\;\czb}$ is degenerate because of the reducibility of gauge generators, so the conventional Faddeev-Popov ghosts $\gC^\cza, \agC_\cza$ themselves become gauge fields with the local symmetry induced by the reducibility generators $\mathcal{Z}^\cza_{\,\cczb}$. The second term of $F^\cza_{\;\czb}$ in (\ref{gRegCond}) plays the role of gauge breaking term for this symmetry, and its effect is the invertibility of $F^\cza_{\;\czb}$. Correspondingly, $F^\ccza_{\;\cczb}$ is the Faddeev-Popov operator of new ghosts $\gC^\ccza, \agC_\ccza$ for original ghosts $\gC^\cza, \agC_\cza$ treated as gauge fields, that is why the fields $\gC^\ccza, \agC_\ccza$ bear the name of ghosts for ghosts.

The situation is trickier in the zero ghost number sector of fields $\phi^\zi,\egC^\ccza$. Their equations of motion read
  \bea{EoM0ghostSector}
    \frac{\var S^{\scriptscriptstyle{FP}}}{\var \phi^\zi}
   & = &
    S_{,\zi} - X^ {\cza}_{,\zi} {\varkappa}_{\cza\czb}  X^ \czb=  0,
   \label{EoMfq}\\
    \frac{\var S^{\scriptscriptstyle{FP}}}{\var \egC^\ccza}
   & = &
    - \sigma^{\cza}_{\,\ccza}  {\varkappa}_{\cza\czb}  X^ \czb=  0.
   \label{EoMaaggfc}
  \eea
Contracting them respectively with $\mathcal{R}^\zi_\czc$ and ${\rho^{-1}}^{\cczb}_{\,\ccza}\omega^{\ccza}_{\,\czc}$ and subtracting from one another we obtain, on account of the Noether identity $S_{,\zi}\mathcal{R}^\zi_\czc=0$, the following on-shell relation:
  \beq{Noether0ghost}
    \Big(X^ {\cza}_{,\zi}\mathcal{R}^\zi_\czc
    - \sigma^{\cza}_{\,\cczb} {\rho^{-1}}^{\cczb}_{\,\ccza}
      \omega^{\ccza}_{\,\czc} \Big)
    {\varkappa}_{\cza\czb}
     X^ \czb
    \;=\;
    F^\cza_{\;\czc}{\varkappa}_{\cza\czb}
    X^ \czb
    \,=\, 0,
  \eeq
which allows one in view of invertibility of $F^\cza_{\;\czc}$ and ${\varkappa}_{\cza\czb}$ to rewrite the equations of motion in the zero ghost number sector as
  \bea{}
    S_{,\zi}
   &=&  0,
   \label{SolutionCorrespondence}
   \\
    X^ \cza\!
   &\equiv&
    \chi^\cza \!+ \sigma^{\cza}_{\,\ccza} \egC^\ccza
    \,=\,  0.
   \label{Solution0ghostXX}
  \eea
This can be interpreted as equations of motion for $\phi^\zi$ supplied by the set of gauge conditions $X^ \cza=0$. The latter look overcomplete because the number of independent equations of motion $S_{,\zi}  =  0$ on $n$ variables $\phi^\zi$ is $n-m_0+m_1$, whereas the number of gauges is $m_0$. The $m_1$ mismatch is, however, corrected by the $m_1$ extra ghost fields $\egC^\ccza$. This goes as follows.

Contracting the rank $m_0$ nondegenerate operator $F^\cza_{\;\czb}$ with the rank $m_1$ full-rank generator $\mathcal{Z}^\czb_{\,\cczc}$ one gets the full-rank condition,
  $ \rank{ (\sigma^{\cza}_{\,\ccza}
       {\rho^{-1}}^{\ccza}_{\,\cczb}\,
       \omega^{\cczb}_{\,\czb} \mathcal{Z}^\czb_{\,\cczc} )}= m_1
  $,
which due to nondegeneracy of $F^\cczb_{\;\cczc}=\omega^{\cczb}_{\,\czb} \mathcal{Z}^\czb_{\,\cczc}$ and ${\rho^{-1}}^{\ccza}_{\,\cczb}$ implies the full-rank conditions $\rank{ \sigma^{\cza}_{\,\ccza}} = m_1$,
$\rank{ \omega^{\cczb}_{\,\czb}} = m_1$. This, in particular, guarantees that all $\egC^\ccza$ are expressible in terms of $\phi^\zi$ from the equations $X^ \cza=0$.

By introducing an arbitrary matrix ${s}^{\ccza}_{\,\cza}(\phi^\zi)$ such that $\det{({s}^\ccza_{\,\cza}\sigma^\czb_{\,\cczb})} \neq 0$ one can express the solution for $\egC^\ccza$ in terms of $\phi^\zi$ from Eq.\,(\ref{Solution0ghostXX})
  \beq{Solution0aaggfc}
    \egC^{\,\ccza}
   \,=\,
     - ({s}^{}_{} \sigma^{}_{})^{-1\,\ccza}_{\;\;\;\;\;\,\cczb}\, {s}^\cczb_{\,\czb} \,\chi^\czb ,
  \eeq
where $({s}^{}_{} \sigma^{}_{})^{-1\,\ccza}_{\;\;\;\;\;\;\,\cczb}$ is the inverse of $(s\sigma)^\cczb_{\,\cczc}={s}^{\cczb}_{\,\czc} \sigma^{\czc}_{\,\ccza}$. On the substitution of this solution in (\ref{Solution0ghostXX}) the rest of relations constitute $m_0-m_1$ independent gauge conditions
  \beq{Solution0fqAdd}
    \Big( \delta^\cza_{\,\czb}  - \sigma^\cza_{\,\ccza} ({s}^{}_{} \sigma^{}_{})^{-1\,\ccza}_{\;\;\;\;\;\,\cczb\,} {s}^{\cczb}_{\,\czb} \Big) \chi^\czb
   \,=\,  0 ,
  \eeq
on $n$ original fields $\phi^\zi$ as it should be. Indeed, what stands here as a matrix coefficient of $\chi^\czb$ is the projector $T^\cza_{\:\czb}(s,\sigma)$ in the space of gauge indices having $m_1$ right zero eigenvalue eigenvectors $\sigma^\czb_{\,\cczb}$ and $m_1$ left zero covectors $s^\ccza_{\;\cza}$,
  \bea{}
   && T^\cza_{\:\czb}(s,\sigma)
   \,=\,
    \delta^\cza_{\,\czb} - \sigma^\cza_{\,\ccza} ({s}^{}_{} \sigma^{}_{})^{-1\,\ccza}_{\;\;\;\;\;\;\,\cczb} {s}^{\cczb}_{\,\czb},
   \label{Tsigmas0}
   \\
   && T^\cza_{\:\czb}(s,\sigma)\,\sigma^\czb_{\,\cczb} \,=\, 0,
    \;\quad\;
    s^\ccza_{\;\cza}T^\cza_{\:\czb}(s,\sigma)
    \,=\, 0,
   \nonumber\\
   && T^\cza_{\:\czb}(s,\sigma)\,T^\czb_{\:\czc}(s,\sigma)
   \,=\,
    T^\cza_{\:\czc}(s,\sigma),\nonumber\\
   &&\rank T^\cza_{\:\czb}(s,\sigma)
   \,=\, m_0-m_1,
   \label{Tsigmas1}
  \eea
which explains why Eq.\,(\ref{Solution0fqAdd}) comprises $m_0-m_1$ gauge conditions rather than $m_0$ conditions. It is remarkable that despite the presence of an auxiliary element --- an arbitrary matrix ${s}^{\cczb}_{\,\czb}$ of rank $m_1$ --- these gauge conditions and expressions for extraghosts (\ref{Solution0aaggfc}) are $s$-independent. This easily follows on shell from the variational equation
  \beq{}
    \var_{{s}}\big(\, ({s}^{}_{} \sigma^{}_{})^{-1\,\ccza}_{\;\;\;\;\;\;\,\cczb} {s}^{\cczb}_{\,\czb} \chi^\czb\,\big)
   \,=\,
    ({s}^{}_{} \sigma^{}_{})^{-1\,\ccza}_{\,\,\;\;\;\;\cczc}\, (\var{{s}^{\cczc}_{\,\czc}})\,T^\czc_{\:\czb}(s,\sigma)\, \chi^\czb
   \,=\, 0.
  \eeq

Finally, consider the on-shell Hessian of the action in the zero ghost number sector of fields $\varPhi_0=(\phi^\zi$, $\egC^\ccza)$ which is actually the Hessian of $S [\,\phi\,] - \tfrac12  X^\cza {\varkappa}_{\cza\czb} X^\czb$. Bearing in mind the on-shell value of $X^\cza=0$ we get the block-matrix operator
 \bea{defB}
  &&\left.\frac{\var^2 S^{\scriptscriptstyle{FP}}}{\var \varPhi_0 \, \var \varPhi_0} \right|_{\,\var S^{\scriptscriptstyle{FP}} /\var \varPhi_{\rm red}=0}\nonumber\\
  &&\qquad =\, \left[
    \begin{array}{cc}
          S_{,\zi\zj}
          {-}   X^ {\cza}_{,\zi} {\varkappa}_{\cza\czb}  X^ {\czb}_{,\zj}
         &\,\, - \sigma^{\cza}_{\,\cczb} \, {\varkappa}_{\cza\czb}  X^ {\czb}_{,\zi}
        \vphantom{\Big|}
        \\
          - \sigma^{\cza}_{\,\ccza} \, {\varkappa}_{\cza\czb}  X^ {\czb}_{,\zi}
         &\,\, - \sigma^{\cza}_{\,\ccza} {\varkappa}_{\cza\czb} \sigma^{\czb}_{\,\cczb}
        \vphantom{\Big|}
    \end{array}
  \right] .
 \eea

In view of invertibility of $\varkappa_{\cza\czb}$ one can introduce
the nondegenerate $m_1 \times m_1$ operator $\kappa_{\ccza\cczb}$ and its inverse
  \bea{def_cSKSinvcSKS}
    \kappa_{\ccza\cczb}
    = \sigma^{\cza}_{\,\ccza} {\varkappa}_{\cza\czb} \sigma^{\czb}_{\,\cczb},
    \;\quad\;
    {\kappa}^{\cczb\ccza} \equiv \big( \kappa_{\ccza\cczb} \big)^{-1}.
  \eea
Then one can factorize the determinant of (\ref{defB}) as the product of determinants of two operators
  \beq{}
    \left. \det\, \frac{\var^2 S^{\scriptscriptstyle{FP}}}{\var \varPhi_0 \, \var \varPhi_0} \right|_{\,\var S^{\scriptscriptstyle{FP}}/\var \varPhi_{\rm red}=0}
    \!\!\!\!
   = \;\det F_{ij} \,
   \det{\big({-}\kappa_{\ccza\cczb} \big)}, \label{det_B}
  \eeq
where the new operator $F_{ij}$ is obviously the gauge-fixed inverse propagator of fields $\phi^\zi$
  \bea{}
   && F_{\zi\zj} \,\equiv\, S_{,\zi\zj}
     -   X^ {\cza}_{,\zi}\, \varPi_{\cza\czb}  X^ {\czb}_{,\zj},
   \label{invPropagator}\\
   &&\varPi_{\cza\czb}
    \,\equiv\,
       {\varkappa}_{\cza\czb}
       -  {\varkappa}_{\cza\czc} \sigma^{\czc}_{\,\ccza}{\kappa}^{\ccza\cczb} \sigma^{\czd}_{\,\cczb} {\varkappa}_{\czd\czb},
   \label{def_cProj}
  \eea
in which the gauge-fixing term is built in terms of the gauge matrices $X^ {\cza}_{,\zi}$ and the gauge-fixing matrix $\varPi_{\cza\czb}$. The latter, however, does not coincide with the original matrix $\varkappa_{\cza\czb}$ in the gauge fermion (\ref{GaugeFermion_1st}), but rather converted to the projector form with the following properties:
 \beq{tSProj_properties}
   \sigma^{\cza}_{\,\ccza} \varPi_{\cza\czb}
   = \varPi_{\cza\czb} \sigma^{\czb}_{\,\cczb}
   = 0,
   \;\quad\;
   \varPi_{\cza\czb} \varkappa^{\czb\czc} \varPi_{\czc\czd} = \varPi_{\cza\czd}.
 \eeq
This projection is fully consistent with the fact that the rank $n-m_0+m_1$ of $S_{,ij}$ should be raised up to $n$ by adding nonzero eigenvalues not in the $m_0$-dimensional subspace, but in the $(m_0-m_1)$-dimensional one. In particular, with the projector (\ref{def_cProj}) one can rewrite on-shell gauge conditions (\ref{Solution0fqAdd}) on fields $\phi^\zi$ as\, $\varPi_{\cza\czb}  \chi^\czb  =  0$\,. The equivalence of this to (\ref{Solution0fqAdd}) is established by choosing ${s}^{\ccza}_{\,\czb} = m^{\ccza\cczb}\sigma^{\cza}_{\,\cczb} {\varkappa}_{\cza\czb}$ and noting that the projector $\delta^\cza_{\,\czb}  - \sigma^\cza_{\,\ccza} ({s}^{}_{} \sigma^{}_{})^{-1\,\ccza}_{\;\;\;\;\;\;\,\cczb} {s}^{\cczb}_{\,\czb}$ does not depend on a nondegenerate matrix factor $m^{\ccza\cczb}$.

    \subsection{One-loop contribution to the generating functional}
       \label{SSect:BV_EA}
From (\ref{redGeneratingFunctional}) the one-loop contribution to the generating functional (its preexponential factor) reads
 \beq{Z_1loop_1st0}
  Z^{1-\rm loop}=
    \,\frac{\det \rho^\ccza_{\,\cczb}}{\big(\! \det \varkappa^{\cza\czb}\big)^{1/2} }\,
   \left(\mathrm{sdet} \,
  \frac{\var^{(l)}\var^{(r)} S^{\scriptscriptstyle{FP}}}{\var \varPhi_{\rm red}\, \var\varPhi_{\rm red}}\,
  \right)^{-1/2}
 \eeq
where the Hessian in the sector of reduced fields has in virtue of (\ref{maSredGF}) the following block matrix structure
\beq{maSGFhessian}
  \frac{\var^{(l)}\var^{(r)} S^{\scriptscriptstyle{FP}}}{\var\varPhi_{\rm red}\, \var \varPhi_{\rm red}}
  \,=\,
  \left[
    \begin{array}{c|c|c}
         \;{ \displaystyle \frac{\var^2 S^{\scriptscriptstyle{FP}}}{\var \varPhi_0 \, \var \varPhi_0} }
         &
        \begin{array}{cc}
          0\,\,\; &\;\,\, 0 \\
          0\,\,\; &\;\,\, 0  \\
        \end{array}
         &\;
        \begin{array}{cc}
          \;0\,\,\; &\;\,\, 0\;\; \\
          \;0\,\,\; &\;\,\, 0\;\; \\
        \end{array}
     \\
     \hline 
        \begin{array}{cc}
          0\,\,\; &\;\,\, 0 \\
          0\,\,\; &\;\,\, 0 \\
        \end{array}
         &
        \hspace{-2pt}
        \begin{array}{cc}
          \;\,0 & \;-F^{\czb}_{\cza}\vphantom{\big|^|}\;\; \\
          \;\;F^{\cza}_{\czb}\vphantom{\big|_|}\!\!\! & 0 \\
        \end{array}
        \hspace{-6pt}
         &
        \begin{array}{cc}
          \;0\,\,\; &\;\,\, 0  \\
          \;0\,\,\; &\;\,\, 0 \\
        \end{array}
     \\
      \hline 
        \begin{array}{cc}
          0\,\,\; &\;\,\, 0 \\
          0\,\,\; &\;\,\, 0 \\
        \end{array}
          &
        \begin{array}{cc}
          0\,\,\; &\;\,\, 0 \\
          0\,\,\; &\;\,\, 0 \\
        \end{array}
         &
        \begin{array}{cc}
          \;0\; & F^{\cczb}_{\;\ccza}\vphantom{\big|^|} \\
          \;F^{\ccza}_{\;\cczb}\vphantom{\big|_|} & 0 \\
        \end{array}
     \\
    \end{array}
  \right]\,.
 \eeq
Block in the left column and upper row here corresponds to the zero ghost number sector $\varPhi_0=(\phi^\zi, \egC^\cczb)$, the middle column and row block corresponds to the odd sector $(\gC^\czb,\agC_\czb)$ of gauge ghost and antighost fields, right bottom block corresponds to the even sector of ghost-for-ghost and antighost-for-ghosts fields $(\gC^\cczb, \agC_\cczb)$.

In virtue of the factorization property in the reduced fields block (\ref{det_B}) the calculation of the full superdeterminant finally gives the one-loop contribution to the generating functional of the first-stage reducible gauge theory
 \bea{Z_1loop_1st}
  \hspace{-1cm}Z^{1-\rm loop}
  &=&
    \,\frac{\det \rho^\ccza_{\,\cczb}}{\big(\!\det \varkappa^{\cza\czb}\big)^{1/2} }\nonumber\\
   &&\times \frac{ \det F^{\cza}_{\;\czb} }{
     \big(\!\det F_{\zi\zj} \big)^{1/2}
     \big(\!\det \kappa_{\ccza\cczb}\big)^{1/2}
     \det F^{\ccza}_{\;\cczb}  }.
 \eea

Let us assemble together all numerous ingredients of this expression which were introduced above in the course of derivation of this formula. Here the inverse propagator of the original gauge field $F_{ij}$ (\ref{invPropagator}), the ghost operator $F^\cza_{\;\czb}$ (\ref{gRegCond}), and the ghosts-for-ghosts operator $F^\ccza_{\;\cczb}$ (\ref{ggRegCond}) are, respectively,
 \bea{}
  F_{\zi\zj}
  &\equiv& S_{,\zi\zj}-X^ {\cza}_{,\zi} \varPi_{\cza\czb}  X^ {\czb}_{,\zj},
 \label{invPropagator1}
 \\
  F^{\cza}_{\;\czb}
  &\equiv&  X^ \cza_{,\zi}\mathcal{R}^\zi_{\czb}
     - \sigma^{\cza}_{\,\ccza} {\rho^{-1}}^{\ccza}_{\,\cczb}\,  \omega^{\cczb}_{\,\czb},
  \label{invgPropagator1}
 \\
  F^{\ccza}_{\;\cczb}
  &\equiv& \omega^\ccza_{\,\czc} \mathcal{Z}^\czc_{\,\cczb}.
  \label{invggPropagator1}
 \eea

Gauge field and ghost operators (\ref{invPropagator1}), (\ref{invgPropagator1}) are both gauge fixed with the aid of gauge conditions matrices $X^\cza_{,\zi}$, $\sigma^{\cza}_{\,\ccza}$, and $\omega^{\cczb}_{\,\czb}$ and the gauge-fixing matrices $\varPi_{\cza\czb}$ and ${\rho^{-1}}^{\ccza}_{\,\cczb}$ --- kernels of bilinear terms in ($X^\cza_{,\zi}$, $\sigma^{\cza}_{\,\ccza}$, $\omega^{\cczb}_{\,\czb}$). Whereas these matrices of gauge conditions for ghosts $\sigma^{\cza}_{\,\ccza}$ and $\omega^{\cczb}_{\,\czb}$ are a part of the originally chosen gauge fermion (\ref{GaugeFermion_1st}), a similar matrix in the gauge fields sector is a special projector $\varPi_{\cza\czb}$ (\ref{def_cProj}), $\varPi_{\cza\czb} \varkappa^{\czb\czc} \varPi_{\czc\czd} = \varPi_{\cza\czd}$ (\ref{tSProj_properties}), on the direction in the space of gauge indices orthogonal to the ``vielbein'' $\sigma^\cza_{\,\ccza}$, $\sigma^{\cza}_{\,\ccza} \varPi_{\cza\czb}
   = \varPi_{\cza\czb} \sigma^{\czb}_{\,\cczb}
   = 0$.
\if{
\bea{}
&&\varPi_{\cza\czb}
  \,\equiv\, 
       {\varkappa}_{\cza\czb}
       -  {\varkappa}_{\cza\czc} \sigma^{\czc}_{\,\ccza}{\kappa}^{\ccza\cczb} \sigma^{\czd}_{\,\cczb} {\varkappa}_{\czd\czb},
       \label{def_cProj1}\\
   &&\kappa_{\ccza\cczb}
     \,=\, \sigma^{\cza}_{\,\ccza} {\varkappa}_{\cza\czb} \sigma^{\czb}_{\,\cczb},
 \qquad
   {\kappa}^{\ccza\cczb} \equiv ( \kappa^{-1} )^{\ccza\cczb}.
   \label{def_cSKSinvcSKS1}
 \eea
}\fi
This type of ``orthogonality'' is determined with respect to the symmetric metric $\varkappa_{\cza\czb}$, which is inverse to $\varkappa^{\czb\cza}$ originally introduced in the gauge fermion (\ref{GaugeFermion_1st}). The projection of this metric onto the space of reducible generator indices $\ccza$ with the aid of the vielbein $\sigma^\cza_{\,\ccza}$ gives rise to the gauge-fixing matrix $\kappa_{\ccza\cczb}$ in the space of these indices. The determinants of all gauge-fixing matrices $\varkappa_{\cza\czb}$, $\rho^\ccza_{\,\cczb}$ and $\kappa_{\ccza\cczb}$ appropriately enter the final algorithm for the one-loop generating functional (\ref{Z_1loop_1st}).

Ranks of all the above gauge matrices are maximal and determined by the range of their indices, and the main criterion of their choice is the invertibility of the full set of gauge and ghost operators (\ref{invPropagator1})-(\ref{invggPropagator1}).

The final important comment is the definition of the on-shell condition for the obtained algorithm (\ref{Z_1loop_1st}). All ghosts, ghosts for ghosts and their antighost fields are zero on shell, $\gC^\cza=  \agC_\cza =  \gC^\ccza  =  \agC_\ccza  =  0$ (\ref{ghostSectorSolutions}). The exception is the only ``non-classical'' field $\egC^{\,a}$ which has a zero ghost number, though originally it was called the extraghost \cite{Batalin:1983ggl}. On shell it is generically nonvanishing and is given by the expression (\ref{Solution0aaggfc}), so that the gauge conditions matrix equals
\beq{Xalphai}
  X^\cza_{,\zi} \,=\, \chi^\cza_{,\zi}
     - \sigma^\cza_{\,\cczc,\zi}\,({s}^{}_{} \sigma^{}_{})^{-1\,\ccza}_{\;\;\;\;\;\;\,\cczb}\, {s}^\cczb_{\,\czb} \,\chi^\czb
 \eeq
and coincides with $\chi^\cza_{,i}$ only for $\phi$-independent $\sigma^\cza_{\,\cczc}$ or vanishing on shell ${s}^\cczb_{\,\czb} \chi^\czb$. For the original gauge field $\phi^\zi$ on-shell restriction and gauge conditions (\ref{SolutionCorrespondence})-(\ref{Solution0ghostXX}) equivalently read $S_{,\zi}  =  0$, \,$\varPi_{\cza\czb}\chi^\czb=0$ (note that the original gauge functions $\chi^\cza(\phi)$ introduced in the gauge fermion (\ref{GaugeFermion_1st}) does not necessarily vanish --- only their projection vanishes on shell).

  \subsection{Ward identities and gauge independence of the effective action}
   \label{SSect:BV_WardIdentities}
The one-loop effective action corresponding to (\ref{Z_1loop_1st}) reads
 \bea{EA_1loop_1st}
  &&\hspace{-1cm}i\varGamma^{1-\rm loop}{} 
   \,=\,
   - \tfrac12 \Tr{\ln{F_{\zi\zj}}}
   + \Tr{\ln{F^{\cza}_{\;\czb}}}
   - \Tr{\ln{F^{\ccza}_{\;\cczb}}}\nonumber\\
   &&\hspace{-1cm}\qquad\qquad \quad
   - \tfrac12 \Tr{\ln\kappa_{\ccza\cczb}}
   - \tfrac12 \Tr{\ln{ \varkappa^{\cza\czb} }}
   + \Tr{\ln{ \rho^{\ccza}_{\,\cczb} }}\;\;
 \eea
and includes gauge field, ghost, ghost-for-ghost contributions and the contribution of three gauge-fixing matrices. On shell this expression should not depend on all gauge-fixing entities $\chi^\cza(\phi)$, $\sigma^{\cza}_{\,\ccza}(\phi)$, $\omega^{\ccza}_{\,\cza}(\phi)$, $\varkappa^{\cza\czb}$, $\rho^{\ccza}_{\,\cczb}$ as it is dictated by the general BV theory. It is worth checking this property and revealing the details of the perturbative mechanism of such gauge independence. In the one-loop approximation this mechanism is based on the on-shell Ward identities for all tree-level propagators of the theory: $F^{-1\,\zi\zj}$, $F^{-1\,\cza}_{\;\;\;\;\;\;\,\czb}$ and $F^{-1\,\ccza}_{\;\;\;\;\;\;\,\cczb}$.

Ward identity for the Green's function of the gauge field operator $F_{ij}$ follows from the sequence of on-shell relations
\bea{}
  \hspace{-0.5cm}\mathcal{R}^{\zi}_{\hspace{0.5pt}\czc}
  \,=\, F^{-1\,\zi\zj} F_{\zj\zk} \mathcal{R}^\zk_{\czc}
    &=&\,-\, F^{-1\,\zi\zj}  X^ {\cza}_{,\zj} \varPi_{\cza\czb}  X^ \czb_{\,,\zk} \mathcal{R}^\zk_{\czc}\nonumber\\
  &=&\, - \, F^{-1\,\zi\zj}  X^ {\cza}_{,\zj} \varPi_{\cza\czb}
     F^{\czb}_{\;\czc},
  \eea
where we used the fact that on shell $\mathcal{R}^\zk_{\czc}$ is a zero vector of $S_{,\zj\zk}$ since $S_{,\zj\zk}\mathcal{R}^\zj_{\czc} = -S_{,\zj} \mathcal{R}^\zj_{\czc,\zk}$, and $\varPi_{\cza\czb}  X^ \czb_{,\zk}\mathcal{R}^\zk_{\czc}= \varPi_{\cza\czb}F^{\czb}_{\;\czc}$ in view of $\varPi_{\cza\czb}\sigma^\czb_{\,\cczb}=0$. Thus, contracting this relation with $F^{-1\,\czc}_{\;\;\;\;\;\;\czd}$, we get on shell
    \bea{}
    F^{-1\,ij}X^ {\cza}_{,\zj} \varPi_{\cza\czb}\,\big|_{\,S_{,\zi}=0}=
    -  \mathcal{R}^{\zi}_{\czc} F^{-1\,\czc}_{\;\;\;\;\;\;\czb},
   \label{WardId1_1st}
 \eea
whence it follows the effective action independence on the choice of gauge matrix $X^\cza_{,\zi}$ and the gauge condition $\chi^\cza(\phi)$
\bea{1loop_XX_indep}
    \hspace{-1cm}i\,\var_\chi \varGamma^{\rm 1-loop}
    &=&i\,\var_{X} \varGamma^{\rm 1-loop}\nonumber\\
  &&\hspace{-1cm}=\big( F^{-1\,\zi\zj}  X^ {\cza}_{,\zi} \varPi_{\cza\czb}
   + \mathcal{R}^\zi_{\cza} F^{-1\,\cza}_{\;\;\;\;\;\;\czb}
   \big)\, \var  X^ \czb_{,\zi}
  \,=\, 0.
 \eea

Similarly, from the relation ${\mathcal Z}^\czc_{\,\cczc}
   = F^{-1\,\czc}_{\;\;\;\;\;\;\,\cza} F^{\cza}_{\;\czb\,}  \mathcal{Z}^{\czb}_{\,\cczc}
   = -  F^{-1\,\czc}_{\;\;\;\;\;\;\,\cza} \sigma^{\cza}_{\,\ccza\,} {\rho^{-1}}^{\ccza}_{\,\cczb}  F^\cczb_{\;\cczc}$
it follows that the Ward identity relating the ghost and ghosts-for-ghosts propagators is
 \bea{}
  &&F^{-1\,\czc}_{\;\;\;\;\;\;\,\cza} \sigma^{\cza}_{\,\ccza\,}
  {\rho^{-1}}^{\ccza}_{\,\cczb} \,\big|_{\,S_{,\zi}=0}
  = \,-\, {\mathcal Z}^\czc_{\,\cczc}  F^{-1\,\cczc}_{\;\;\;\;\;\;\,\cczb},
   \label{Ward10}
 \eea
and the effective action turns out to be on-shell independent of $\omega^\cczb_{\,\cza}$,
 \bea{}
  &&\hspace{-1cm} i\,\var_{\omega}\varGamma^{\rm 1-loop}\nonumber\\
  &&\hspace{-1cm}\qquad= - \big( F^{-1\,\czb}_{\;\;\;\;\;\;\,\cza}
  \sigma^{\cza}_{\,\ccza}\, {\rho^{-1}}^{\ccza}_{\,\cczb}
       + \mathcal{Z}^{\czb}_{\,\ccza}\,
       F^{-1\,\ccza}_{\;\;\;\;\;\;\,\cczb}\big)\,
       \var \omega^\cczb_{\,\czb} \,=\, 0.    \label{1loop_cO_indep}
 \eea

The dependence of $\varGamma^{\rm 1-loop}$ on $\sigma^\cza_{\,\ccza}$ involves three terms of (\ref{EA_1loop_1st}) the sum of which also turns out to be zero on shell
  \bea{}
    i\,\var_{\sigma} \varGamma^{\rm 1-loop}
    &=& - \tfrac12\,\var_{\sigma}{\rm Tr}\,\ln F_{ij}+
      \var_{\sigma}{\rm Tr}\ln F^\cza_{\;\czb}
    \nonumber\\
    && - \tfrac12\,\var_{\sigma}{\rm Tr}\ln\kappa_{\ccza\cczb}
    \;=\; 0.
  \eea
To prove this one should use in the variation of the operator (\ref{invPropagator1}) the variation of the projector $\varPi_{\cza\czb}$, $\var_\sigma\varPi_{\cza\czb}=
-2\,\var\sigma^\czc_{\,\ccza}\,\kappa^{\ccza\cczb}\,\varPi_{\czc(\cza}\,
\varkappa_{\czb)\czd}\,\sigma^\czd_{\,\cczb}$, the variation of $\kappa_{\ccza\cczb}$, $\var_\sigma \kappa_{\ccza\cczb}=2\,
\varkappa_{\cza\czb}\,\sigma^\czb_{(\ccza} \var\sigma^\cza_{\,\cczb)}$, and the corollaries of (\ref{WardId1_1st}) and (\ref{Ward10}),
\bea{}
    &&\hspace{-0.5cm}X^ \czc_{\,,\zi}\,F^{-1\,\zi\zj}X^ {\czb}_{\,,\zj}\, \varPi_{\czb\cza}\,\big|_{\,S_{,\zi}=0}
    =-X^ \czc_{\,,\zi}\mathcal{R}^\zi_\czb\,
    F^{-1\,\czb}_{\;\;\;\;\;\;\,\cza},\\
    &&\hspace{-0.5cm}X^ \czc_{\,,\zi}\mathcal{R}^\zi_\czb\,
    F^{-1\,\czb}_{\;\;\;\;\;\;\,\cza}
    =\big(F^\czc_{\;\czb}
    +\sigma^{\czc}_{\,\ccza\,}
    {\rho^{-1}}^{\ccza}_{\,\cczb\,}
      \omega^{\cczb}_{\,\czb}\big)\,
      F^{-1\,\czb}_{\;\;\;\;\;\;\,\cza}\nonumber\\
    &&\hspace{-0.5cm}\qquad\qquad\qquad=\delta^{\czc}_{\,\cza}
     + \,\sigma^{\czc}_{\,\ccza\,}
     {\rho^{-1}}^{\ccza}_{\,\cczb\,}
     \omega^{\cczb}_{\,\czb\,} F^{-1\,\czb}_{\;\;\;\;\;\;\,\cza},               \label{Ward100}\\
    &&\hspace{-0.5cm}\omega^\cczb_{\,\czb\,}
    F^{-1\,\czb}_{\;\;\;\;\;\;\,\cza}
    \sigma^{\cza}_{\,\ccza}\,\big|_{\,S_{,\zi}=0}
    =-\rho^\cczb_{\,\ccza} .         \label{Ward101}
     \eea

The $\varkappa^{\cza\czb}$-variation of the effective action also vanishes on shell because of the variations $\var_\varkappa\varPi_{\cza\czb}=-\varPi_{\cza\czd} \var\varkappa^{\czd\czc}\varPi_{\czc\czb}$, \,$\var_{\!\varkappa}\kappa_{\ccza\cczb}= \sigma^{\cza}_{\,\ccza} \var{\varkappa}_{\cza\czb} \sigma^{\czb}_{\,\cczb}$, Eq.\,(\ref{Ward100}) and its corollary $\varPi_{\cza\czc}X^{\czc}_{\,,\zi} F^{-1\,\zi\zj} X^{\czd}_{\,,\zj}\varPi_{\czd\czb}= -  \varPi_{\cza\czb}$ (which in its turn is provided by the orthogonality relation $\varPi_{\cza\czb}\sigma^\czb_{\,\cczb}=0$),
  \bea{1loop_var_cK1}
   i\,\var_{\varkappa} \varGamma^{\rm 1-loop}
    &\;=&\;
   - \tfrac12 F^{-1\,\zj\zi} \var_{\!\varkappa} F_{\zi\zj}
   - \tfrac12 {\kappa}^{\cczb\ccza} \var_{\!\varkappa} \kappa_{\ccza\cczb}\nonumber\\
   &&- \tfrac12 {\varkappa}_{\czb\cza} \var \varkappa^{\cza\czb}
   \;=\;0.
  \eea

Finally, on-shell independence of the gauge-fixing matrix $\rho^\ccza_{\,\cczb}$ follows by direct variation and the use of the corollary (\ref{Ward101}) of the Ward identity (\ref{Ward10}) for the ghost propagator,
 \bea{1loop_var_cR}
  \hspace{-1cm}i\,\var_{\rho} \varGamma^{\rm 1-loop}
  \nonumber\\
  &&\hspace{-1cm}=
   - \big(\omega^{\cczb}_{\,\czb}\, F^{-1\,\czb}_{\;\;\;\;\;\;\cza}  \sigma^{\cza}_{\,\ccza}
   +\rho^{\cczb}_{\,\ccza}\big)\,
   \var{{\rho^{-1}}^{\ccza}_{\,\cczb}}
  \;=\; 0.
 \eea

 \section{Reducible gauge structure of restricted gauge theories}
  \label{Sect:Restricting_and_Reducibility}
Reducible structure of a gauge theory can be induced by the procedure of {\em restriction} of the originally irreducible gauge theory. To see this consider a generic gauge theory with the action $\hat S[\,\varphi^\zI\,]$ subject to a closed gauge algebra of \emph{irreducible} generators $\hat{\mathcal R}^\zI_\cza$,
\bea{gRprnt_commutator}
   \hat{S}_{,\zI}\hat{\mathcal{R}}^\zI_\cza
  = 0,
  \;\quad\;
  \hat{\mathcal{R}}^\zI_{\cza,\zJ} \hat{\mathcal{R}}^\zJ_{\czb} - \hat{\mathcal{R}}^\zI_{\czb,\zJ} \hat{\mathcal{R}}^\zJ_{\cza}
  =  \hat{\mathcal{R}}^\zI_{\czc} \hat{C}^{\czc}_{\cza\czb}.
 \eea
Irreducibility of the generators $\hat{\mathcal R}^\zI_\cza$  implies that the rank of their matrices coincides with the range of the indices $\cza$ which in its turn is lower than the range of the indices $\zI$ enumerating the original gauge fields $\varphi^\zI$, $\rank {\hat{\mathcal{R}}^\zI_\cza} = {\rm range}\: \cza = m_0 < \hat{n} = {\rm range}\: \zI$. We will call such a theory the {\em parental} one.

{\em Restricted} gauge theory, originating from the parental one, is the theory whose configuration space variables are kinematically constrained by the equations
 \beq{rSpace_rC}
  \theta^\ccza(\varphi^\zI) \,=\, 0,
 \eeq
    where we will consider the functions $\theta^\ccza(\varphi^\zI)$ to be functionally independent, that is characterized by the full-rank condition of their gradient matrix,
 \beq{rC_prnt_Irreducibility}
  \rank {\theta^\ccza_{,\zI}} \,=\, m_1 \,\equiv\, {\rm range}\:\ccza.
 \eeq
For simplicity we assume that these functions are either ultralocal (algebraic) in spacetime or the surface of these constraint functions can be parametrized in terms of local independent fields which will be denoted in what follows by $\phi^\zi$.

   \subsection{Two representations of a restricted theory}
    \label{SSect:Restricted_and_Reduced_Reps}
Such a restricted theory can be described in two equivalent ways. One way is to represent it in terms of the Lagrange multipliers action
 \beq{S_fixed}
   S^\lambda[\,\varphi,\lambda\,] \,=\, \hat{S}[\,\varphi\,] - \lambda_\ccza\theta^\ccza(\varphi),
 \eeq
whose classical equations of motion obtained by varying both its gauge fields $\varphi^\zI$ and Lagrange multipliers $\lambda_\ccza$ read
 \beq{lambdaEoM}
  \hat S_{,\zI}-\lambda_\ccza\,\theta^\ccza_{,\zI}=0, \;\quad\; \theta^\ccza=0.
 \eeq

Another representation is the {\em reduced} theory, when one solves first the constraints (\ref{rSpace_rC}) with respect to $\varphi^\zI$ as functions of the reduced set of fields $\phi^\zi$ and formulates the theory in terms of the {\em reduced} action $S^{\rm red}[\,\phi\,]$. The latter is obtained by substituting in the parental theory action the functions $e^\zI(\phi)$ of embedding the $\phi^\zi$-subspace into the space of original gauge fields $\varphi^\zI$,
 \bea{}
  &&\varphi^\zI=e^{\zI}(\phi),
  \;\quad\;
  \theta^\ccza\big(e^\zI(\phi)\big)\equiv0, \label{embedding}\\
  &&S^{\rm red}[\,\phi\,]=\hat S[\,e(\phi)\,].  \label{embeddingS}
 \eea

The equations of motion in both formulations are obviously equivalent\footnote{Classical equivalence of two theories can be defined by equivalence of spaces of solutions of their equations of motion. In particular it can be established by the local correspondence (\ref{tang_lambda_EoM}) and the statement that reduced fields do not carry additional degrees of freedom. This statement means that the above assumptions on $\theta^\ccza(\varphi)$ imply the existence of such reducibility, that is the possibility of excluding auxiliary variables in terms of the reduced ones.} because the contraction of the first of Eqs.\,(\ref{lambdaEoM}) results in
 \beq{tang_lambda_EoM}
  e^\zI_{,\zi}\,\big(\hat S_{,\zI}-\lambda_\ccza\,\theta^\ccza_{,\zI}\big)=
  e^\zI_{,\zi}\,\hat S_{,\zI}\equiv\frac{\var S^{\rm red}[\,\phi\,]}{\var\phi^\zi},
  \;\quad\;
  e^\zI_{,\zi}\equiv\frac{\var e^\zI(\phi)}{\var\phi^\zi},
 \eeq
where we took into account that in view of (\ref{embedding}) the covector $\theta^\ccza_{,\zI}$ is orthogonal to the surface of constraints $\theta^\ccza=0$,
 \beq{orthogonality}
  \frac{\var}{\var\phi^\zi}\theta^\ccza\big(\varphi(\phi)\big)
  \,=\,
  \theta^\ccza_{,\zI}\,\big|_{\,\theta=0}\,e^\zI_{,\zi}
  \,=\, 0.
 \eeq

An important question is the relation of the original \emph{parental} theory and the \emph{restricted} one from the point of view their physical equivalence. Solutions of the system of equations (\ref{lambdaEoM}) are obviously inequivalent to those of the parental theory, $\hat{S}_{,\zI}=0$, when on shell the Lagrange multipliers $\lambda_\ccza$ are nonvanishing. Note that in view of linearity of the $\lambda$-action (\ref{S_fixed}) in $\lambda_\ccza$ the variational equations with respect to Lagrange multipliers do not allow one to express them from the full set of equations of motion. There is another way to write down their equations of motion by contracting the first set of equations (\ref{lambdaEoM}) with the parental theory generators $\hat{\mathcal{R}}^\zI_\cza$ and use the Noether identities (\ref{gRprnt_commutator}). Then we get
  \bea{lambdaEoMQconsequence}
    \lambda_\ccza Q^\ccza_{\,\cza} \,=\, 0,
  \eea
where $Q^\ccza_{\,\cza}=Q^\ccza_{\,\cza}(\varphi)$ is the \emph{gauge-restriction operator} which will be very important in what follows,
  \beq{def_rQ}
    Q^\ccza_{\,\cza} \,\equiv\, \theta^\ccza_{,\zI\,}\hat{\mathcal{R}}^{\zI}_\cza.
  \eeq

This equation for $\lambda_\ccza$ has a unique solution $\lambda_\ccza=0$ only when the rank of $Q^\ccza_{\,\cza}$ is maximal, that is it coincides with the range of the index $\ccza$ enumerating the constraint functions $\theta^\ccza$. In this case the meaning of the constraints $\theta^\ccza(\varphi)=0$ is nothing but a partial gauge fixing of gauge invariance of the parental theory (${\rm range}\: \ccza=m_1 < m_0={\rm range}\:\cza$).

On the contrary, when the rank of gauge-restriction operator $Q^\ccza_{\,\cza}$ is lower than $m_1$
 \bea{rankQphys}
  \rank Q^\ccza_{\,\cza} \,=\, m_1-m_2
 \eea
then $m_2$ Lagrange multipliers can be freely specified, the rest $m_1-m_2$ of them being fixed as unique functions of the former free ones. This implies, in particular, the existence of $m_2$ gauge invariants of the parental theory constrained to be zero in restricted theory. To see this, note that the rank deficiency implies that operator $Q^\ccza_{\,\cza}$ has $m_2$ left zero vectors $Y^{\!\zA}_{\,\ccza}$,
 \beq{invariant}
  Y^{\!\zA}_{\,\ccza\,}Q^\ccza_{\,\cza}=0,
  \;\quad\; {\rm range}\: \zA = m_2 \leq m_1,
 \eeq
and this, according to the definition of this operator, implies the existence of $m_2$ parental gauge invariants
 \beq{invariant1}
  \theta^\zA(\varphi)\equiv Y^{\!\zA}_{\,\ccza}(\varphi)\,\theta^\ccza(\varphi),
  \;\quad\;
   \theta^\zA_{,\zI\,}\hat{\mathcal{R}}^{\zI}_\cza \big|_{\,\theta=0}=0,
 \eeq
which are constrained to vanish. So generically the restricted theory (\ref{S_fixed}) is \emph{inequivalent} to the parental gauge theory $\hat S[\,\varphi^\zI\,]$, because the latter does not \emph{a priori} impose any restrictions on its gauge-invariant objects. In what follows we will consider such restricted theories which incorporate new physics beyond their parental ones.

There is another representation of the solution for Lagrange multipliers. If one constructs the set $\theta^\zI_\cczb$ dual to covariant vectors $\theta^\ccza_{,\zI}$,
 \beq{theta_dual}
  \theta^\ccza_{,\zI}\,\theta^\zI_\cczb
  \,=\, \delta^\ccza_{\,\cczb},
 \eeq
then, contracting the equation of motion (\ref{lambdaEoM}) with $\theta^\zI_\cczb$ one obtains
 \beq{lambda_solution}
  \lambda_\ccza=\hat S_{,\zI} \theta^\zI_\ccza .
 \eeq

For local in spacetime restriction functions\footnote{We consider either ultralocal (algebraic) restriction functions $\theta^\ccza(\varphi)$ or those for which $m_1(={\rm range}\: \ccza)$ ultralocal independent combinations of $\varphi^\zI$ can be expressed from the restriction conditions  $\theta^\ccza(\varphi)=0$. In such a case a useful criterion for the existence of local (in fact ultralocal) vectors $\theta^\zI_\cczb$  dual to $\theta^\ccza_{,\zI}$, (\ref{theta_dual}) is the existence of the ultralocal nondegenerate linear combination  $\theta^\ccza_{,\zI} c^\zI_\cczb$ with local vectors $c^\zI_\cczb$ (in particular, the presence of the ultralocal minor of maximal $m_1$ rank in the matrix $\theta^\ccza_{,\zI}$).}  $\theta^\ccza$ this is a local expression for Lagrange multipliers in terms of $\hat S_{,\zI} $ --- the left-hand side of equations of motion of the {\em parental} theory. Resolving Lagrange multipliers $\lambda_\ccza$ in restricted equations of motion (\ref{lambdaEoM}) leads to the projected set of equations,
 \beq{lambda_eq_lin}
  \hat S_{,\zJ} (\delta^\zJ_{\,\zI} - \theta^\zJ_\ccza\,\theta^\ccza_{,\zI}) =0,
 \eeq
equivalent to $\hat S_{,\zJ} e^\zI_{,\zi} =0$. The projector $(\delta^\zJ_{\,\zI} - \theta^\zJ_\ccza\,\theta^\ccza_{,\zI})$ enforces only a part of parental equations of motion $\hat S_{,\zI}$ ``tangential'' to restriction surface $\theta^\ccza=0$. Which is equivalent to (\ref{tang_lambda_EoM}) and reinstates the observation that along tangential directions equations of motion for restricted theory coincides with that for parental theory.

On the contrary, projected on complementary (``normal'') directions, equations of motion for these two theories differ. Contraction of the parental equations $\hat S_{,\zI} =0$ and restricted theory equations $\hat S_{,\zI}-\lambda_\ccza\,\theta^\ccza_{,\zI}=0$ with $\theta^\zI_\ccza$ gives correspondingly $\hat S_{,\zI}\theta^\zI_\ccza =0$ and $\hat S_{,\zI}\theta^\zI_\ccza = \lambda_\ccza$. The normal subset of the parental equations in addition to equations (\ref{lambda_eq_lin}) further restricts classical configurations of fields $\varphi^\zI$ in parental theory. While the normal subset of the restricted theory equations merely expresses $\lambda_\ccza$ in terms of $\varphi^\zI$ and thus does not additionally constrain the latter.\footnote{
 Comparing the sets of equations of motion which define the dynamics of $\varphi^\zI$ in parental and restricted theories one finds that in the restricted theory $m_1$ constraints $\hat S_{,\zI} \theta^\zI_\ccza(\varphi)=0$ are ``removed'' while  $m_1$ new constraints $\theta^\ccza (\varphi)=0$ are applied. As it is this does not, however, predict the number of degrees of freedom in restricted theory, since this requires a deeper analysis of interrelation between the restriction conditions (\ref{rSpace_rC}) and dynamical and gauge structures of the parental theory.
}
However due to gauge invariance of the parental theory (\ref{gRprnt_commutator}) not all dynamical restrictions on $\varphi^\zI$ in the right-hand side of (\ref{lambda_solution}) are effectively removed. The linear dependence $\hat{S}_{,\zI}\hat{\mathcal{R}}^\zI_\cza=0$ between $\hat{S}_{,\zI}$ may express a certain part of $\hat S_{,\zI} \theta^\zI_\ccza$ in (\ref{lambda_solution}) as linear combinations of tangential equations (\ref{lambda_eq_lin}) which are still frozen to $0$ within the restricted equations of motion. Obviously such linear combinations are found by first contracting (\ref{lambda_solution}) with $\theta^\ccza_{,\zI}$ and then with $\hat{\mathcal{R}}^\zI_\cza$ which leads to $\lambda_\ccza Q^\ccza_{\cza}=\hat S_{,\zI} \theta^\zI_\ccza  Q^\ccza_{\cza}$. The same structure $\hat S_{,\zI} \theta^\zI_\ccza  Q^\ccza_{\cza}$ appears in the left-hand side of tangential equations (\ref{lambda_eq_lin}) after contracting with  $\hat{\mathcal{R}}^\zI_\cza$ and thus {vanishes} on-shell. This reinstates constraints on Lagrange multipliers (\ref{lambdaEoMQconsequence}), $\lambda_\ccza Q^\ccza_{\,\cza}=0$.

   \subsection{Gauge symmetry and reducibility}
    \label{SSect:Restricted_Gauge_Reducibility}
We assume that the restricted theory does not acquire local gauge symmetries beyond those of the original parental action. On the other hand, not all gauge transformations of the the parental theory $\var_\xi\varphi^\zI=\hat{\mathcal{R}}^{\zI}_\cza \xi^\cza$ generate symmetries of the reduced action (\ref{S_fixed}). These symmetries are only those which {preserve} the constraints
 \beq{residual_gauge structure}
  \var_\xi \theta^\ccza\,\big|_{\,\theta=0}
  \equiv\,
  \theta^\ccza_{,\zI}\hat{\mathcal{R}}^{\zI}_\cza \xi^\cza\,\big|_{\,\theta=0}
  \,=\, 0,
 \eeq
so that with the definition (\ref{def_rQ}) the allowed gauge transformation parameters $\xi^\cza_{\rm red}$ in the restricted theory should be solutions of the linear equation $Q^\ccza_{\,\cza}\xi^\cza_{\rm red}=0$, i.e. right zero vectors of the matrix $Q^\ccza_{\,\cza}$. The subspace of reduced gauge parameters can be obtained by projecting gauge parameters of the parental theory $\xi^\cza$ with a projector $T^\cza_{\:\czb}$ which is orthogonal in the space of gauge indices to $Q^\ccza_{\,\cza}$,
 \beq{tkQProj_left kernel}
  \xi^\cza_{\rm red}=T^{\cza}_{\:\czb} \, \xi^\czb,
   \;\quad\;
  Q^\ccza_{\,\cza\,} T^{\cza}_{\:\czb}= 0.
 \eeq

On the other hand, gauge symmetries of the restricted theory can be formulated in the parental field space of $\varphi^\zI$ with free (nonreduced) gauge parameters $\xi^\cza$. This happens if instead of projecting $\xi^\cza$ we project with respect to the gauge index $\cza$ the original parental generators $\hat{\mathcal{R}}^{\zI}_{\cza}$. Thus we get the \emph{reducible} set of gauge generators
 \beq{gRprnt_projected}
  \mathcal{R}^{\zI}_{\czb}
  \,\equiv\,
  \hat{\mathcal{R}}^{\zI}_{\cza\,} T^{\cza}_{\:\czb}.
 \eeq
This formulation is covariant from the viewpoint of the parental gauge theory because the original multiplets of parental field indices $\zI$ and gauge indices $\cza$ remain unsplitted, but the price paid for this covariance is the reducibility of the generators $\mathcal{R}^{\zI}_{\czb}$. They are indeed reducible because the projector $T^{\cza}_{\:\czb}$ in view of rank deficiency (following from (\ref{tkQProj_left kernel})) possess right zero vectors $k^{\czb}_{\,\cczb}$ which become also the right zero vectors of $\mathcal{R}^{\zI}_{\czb}$
 \beq{tkQProj_right kernel}
  T^{\cza}_{\:\czb\,} k^{\czb}_{\,\cczb} = 0, \;\quad\; \mathcal{R}^{\zI}_{\czb\,} k^{\czb}_{\,\cczb} = 0.
 \eeq

Despite reducibility an important advantage of such projected representation is that it generates gauge transformations with arbitrary \emph{unrestricted} gauge parameters $\xi^\cza$. In the course of subsequent Batalin-Vilkovisky extension of configuration space it will give rise to gauge ghosts equivalent to that of the parental theory. This property is achieved at the cost of \emph{reducibility} of the projected gauge structure (\ref{gRprnt_projected}) and introduction of ghost for ghosts fields.

The projector $T^{\cza}_{\:\czb}$ may be explicitly constructed in terms of the left- and right-kernel bases (\ref{tkQProj_left kernel}),(\ref{tkQProj_right kernel})
 \beq{tkQProj}
  T^{\cza}_{\:\czb}
  \,=\,
  \delta^{\cza}_{\,\czb}
   - k^{\cza}_{\,\ccza}\, {(Qk)^{-1}} \vphantom{|} ^\ccza_{\,\cczb}\, Q^{\cczb}_{\,\czb}
  \,\equiv\,
  T^{\cza}_{\:\czb}(Q,k).
 \eeq
Left kernel --- $Q^\ccza_{\,\cza}$ (\ref{def_rQ}) --- is fixed by the restriction conditions and the choice of parental gauge generators. Thus (\ref{tkQProj}) may be considered as the family  {parametrized} by its right kernel basis $k^{\czb}_{\,\cczb}$. The structure of  $T^{\cza}_{\:\czb}(Q,k)$, which in what follows will be referred without specification of kernels, is analogous to the set of projectors $T^\cza_{\:\czb}(s,\sigma)$ (\ref{Tsigmas0}) introduced above.

Here we emphasize the property that the matrix $(Qk)^\ccza_{\,\cczb}\equiv Q^\ccza_{\,\czc}\, k^\czc_{\,\cczb}$ is not directly invertible. Rather it could have rank equal or lower to that of $\rank Q^\ccza_{\,\czc}=m_1-m_2$. The rank deficiency is dictated by the requirement of new physics incorporated by the restricted theory (\ref{rankQphys}) --- physical inequivalence to the parental gauge theory. As discussed in Appendix \ref{ASect:Projectors_Variation}, this extra difficulty can be circumvented by the Moore-Penrose construction of the generalized inverse \cite{MoorePenrose}, provided the following rank restriction conditions (\ref{g_rank}) are satisfied,
 \beq{g_rank}
  \rank{(Qk)^\ccza_{\,\cczb}}
  = \rank{ Q^{\ccza}_{\,\cza}}
  = \rank{ k^{\czb}_{\,\cczb}}.
 \eeq
This guarantees unambiguous definition of projector (\ref{tkQProj}) and its correct rank property, $\rank {T^{\cza}_{\:\czb}}= m_0-m_1+m_2$.

Rank restriction requirements (\ref{g_rank}) in particular are guaranteed when $k^{\czb}_{\,\cczb}$ is parametrized in terms of nondegenerate two-forms $B^{\czb\cza}(\varphi)$ and $c_{\ccza\cczb}(\varphi)$ --- metrics in spaces of $\cza$-indices and $\ccza$-indices respectively,
 \beq{vK_from_rQ}
  k^{\czb}_{\,\cczb} = B^{\czb\cza} Q^{\ccza}_{\,\cza} c_{\ccza\cczb},
  \,\quad\,
  T^{\cza}_{\:\czb}=
    \delta^{\cza}_{\,\czb}
    - B^{\cza\czc}Q^\ccza_{\,\czc} \big(\,QBQ^T\,\big)^{-1}_{\ccza\cczb}Q^\cczb_{\,\czb},
 \eeq
which will be used in unimodular gravity calculations in Sec.\,\ref{Sect:UMG_EA}. In fact in this case the projector $T^{\cza}_{\:\czb}$ is defined by $Q^{\ccza}_{\,\cza}$ and $B^{\czb\cza}$ only because the dependence on $c_{\ccza\cczb}$ completely cancels out.\footnote{
 The ``symmetric'' form (\ref{vK_from_rQ}) has the structure analogous to that of symmetric projector $\varPi_{\cza\czb}$ (\ref{def_cProj}) (with the left index raised) when  $\varkappa_{\cza\czb}\,\leftrightarrow\,(B^{\czb\cza})^{-1}$ and $\sigma^\cza_{\,\cczb}\,\leftrightarrow\, B^{\cza\czb}Q^{\ccza}_{\,\czb} c_{\ccza\cczb}$.
}


The generators (\ref{gRprnt_projected}) of gauge transformations in restricted theory\footnote{
 In this section we focus on the gauge transformations (\ref{gRprnt_projected}) of fields $\varphi^\zI$ in the restricted theory (\ref{S_fixed}). However, the configuration space of the latter also contains fields $\lambda_\ccza$. The Noether identities for the equations of motion (\ref{lambdaEoM}) show that the gauge transformation of Lagrange multipliers should be zero, $\var_\xi\lambda_\ccza\equiv \mathcal{R}_{\ccza\cza}\xi^\cza =0$. This is so when the projector $T^{\cza}_{\:\czb}(Q,k)$ (\ref{tkQProj}) in the gauge generators  (\ref{gRprnt_projected}) is constructed with respect to the operator $Q^\ccza_{\,\cza}$ (\ref{def_rQ}) in the whole neighborhood of the restriction surface $\theta^\ccza(\varphi)=0$. This, in particular, implies that $Q^\ccza_{\,\cza}$ has a constant rank in this neighborhood. We assumed this property and so the gauge symmetry representation $\mathcal{R}^{\zI}_{\czb}= \hat{\mathcal{R}}^{\zI}_{\cza\,} T^{\cza}_{\,\czb}(Q,k) $ (\ref{gRprnt_projected}) and $\mathcal{R}_{\ccza\cza} =0$ is enough for the purpose of this paper. In generic case when one uses the projector (\ref{tkQProj}) with some other \emph{constant rank} left kernel, $Q'^\ccza_{\,\cza}=Q^\ccza_{\,\cza} + \Gamma^\ccza_{\cczb\cza}\theta^\cczb$, differing from $Q^\ccza_{\,\cza}$  (\ref{def_rQ}) off the restriction surface, Noether identities imply nonzero generators $\mathcal{R}_{\cczb\cza} =\lambda_\ccza  \Gamma^\ccza_{\cczb\czb} T^{\czb}_{\,\cza}(Q',k)$. This is inevitable when the rank of $Q^\ccza_{\,\cza}$ (\ref{def_rQ}) jumps outside of the restriction surface.
}
are thus far defined in the parental theory field space of $\varphi^\zI$. They determine the gauge transformations $\var_\xi^{\rm red}\varphi^\zI = \mathcal{R}^{\zI}_{\czb}\xi^\czb$ {tangential} to the surface of $\theta^\ccza(\varphi)=0$. As any field on this surface can be {parametrized} by $\phi^\zi$ these gauge transformations can be expressed via the gauge transformations of the {\em reduced restricted} gauge theory $\var_\xi^{\rm red}\phi^\zi=\mathcal{R}^\zi_{\czb}\xi^\czb$, $\var_\xi^{\rm red}\varphi^\zI=e^\zI_{,\zi}\,\var_\xi^{\rm red}\phi^\zi$, so that the corresponding generators $\mathcal{R}^\zi_{\czb}$ of reduced space gauge fields $\phi^\zi$ are related to $\mathcal{R}^{\zI}_{\czb}$ by obvious pushforward relations
 \beq{gR_restricted}
  \mathcal{R}^{\zI}_{\czb}(e^\zI(\phi))
  \,=\,
  \frac{\var e^\zI(\phi)}{\var \phi^\zi} \mathcal{R}^{\zi}_{\czb} (\phi^\zi)
  \,\equiv\,
  e^\zI_{,\zi}\mathcal{R}^{\zi}_{\czb}.
 \eeq

Thus, the reduced-space representation of the restricted gauge theory has reducible generators $\mathcal{R}^{\zi}_{\czb}$ and their first-stage reducibility generators $\mathcal{Z}^\czb_{\,\ccza}$,
 \beq{Reducibility}
  \frac{\var S^{\rm red}[\,\phi\,]}{\var\phi^\zi}\,\mathcal{R}^{\zi}_{\czb}(\phi)
  =0,
  \;\quad\;
  \mathcal{R}^{\zi}_{\czb\,} \mathcal{Z}^\czb_{\,\ccza} = 0
 \eeq
where $\mathcal{Z}^\czb_{\,\ccza}$ exhaust linear combinations of right zero vectors of the projector $T^{\cza}_{\:\czb}$ and have the form
 \beq{ggZ_initial}
   \mathcal{Z}^{\czb}_{\,\ccza}
   = k^{\czb}_{\,\cczb\,} \mu^{\cczb}_{\,\ccza}
 \eeq
with arbitrary\HIDE{ invertible} factor $\mu^{\cczb}_{\,\ccza}$, so that  $\rank { k^{\czb}_{\,\cczb} \mu^{\cczb}_{\,\ccza} } = \rank{ k^{\czb}_{\,\cczb} }$. Such natural choice imply off-shell first-stage reducibility (\ref{Reducibility}) of projected gauge generators.

Note that the projectors (\ref{tkQProj}) generically are nonlocal depending on structure of restriction functions and gauge generators involved in the construction of gauge-restriction operator $Q^\ccza_{\,\cza}$.

   \subsection{Gauge algebra of a restricted theory}
     \label{SSect:Restricted_Gauge_Algebra}
To proceed further we have to know the gauge algebra of the restricted theory generators. The \emph{a priori} maximum that can be stated from the first of their relations (\ref{Reducibility}) (under certain assumptions of regularity of $\mathcal{R}^{\zi}_{\czb}(\phi)$ \cite{BV3}) is that they satisfy an open algebra (\ref{Jacobi_Id_Lagrange}) with some structure functions and $S_{,\zi}$ replaced by $S^{\rm red}_{,\zi}$. Quite remarkably, we will be able to derive this algebra first by expressing via Eq.\,(\ref{gR_restricted}) $\mathcal{R}^{\zi}_{\czb}$ in terms of $\mathcal{R}^{\zI}_{\czb}$ and using the algebra of the projected (full space) generators (\ref{gRprnt_projected}), $\mathcal{R}^{\zI}_{\czb}= \hat{\mathcal{R}}^{\zI}_{\cza} T^{\cza}_{\:\czb}$. The latter, in its turn, follows from the gauge algebra of the parental theory and turns out to be also closed in the case of closure of the algebra of $\hat{\mathcal{R}}^\zI_{\cza}$.  Thus, finally we will show that the resulting algebra is also closed as long as we start with the closed algebra of the parental theory.

Expressing $\mathcal{R}^{\zi}_{\czb}$ in terms of $\mathcal{R}^{\zI}_{\czb}$ goes with the aid of covariant vectors $e^\zi_\zI$ dual to $e^\zJ_{,\zj}$
  \beq{gR_restricted_inverse}
    \mathcal{R}^{\zi}_{\cza}(\phi)
    \,=\, e^\zi_\zI\mathcal{R}^\zI_{\cza}\big(e(\phi)\big).
  \eeq
These vectors satisfy the biorthogonality relation $e^\zi_\zI\,e^\zI_{,\zj}=\delta^\zi_{\,\zj}$
and represent a part of the complete local basis $(e^\zi_\zI,\theta^\ccza_{,\zI})$ in the cotangent bundle of the manifold of $\varphi^\zI$. Vectors $e^\zi_\zI$ may be chosen orthogonal to the set $\theta^\zI_\cczb$ dual to $\theta^\ccza_{,\zI}$ (\ref{theta_dual}), so that $(e^\zI_{,\zi},\theta^\zI_\ccza)$ and  $(e^\zi_\zI,\theta^\ccza_{,\zI})$ form biorthogonal pair of tangent and cotangent bases
  \bea{}
   &&\hspace{-1cm}e^\zi_\zI\,e^\zI_{,\zj}=\delta^\zi_{\,\zj},
    \;\;\;
    \theta^\ccza_{,\zI}\,\theta^\zI_\cczb=\delta^\ccza_{\,\cczb},
    \;\;\;
    e^\zi_\zI\,\theta^\zI_{\cczb}=0,
    \;\;\;
    \theta^\ccza_\zI\,e^\zI_{,\zj}=0,       \label{basis_biorthogonality}
   \\
   &&\hspace{-1cm}e^\zI_{,\zi}\,e^\zi_\zJ+\theta^\zI_\ccza\,
    \theta^\ccza_{,\zJ}=\delta^\zI_{\,\zJ},
   \label{basis_completeness}
  \eea
the last equation here expressing completeness of this basis.\footnote{\label{foot11}
 A particular ``{symmetric}'' biorthogonal basis may be constructed with the aid of the nondegenerate metric $G_{\zI\zJ}$ on the manifold of $\varphi^\zI$. This induces the metric $G_{\zi\zj}\equiv e^\zI_{,\zi}\,G_{\zI\zJ}\,e^\zJ_{,\zj}$, $G^{\zi\zj}\equiv(G_{\zj\zi})^{-1}$, on the reduced $\phi^\zi$-space --- the surface of $\theta^\ccza(\varphi)=0$, and its inverse $G^{\zI\zJ}\equiv(G_{\zJ\zI})^{-1}$ induces the metric in the directions normal to this surface, $G^{\ccza\cczb}\equiv \theta^\ccza_{,\zI}\,G^{\zI\zJ}\,\theta^\cczb_{,\zJ}$,
   $G_{\ccza\cczb}\equiv (G^{\cczb\ccza})^{-1}$.
 Starting from mutually orthogonal $e^\zI_{,\zi}$ and $\theta^\ccza_{,\zI}$  one can construct complementary basis elements $e^\zi_\zI$ and $\theta^\zI_\ccza$ satisfying (\ref{basis_biorthogonality}),(\ref{basis_completeness}) as
 $e^\zi_\zI=G^{\zi\zj}e^\zJ_{,\zj} G_{\zJ\zI}$ and
  $\theta^\zI_\ccza = G^{\zI\zJ}\theta^\cczb_{,\zJ}G_{\cczb\ccza}$.
}
For fixed $e^\zI_{,\zj}$ and $\theta^\ccza_{,\zI}$ there is a freedom of choosing complementary basis elements
 \beq{}
  e^{\zi}_{\zI} \,\to\,  e^{\zi}_{\zI} + b^\zi_{\,\ccza} \theta^\ccza_{,\zI},
  \;\,\quad\;\,
  \theta^\zI_{\ccza} \,\to\,  \theta^\zI_{\ccza} - e^\zI_{,\zi} b^\zi_{\,\ccza}
 \eeq
which preserves (\ref{basis_biorthogonality}),(\ref{basis_completeness}). However this freedom does not spoil expression (\ref{gR_restricted_inverse}) since the shift ambiguity term in $e^{\zi}_{\zI} $ is killed due to defining property of restricted generators, (\ref{gRprnt_projected}) \,$ \theta^\ccza_{,\zI}\mathcal{R}^{\zI}_{\czb} = \theta^\ccza_{,\zI}  \hat{\mathcal{R}}^{\zI}_{\cza\,} T^{\cza}_{\:\czb} = Q^\ccza_{\,\cza} T^{\cza}_{\:\czb} =0$, thus making expression (\ref{gR_restricted_inverse}) for $ \mathcal{R}^{\zi}_{\cza}(\phi)$ unambiguous.

Using (\ref{gR_restricted_inverse}) we have
 \bea{10000}
  \mathcal{R}^{\zi}_{[\,\cza,\,\zj}\mathcal{R}^\zj_{\czb\,]} &\;=\;&
  e^\zi_\zI\,\mathcal{R}^\zI_{[\,\cza,\,\zJ}\mathcal{R}^\zJ_{\czb\,]} +
  \mathcal{R}^\zk_{[\,\cza}\,\mathcal{R}^\zj_{\czb\,]} e^\zi_{\zJ,\zj}\,e^\zJ_{,\zk}\nonumber\\
   &\;=\;&
  e^\zi_\zI\,\mathcal{R}^\zI_{[\,\cza,\,\zJ}\mathcal{R}^\zJ_{\czb\,]} ,
 \eea
where square brackets denote pairs of antisymmetrized indices and the second term in the middle part vanishes because for dual $e^\zi_{\zJ}$ and $e^\zJ_{,\zk}$ (\ref{basis_biorthogonality})\, $e^\zi_{\zJ,\zj}e^\zJ_{,\zk}=-e^\zi_\zJ e^\zJ_{,\zk\zj}\equiv
-e^\zi_\zJ \,\var^2 e^\zJ(\phi)/\var\phi^\zk\var\phi^\zj$ and so antisymmetrization in indices kills it. Thus the gauge algebra of $\mathcal{R}^\zi_{\cza}$ directly follows from that of $\mathcal{R}^\zI_{\cza}$.

As shown in Appendix \ref{ASect:Gauge_Struct_Restriction} the commutator of projected generators $\mathcal{R}^\zI_{\cza}$ replicates the algebraic relation of the parental theory. If the latter is prescribed by
 \beq{gRprnt_commutator_COPY}
   \hat{\mathcal{R}}^\zI_{\cza,\zJ} \hat{\mathcal{R}}^\zJ_{\czb}
   \,-\,
   \hat{\mathcal{R}}^\zI_{\czb,\zJ} \hat{\mathcal{R}}^\zJ_{\cza}
  \;=\;  \hat{\mathcal{R}}^\zI_{\czc} \hat{C}^{\czc}_{\cza\czb} +  \hat{E}^{\zI\zJ}_{\cza\czb}\hat{S}_{,\zJ},
 \eeq
then the projected generators of the restricted gauge theory $\mathcal{R}^{\zI}_{\czb}= \hat{\mathcal{R}}^{\zI}_{\cza} T^{\cza}_{\:\czb}$, which are defined by Eqs.\,(\ref{gRprnt_projected})--(\ref{tkQProj}) both on and {\em outside} of the constraint surface $\theta^\ccza=0$,  are subject to a similar\HIDE{ gauge algebra} relation
  \beq{Jacobi_Id_Lagrange_projected_COPY}
   \mathcal{R}^\zI_{\cza,\zJ} \mathcal{R}^\zJ_{\czb}
   \,-\,
     \mathcal{R}^\zI_{\czb,\zJ} \mathcal{R}^\zJ_{\cza}
  \;=\;  \mathcal{R}^\zI_{\czc} C^{\czc}_{\cza\czb}
  + {E}^{\zI\zJ}_{\cza\czb} \hat{S}_{,\zJ}
  \eeq
with new structure functions $C^{\czc}_{\cza\czb}$ and ${E}^{\zI\zJ}_{\cza\czb}$
 \bea{}
  {C}^{\czc}_{\cza\czb}
  &\equiv&
   T^{\czc}_{\:\czz} \hat{C}^{\czz}_{\czd\cze}  T^{\czd}_{\:\cza}  T^{\cze}_{\:\czb}
   + N^{\czc}_{\,\czb\cza}
   - N^{\czc}_{\,\cza\czb},\label{cT_restricted}
 \\
  {E}^{\zI\zJ}_{\cza\czb} 
  &\equiv&
  D^{\zI}_{\,\zK}
  D^{\zJ}_{\,\zL}
   \hat{E}^{\zK\zL}_{\czc\czd} \,
   T^{\czc}_{\:\cza}
   T^{\czd}_{\:\czb},
  \label{cE_restricted}
 \eea
where
\bea{}
    N^{\czc}_{\,\cza\czb}\!
  &\equiv&  T^{\czc}_{\:\czd} k^{\czd}_{\,\ccza,\zJ}
  {(Qk)^{-1}} \vphantom{|} ^\ccza_{\,\cczb} Q^{\cczb}_{\,\cza} \mathcal{R}^\zJ_{\,\czb}, \label{N}\\
  D^{\zI}_{\,\zJ}\,
   &\equiv& \delta^{\zI}_{\,\zJ}
  -\hat{\mathcal{R}}^\zI_{\,\cze}k^{\cze}_{\,\ccza}
  {(Qk)^{-1}} \vphantom{|} ^\ccza_{\,\cczb}
   \theta^{\cczb}_{,\zJ}.                    \label{D}
\eea

Thus, in view of (\ref{10000}) for a \emph{closed} algebra of the parental theory with $\hat{E}^{\zI\zJ}_{\cza\czb} =0$ one gets a \emph{closed} gauge algebra of reduced fields representation of the restricted theory with ${E}^{\zI\zJ}_{\cza\czb} =0$ and the same structure functions (\ref{cT_restricted})
\bea{reduced_algebra}
   \mathcal{R}^\zi_{\cza,\zj} \mathcal{R}^\zj_{\czb} \,-\,  \mathcal{R}^\zi_{\czb,\zj} \mathcal{R}^\zj_{\cza}
  &=&  \mathcal{R}^\zi_{\czc} C^{\czc}_{\cza\czb}.
  \eea
These structure functions can be nonlocal and with respect to its lower indices have both transversal and longitudinal nature regarding the projector $T^{\cza}_{\:\czb}$ because of the properties of $N^{\czc}_{\,\cza\czb}$-tail of the expression (\ref{cT_restricted}),
$N^{\czc}_{\,\cza\czb}
 = T^{\czc}_{\:\czd} N^{\czd}_{\,\cza\czb}
 = N^{\czc}_{\,\cza\czd} T^{\czd}_{\:\czb}$ and $N^{\czc}_{\,\czd\czb} T^{\czd}_{\:\cza}=0$.

 \section{One-loop effective action of restricted gauge theory}
  \label{Sect:1loop_EA_Restricted}
   \subsection{Two representations of the one-loop effective action}
    \label{SSect:1loop_EA_Restricted}
The reduced representation of the restricted gauge theory, which was constructed above, is subject to the BV formalism of the first-stage reducible model with first-stage reducibility generators $\mathcal{Z}^{\cza}_{\,\ccza} \propto k^\cza_{\,\cczb}$. Therefore, one can use Eq.\,(\ref{Z_1loop_1st}) or Eq.\,(\ref{EA_1loop_1st}) along with Eqs.\,(\ref{invPropagator1})--(\ref{invggPropagator1}) and the replacement of $S[\,\phi\,]$ by $S^{\rm red}[\,\phi\,]$ in order to build its one-loop effective action. However, as it was discussed above, our goal is to perform quantization in the representation of the original parental fields $\varphi^\zI$, rather than in terms of the reduced variables $\phi^\zi$. Hence, an interesting task arises --- to convert these algorithms into this representation.

Such a conversion is, of course, based on the relations of embedding the restricted theory into the parental one (\ref{embedding})--(\ref{embeddingS}) and on the classical equation of motion (\ref{lambdaEoM}) which determines the background on top of which the semiclassical expansion is built. These relations imply that all field-dependent entities of the restricted theory, including all gauge-fixing elements, result from the embedding of $\phi^\zi$ into the space of $\varphi^\zI$. Conversely, the objects of the restricted theory are the functions on the parental configuration space. Therefore, like the relation (\ref{embeddingS}) for the classical parental and reduced actions $S^{\rm red}[\,\phi\,]=\hat S[\,e(\phi)\,]$, we have
  \bea{}
   &&\hspace{-0.5cm}
    \chi^\cza, \sigma^{\cza}_{\,\ccza}, \omega^{\ccza}_{\,\cza}, \varkappa^{\cza\czb}, \rho^{\ccza}_{\,\cczb}
   \;=\;
    \chi^\cza(e(\phi)),\, \sigma^{\cza}_{\,\ccza}(e(\phi)),\,\omega^{\ccza}_{\,\cza}(e(\phi)),\,
   \nonumber\\
   &&\qquad\qquad\qquad\qquad\;\;\;\;
    \varkappa^{\cza\czb}(e(\phi)),\, \rho^{\ccza}_{\,\cczb}(e(\phi)),
  \eea
where, to avoid messy notation, we do not supply $\chi^\cza(\varphi)$ and other gauge-fixing quantities of the parental theory by hats. In particular, this means that the gauge condition matrices $X^\cza_{,\zi}$, (\ref{Xalphai}), and $X^\cza_{,\zI}$ in both representations are related by the embedding formula for a covector,
  \beq{Solution0aaggfc1}
    X^\cza_{,\zi}=e^\zI_{,\zi}X^\cza_{,\zI},\quad
    X^\cza_{,\zI}=\chi^\cza_{,\zI}
     - \sigma^\cza_{\cczc,\zI}\,({s}^{}_{} \sigma^{}_{})^{-1\,\ccza}_{\;\;\;\;\;\;\,\cczb}\,
     {s}^\cczb_{\,\czb} \,\chi^\czb.
  \eeq

The relation between $S^{\rm red}_{,\zi\zj}$ and $\hat S_{,\zI\zJ}$ is trickier. From (\ref{embeddingS}) it follows that on shell
  \bea{}
    S^{\rm red}_{,\zi\zj}
   &=&
    \big(e^\zI_{,\zi}e^\zJ_{,\zj}\hat S_{,\zI\zJ}+\hat S_{,\zI}e^\zI_{,\zi\zj} \big)\big|_{\,\theta^\ccza=0}
   \nonumber\\
   &=&
    e^\zI_{,\zi}e^\zJ_{,\zj}\big(\hat S_{,\zI\zJ}-\lambda_\ccza\,\theta^\ccza_{,\zI\zJ}\big) \big|_{\,S^{\lambda}_{,\zI}=0,\;\;\theta^\ccza=0}
   \nonumber\\
   &\equiv&
    e^\zI_{,\zi}e^\zJ_{,\zj} S^{\lambda}_{,\zI\zJ}\,\big|_{\,S^{\lambda}_{,\zI}=0,\;\;\theta^\ccza=0}\,,
  \eea
where we took into account the equation of motion $\hat S_{,\zI}=\lambda_\ccza\theta^\ccza_{,\zI}$ and the corollary of Eq.\,(\ref{embedding}) $\theta^\ccza_{,\zI\zJ}e^\zI_{,\zi}e^\zJ_{,\zj}+\theta^\ccza_{,\zI}e^\zI_{,\zi\zj}=0$. Therefore, with the inclusion of the gauge-breaking term we have for the operator (\ref{invPropagator1})
  \beq{invPropagator2}
    F_{\zi\zj}
   \,=\,
    e^\zI_{,\zi}e^\zJ_{,\zj} F_{\zI\zJ},
   \;\quad\;
    F_{\zI\zJ}
   \,=\,
    \hat S_{,\zI\zJ}\!-\!\lambda_\ccza\,\theta^\ccza_{,\zI\zJ}
    \!-\!X^\cza_{,\zI}\varPi_{\cza\czb}X^\czb_{,\zJ}.
  \eeq

What remains now is to convert the functional determinant of $F_{\zi\zj}$ on the space of $\phi^\zi$ to the functional determinant of $F_{\zI\zJ}$ on the space of $\varphi^\zI$. This can be done by comparing two expressions for one and the same Gaussian integral with the quadratic part of the action $S^\lambda[\,\varphi,\lambda\,]$ in the exponential. On the one hand it equals
  \begin{eqnarray}
    &&\int Dh\,D\lambda\; e^{\,i\left(\frac12\,h^\zI F_{\zI\zJ}h^\zJ - \lambda_\ccza\theta^\ccza_{,\zI}h^\zI\right)}
   \nonumber\\
   &&\qquad\qquad=\left(\det
    \left[\begin{array}{c|c} 
    F_{\zI\zJ}\vphantom{\big|_|}&
    -\theta^\cczb_{,\zJ}\\
    \hline
    -\theta^\ccza_{,\zI}\vphantom{\big|^|}&0
    \end{array}
    \right]\right)^{-1/2}\nonumber\\
   &&\qquad\qquad=
    \big(\!\det F_{\zI\zJ}\big)^{-1/2}\,
    \big(\!\det \varTheta^{\ccza\cczb}\big)^{-1/2},                                    \label{detF0}\\
    &&\varTheta^{\ccza\cczb} \,\equiv\,
    \theta^\ccza_{,\zI}F^{-1\,\zI\zJ}\theta^\cczb_{,\zJ},
    \quad \quad
    \label{Theta}
  \end{eqnarray}
where $h^\zI$ denotes perturbations of $\varphi^\zI$. On the other hand, this integral can be calculated in the parametrization of reduced fields $\phi^\zi$ with perturbations $h^\zi$ and specially chosen set of remaining ``reducibility'' fields $\theta^\ccza$ with perturbations $h^\ccza$,
 \beq{reparametrization}
  h^\zI = e^\zI_{,\zi}h^\zi_{(e)}+\theta^\zI_{\ccza}h^\ccza_{(\theta)},
  \;\quad\;
  h^\zi_{(e)} = e^\zi_{,\zI} h^\zI,
  \;\quad\;
  h^\ccza_{(\theta)} = \theta^\ccza_{,\zI} h^\zI,
 \eeq
where $e^\zI_{,\zi}$, $\theta^\ccza_{,\zI}$, $e^\zi_{\zI}$, $\theta^\zI_{\ccza}$ are calculated on the classical background and chosen to satisfy the biorthogonality relations (\ref{basis_biorthogonality}),(\ref{basis_completeness}). Thus, making the change of integration variables $h^\zI \to (h^\zi_{(e)}, h^\ccza_{(\theta)})$ we have for the same integral
 \bea{}
   &&\int Dh\; e^{\,i\,\frac12 h^\zI F_{\zI\zJ}h^\zJ} \delta(\theta^\ccza_{,\zI}h^\zI)
   \nonumber\\
   &&\qquad=\;\det\big[\,e^\zI_{,\zi}\,\,\theta^\zI_{\ccza}\,\big]\int D h_{(e)}\,D h_{(\theta)}\; e^{\,i\,\frac12 h^\zi_{(e)} F_{\zi\zj} h^\zj_{(e)}} \delta( h^\ccza_{(\theta)})\nonumber
   \\
   &&\qquad= \;\det\big[\,e^\zI_{,\zi}\,\,\theta^\zI_{\ccza}\,
   \big]\,\big(\!\det F_{\zi\zj}\big)^{-1/2},   \label{detF1}
 \eea
whence the comparison of these two expressions gives
 \beq{}
    \det F_{\zi\zj} \,=\, \det F_{\zI\zJ}\, \det \varTheta^{\ccza\cczb}\,
    \big(\!\det\big[\,e^\zI_{,\zi}\,\,\theta^\zI_{\ccza}\,
    \big]\big)^2.    \label{detF}
 \eeq

For ultralocal restriction constraint (\ref{rSpace_rC}) the reparametrization (\ref{reparametrization}) is also ultralocal in spacetime, $e^\zI_{,\zi}, \theta^\ccza_{,\zI}, e^\zi_{\zI}, \theta^\zI_{\ccza} \sim \delta(x,y)$, and the last squared determinant here contributes to the effective action $\delta(0)$ terms. Repeating the derivation procedure of Eqs.\,(\ref{reparametrization})--(\ref{detF}) with $F_{\zI\zJ}$ replaced by some ultralocal symmetric matrix $G_{\zI\zJ}$ proportional to undifferentiated delta function of spacetime coordinates --- the metric on the configuration space of $\varphi^\zI$,
 \beq{}
  G_{\zI\zJ}\sim\delta(x_\zI,x_\zJ),
 \eeq
one finds the expression for Jacobian $\big[\,e^\zI_{,\zi}\,\,\theta^\zI_{\ccza}\,\big]$ in the right-hand side of (\ref{detF}) in terms of the corresponding functional determinants of this metric, its inverse $G^{\zI\zJ}\equiv(G_{\zJ\zI})^{-1}$, the induced metric $G_{\zi\zj}$ of the reduced $\phi^\zi$-space and the metric $G_{\ccza\cczb}$ in the $\theta^\ccza$-directions (see footnote \ref{foot11})
 \bea{}
  &&G_{\zi\zj}\equiv e^\zI_{,\zi}\,G_{\zI\zJ}\,e^\zJ_{,\zj},
  \;\quad\; G^{\zi\zj}
  \equiv(G_{\zj\zi})^{-1},\nonumber\\
  &&G^{\ccza\cczb}\equiv \theta^\ccza_{,\zI}\,G^{\zI\zJ}\,\theta^\cczb_{,\zJ},
   \;\quad\;
   G_{\ccza\cczb}\equiv (G^{\cczb\ccza})^{-1}.
   \label{metrics}
 \eea
The square of this Jacobian therefore reads
 \beq{}
  \big(\!\det\big[\,e^\zI_{,\zi}\,\,\theta^\zI_{\ccza}\,\big]\big)^2
  \,=\,
  \frac{\det G_{\zi\zj}}{\det G_{\zI\zJ}}\,\det G_{\ccza\cczb}
  \,\sim\, \exp\big(\delta(0)(...)\big).
 \eeq
Note that $G^{\ccza\cczb}$ is ultralocal (for ultralocal restrictions) in contrast to the operator $\varTheta^{\ccza\cczb}$ defined by Eq.\,(\ref{Theta})

Thus, this is an inessential normalization factor of the generating functional or, if treated seriously, it can be absorbed into the definition of the local path integral measure provided one identifies the ultralocal matrix $G_{\zi\zj}$ with the Hessian of the gauge-fixed reduced action $S^{\rm red}_{\rm gf}[\,\phi\,]$ with respect to the configuration space velocities $\dot\phi$,\footnote{
 Here, of course, the functional dependence of the reduced and parental actions on velocities is mediated by their gauge-fixed Lagrangians, i.e. $S^{\rm red}_{\rm gf}=\int dt\, L^{\rm red}_{\rm gf}(\phi,\dot\phi)$ and $S^{\lambda}_{\rm gf}=\int dt\, L^{\lambda}_{\rm gf}(\varphi,\dot\varphi)$. The gauge-fixed action is just the corresponding classical action with added gauge breaking term. Also remember that we consider relativistic gauges, so that the ``kinetic'' metric (\ref{kin_metric}) in reduced theory is nondegenerate. Extending $e^\zi_{\zI} G_{\zi\zj} e^\zj_{\zJ}$ with linear combinations $b_{\zI\ccza} \theta^\ccza_{,\zJ} + \theta^\ccza_{,\zI}b_{\zJ\ccza}$ one acquires the nondegenerate metric $G_{\zI\zJ}$.
}
  \bea{kin_metric}
   &&\hspace{-1cm}
    \int D\phi \,\,\to\,\, \int D\phi\, \big(\!\det G_{\zi\zj} \big)^{1/2},\nonumber\\
   &&\hspace{-1cm}
    G_{\zi\zj}=\frac{\var^2 S^{\rm red}_{\rm gf}}{\var\dot\phi^\zi \var\dot\phi^\zj}=
    \frac{\var^2 S^{\lambda}_{\rm gf}}{\var\dot\varphi^\zI\var\dot\varphi^\zJ}\,
    e^\zI_{,\zi}e^\zJ_{,\zj}\equiv e^\zI_{,\zi}G_{\zI\zJ}e^\zJ_{,\zj}.
  \eea
Then $\det F_{\zi\zj}$ in the one-loop expressions (\ref{Z_1loop_1st}) or (\ref{EA_1loop_1st}) get replaced by
  \beq{}
    \det F^\zi_{\:\zj} \,=\, \det F^\zI_{\:\zJ}\,
    \det\varTheta^\ccza_{\:\cczb},
   \label{detFfin}
  \eeq
where raising the indices of operator {\em forms} $F_{ij}$ and $F_{\zI\zJ}$ is done by the corresponding local metrics of Eq.\,(\ref{metrics}) and the second index in the operator of the last determinant is analogously lowered by the ``induced'' metric $G_{\ccza\cczb}$ along normal directions,
 \beq{}
  F^\zi_{\:\zj}\equiv G^{\zi\zk}F_{\zk\zj},
  \,\quad\,
  F^\zI_{\:\zJ}\equiv G^{\zI\zK}F_{\zK\zJ},
  \,\quad\,
  \varTheta^\ccza_{\:\cczb}\equiv \varTheta^{\ccza\cczc}G_{\cczc\cczb}.
 \eeq

Replacement of $F_{\zi\zj}$ by (\ref{detFfin}) in (\ref{Z_1loop_1st}) finally gives the one-loop generating functional of a restricted gauge theory fully in terms of the parental theory structures
 \bea{Z_1loop_1st1}
  \hspace{-1cm}Z^{1-\rm loop}_{\rm restricted}
  &=&
    \;\frac{\det \rho^\ccza_{\,\cczb}}{\big(\!\det \varkappa^{\cza\czb}\big)^{1/2} \big(\!\det \kappa_{\ccza\cczb}\big)^{1/2}}\nonumber\\
   &&\times\;\frac{\det F^{\cza}_{\;\czb} }{
     \big(\!\det F^{\zI}_{\:\zJ} \big)^{1/2}
     \big(\!\det\varTheta^\ccza_{\:\cczb}\big)^{1/2}
     \det F^{\ccza}_{\;\cczb}  }\,.
 \eea
Here the gauge-fixed ghost operator $F^{\cza}_{\;\czb}$ is, of course, given by the unreduced theory version of (\ref{invgPropagator1}) with the projected generators ${\mathcal R}^\zI_\cza=\hat{\mathcal R}^\zI_\czc\, T^{\czc}_{\:\czb}(Q,k)$ (\ref{gRprnt_projected}),
 \beq{}
  F^{\cza}_{\;\czb}
  \,=\,X^ \cza_{,\zI}\hat{\mathcal R}^\zI_\czc\, T^{\czc}_{\:\czb}
     - \sigma^{\cza}_{\,\ccza} {\rho^{-1}}^{\ccza}_{\,\cczb}\,  \omega^{\cczb}_{\,\czb}.
  \label{invgPropagator3}
 \eeq

Note that despite ultralocal reparametrization (\ref{reparametrization}) the transition to parental theory representation leads to the additional nontrivial factor --- the determinant of (\ref{Theta}). The complexity of this factor follows from the fact that this is no longer a determinant of the local differential operator. Rather, this is the determinant of a {\em nonlocal} object $\varTheta^\ccza_{\:\cczb}$ --- the Green's function of the local operator $F_{\zI\zJ}$ sandwiched between two normal covariant vectors $\theta^\ccza_{,\zI}$ and $\theta^\cczb_{,\zJ}$.

    \auxtitle{Ambiguity in the choice of generator bases}

The BV generating functional (\ref{GeneratingFunctional})  is not invariant under the change of generator bases of $\hat{\mathcal{R}}^\zI_\cza$ and $\mathcal{Z}^\cza_{\,\cczb}$ as obvious from one-loop expressions (\ref{Z_1loop_1st}) or (\ref{Z_1loop_1st1}). When linearly transformed with respect to their gauge index $\cza$ gauge generators $\hat{\mathcal{R}}^\zI_\cza$ still remain the generators of gauge transformations of the parental theory. A unique representations of $\hat{\mathcal{R}}^\zI_\cza$ is not \emph{a priori} fixed in Lagrangian theory. The linear transformations of $\mathcal{Z}^\cza_{\,\cczb}$ with respect to its reducibility index $\cczb$ are also not fixed and this ambiguity reveals itself in the (yet) arbitrary matrix $\mu^\ccza_{\,\cczb}$ (\ref{ggZ_initial})
  \beq{Zmu}
    \mathcal{Z}^\cza_{\,\cczb} \,=\, k^\cza_{\,\ccza} \, \mu^\ccza_{\,\cczb}.
  \eeq
Ambiguity of the linear reparametrizations of generators results in ``normalizing'' factors in the generating functional and is the well-known feature of the BV formalism which will be addressed later in this section.

Another ambiguity is the choice of a projector parameter $k^\cza_{\,\ccza}$ which implicitly enters restricted theory gauge generators $\mathcal{R}^\zI_\cza$ (\ref{gRprnt_projected}) via projectors  (\ref{tkQProj}) and reducibility generators $\mathcal{Z}^\cza_{\,\cczb} \propto k^\cza_{\,\ccza}$. The only restriction on the choice of $k^\cza_{\,\ccza}$ is that it should satisfy the rank conditions (\ref{g_rank}). Otherwise it is arbitrary, and this arbitrariness may extend to the effective action (\ref{Z_1loop_1st1}).
The problem of potential dependence of the latter on $k^\cza_{\,\ccza}$ is a specific issue of reducible gauge structure approach to restricted theory and it should be fixed independently of the generators normalization issue. Below we show that the requirement of {independence} of the generating functional (\ref{Z_1loop_1st1}) on the choice of $k^\cza_{\,\ccza}$ may be satisfied by a special choice of the factor $\mu^\ccza_{\,\cczb}$ in (\ref{Zmu}).

   \subsection{Independence of projector parameter}
    \label{SSect:Projector_Param_Independence}
The variational equation
 \bea{vareq}
  &&\!\!-\var_k\varGamma^{1-\rm loop} =\, \var_k\ln\frac{\det F^\cza_{\;\czb}}{\det F^\ccza_{\;\cczb}}\nonumber\\
  &&\quad=\, F^{-1\,\cza}_{\;\;\;\;\,\,\,\czb} X^\czb_{,\zI}\,\var_k\mathcal{R}^\zI_\cza
  -(\omega\mathcal{Z})^{-1\,\ccza}_{\;\;\;\;\,\,\,\cczb} \,\omega^\cczb_{\,\czb} \var_k \mathcal{Z}^\czb_{\,\ccza}\,=\,0
  \qquad\quad
 \eea
can be transformed by using the relation $\var_k\mathcal{R}^\zI_\cza= \hat{\mathcal{R}}^\zI_\czb\,\var_k T^\czb_{\:\cza}(Q,k)=\mathcal{R}^\zI_\czb\,\var_k T^\czb_{\:\cza}(Q,k)$ (note that $\var_k T^\czb_{\:\cza}(Q,k)\propto T^\czb_{\:\czc}(Q,k)$), the analog of the Ward identity (\ref{Ward100}) and the identity (\ref{Ward10}), so that
 \bea{}
  F^{-1\,\cza}_{\;\;\;\;\;\;\,\czc}\,(X^\czc_{,\zi}\mathcal{R}^\zi_\czb)
  &=&
  F^{-1\,\cza}_{\;\;\;\;\;\;\,\czc}\big(F^\czc_{\;\czb}
    +\sigma^{\czc}_{\,\ccza}\,{\rho^{-1}}^{\ccza}_{\,\cczb}\,  \omega^{\cczb}_{\,\czb}\big)\nonumber\\
    &=&
    \delta^\cza_{\,\czb} - \mathcal{Z}^\cza_{\,\ccza} (\omega\mathcal{Z})^{-1\,\ccza}_{\;\;\;\;\;\,\cczb}\omega^\cczb_{\,\czb}.
  \label{Ward102}
 \eea
For the first term in the right-hand side of (\ref{vareq}) this gives
 \beq{}
  F^{-1\,\cza}_{\;\;\;\;\;\;\czb} \,X^\czb_{,\zI}\,\var_k\mathcal{R}^\zI_\cza
  \,=\,
   (\omega\mathcal{Z})^{-1\,\ccza}_{\;\;\;\;\;\,\cczb}\, \omega^\cczb_{\,\czb}\,T^\czb_{\:\cza}(Q,k)\,\var_k \mathcal{Z}^\cza_{\,\ccza},
 \eeq
where we used the fact that $\var_k T^\cza_{\:\cza} = \var_k(\delta^\cza_{\,\cza}-\delta^\ccza_{\,\ccza})=0$ and $\var_k T^\czb_{\:\cza} \mathcal{Z}^\cza_{\,\ccza} =-T^\czb_{\:\cza}\var_k \mathcal{Z}^\cza_{\,\ccza}$ in view of $T^\czb_{\:\cza}(Q,k) \mathcal{Z}^\cza_{\,\ccza}=0$. Thus Eq.\,(\ref{vareq}) takes the form
 \bea{}
  &&\hspace{-1cm}(\omega\mathcal{Z})^{-1\,\ccza}_{\;\;\;\;\;\,\cczb}\, \omega^\cczb_{\,\czb}\,\big(T^\czb_{\:\cza}(Q,k)-\delta^\czb_{\,\cza}\big)\,\var_k \mathcal{Z}^\cza_{\,\ccza}\nonumber\\
  &&\hspace{-1cm}\qquad\,=\,(\omega\mathcal{Z})^{-1\,\ccza}_{\;\;\;\;\;\,\cczb}\,(\omega k)^\cczb_{\,\cczc}\,(Qk)^{-1\,\cczc}_{\;\;\;\;\;\,\cczd}\,Q^{\cczd}_{\,\cza}\,\var_k \mathcal{Z}^\cza_{\,\ccza} \,=\, 0.
  \quad
 \eea
With the expression (\ref{Zmu}) this equation reads in condensed matrix notations
 \bea{}
  &&\hspace{-1cm}{\rm tr}\, \big(\mu^{-1}\var_k\mu+(Qk)^{-1}\var_k(Qk)\big)\nonumber\\
  &&\hspace{-1cm}\qquad=\,\var_k\ln\big(\det\mu\,\det(Qk)\big)\,=\,0,
 \eea
where ${\rm tr}$ denotes the trace over indices of the matrices $\mu=\mu^\ccza_{\,\cczb}$ and $(Qk)=(Qk)^\ccza_{\,\cczb}$. Without loss of generality this equation can be solved by $\mu^\ccza_{\,\cczb}=(Qk)^{-1\,\ccza}_{\;\;\;\;\;\,\cczb}$, so that finally
 \beq{Zmu1}
  \mathcal{Z}^\cza_{\,\cczb} \,=\, k^\cza_{\,\ccza}\,(Qk)^{-1\,\ccza}_{\;\;\;\;\;\,\cczb}.
 \eeq


   \subsection{Canonical normalization of generators}
    \label{SSect:Canon_Measure_Normalization}
Specification of the gauge generators basis $\hat{\mathcal{R}}^\zI_\cza$ is a more complicated issue. However, we will give brief and, perhaps, not so exhaustive arguments in favor of a concrete choice which, in particular, will confirm the form (\ref{Zmu1}) of $\mathcal{Z}^\cza_{\,\cczb}$. Conventional Lagrangian quantization in the form of the Faddeev-Popov integral (or the BV integral in more complicated models with open and reducible algebras) is not intrinsically closed. It does not provide uniquely the concrete form of the local measure and does not resolve the associated problem of the choice of the generators bases. In order to fix them one should appeal to the canonical form of the path integral which for a general class of relativistic gauge conditions comprises the BFV formalism\HIDE{\cite{Henneaux:1992ig,FV:1975}}. For simple gauge systems\footnote{
 The simplification used here is that the Lagrangian gauge theory (which later will be the parental gauge system) admits the so-called one-step Hamiltonization. That is, when performing the Legendre transform only with respect to the fields with velocities in the Lagrangian action, one obtains the constrained canonical action having only first-class constraints, and Dirac consistency equations for these ``primary'' constraints do not generate new ``secondary'' constraints. This is the case of Einstein general relativity, Yang-Mills theory, and many others.
}
fixing the measure and the basis of generators looks as follows.

The starting point is the canonical formalism of the gauge theory whose canonical action $S_{\rm can}=\int dt\,(p\,\dot q\,-H - v^\cza \gamma_\cza)$ explicitly contains first-class constraints $\gamma_\cza$ dual to the Lagrange multipliers $v^\cza$ which are a part of the Lagrangian configuration space of the theory $\varphi^\zI=(q,v)$. This action is invariant with respect to gauge transformations which are canonical (ultralocal in time and generated by Poisson brackets with constraints) in the sector of phase space variables $\var^\xi(q,p)=\{(q,p),\gamma_\cza\}\xi^\cza$, but contain the first-order time derivative of the gauge transformation parameter in the sector of Lagrange multipliers, $\var^\xi v^\cza=\dot\xi^\cza+...\,$. Here dots denote ultralocal in time terms containing structure functions of the Poisson bracket algebra of first-class constraints and the Hamiltonian $H$, their explicit form being unimportant for us in what follows. The corresponding canonical path integral in the class of {\em canonical} gauges $\chi^\cza=\chi^\cza(q,p)$ reads as
\bea{}
 &&Z=\int Dq\,Dp\,Dv\, e^{\,i\,S_{\rm can}[q,p,v]}\prod_t \delta(\chi^\cza)\,\det\{\chi^\cza,\gamma_\czb\}\nonumber\\
 &&\quad=\int Dq\,Dp\, e^{\,i\int dt(p\dot q - H)} 
  \prod_t \delta(\chi^\cza)\,\delta(\gamma_\cza)
  \det\{\chi^\cza,\gamma_\czb\},
  \nonumber\\
\eea
and it is obviously invariant under the linear changes of the basis of constraints, $\gamma_\cza \,\to\, \gamma'_\cza=\gamma_\czb\varOmega^\czb_{\,\cza}$, with any ultralocal in time and nondegenerate matrix $\varOmega^\czb_{\,\cza}$. Subsequent integration over momenta \,$p$\, converts this integral into the Lagrangian form which, modulo corrections associated with the transition to the Lagrangian expressions for momenta, takes the form of the Faddeev-Popov integral
 \beq{Lagr_int}
  Z \,= \int D\varphi\,\mu[\,\varphi\,]\,\det Q^\cza_{\,\czb}
   \;\delta{(\chi^{\cza})}\,e^{\,i\,S[\,\varphi\,]}.
 \eeq
Here the local measure $\mu[\,\varphi\,]\sim\exp\big[\delta(0)(...) \big]$ absorbs ultralocal in time factors associated with the above corrections, and the Faddeev-Popov operator $Q^\cza_{\,\czb}$ is built in terms of the gauge generators ${\mathcal R}^\zI_{\czb}$ --- the Lagrangian version of the above transformations in the canonical formalism, $\var^\xi\varphi^\zI={\mathcal R}^\zI_{\cza}\xi^\cza\propto \var^\xi(q,v)$,
  \beq{Q}
    Q^\cza_{\,\czb} \,=\, \chi^{\cza}_{,\zI} {\mathcal R}^\zI_{\czb}.
  \eeq
Starting from the BFV canonical quantization\footnote{
 The Faddeev-Popov-type expression (\ref{Lagr_int}) is obtained by neglecting ghost vertices which contribute to higher loops and are irrelevant in the context of the present discussion. In the same manner we neglect possible higher-order loop contributions to the Lagrangian generating functional, which appear when the classical relation between momenta and velocities acquire ghost corrections.
} one can arrive at the same expression for this wider class of gauge conditions \cite{Henneaux:1992ig,FV:1975} including relativistic (or ``dynamical'') gauges so that gauges $\chi^\cza$ involve linear dependence on the time derivatives of Lagrange multipliers $v^\cza$.

The integral (\ref{Lagr_int}) is not explicitly invariant under the rotation of the generator basis, ${\mathcal R}^\zI_\cza\to {\mathcal R'}^\zI_\cza={\mathcal R}^\zI_\czb\varOmega^\czb_{\,\cza}$, but implicitly the choice of this basis is fixed by the requirement that it should match with the basis of canonical gauge transformations $\var^\xi\varphi=\var^\xi(q,v)$ implying that $\var^\xi v^\cza=\dot\xi^\cza+...$ \,or\, ${\mathcal R}^\cza_\czb =\delta^\cza_{\,\czb}(d/dt)+...\,$. The ambiguity of splitting the configuration space of gauge fields $\varphi^\zI\sim(q,v)$ into canonical coordinates and Lagrange multipliers (which is ultralocal in time) can only lead to some inessential local matrix $\mu^\cza_{\,\czb}$ in the last equation, $\delta^\cza_{\,\czb}(d/dt)\to\mu^\cza_\czb(d/dt)$, and some extra factor in the local measure $\mu[\,\varphi\,]\sim\exp\big[ \delta(0)(...)\big]$ which is again unimportant in concrete applications. Thus, the ambiguity in the choice of generator basis ${\mathcal R}^\zI_\czb$ coming from the canonical formalism reduces to its rotation by ultralocal in time matrices, which is equivalent to changing the basis of first-class constraints in the canonical formalism. This ambiguity is physically inessential, so that finally the choice of generator basis is fixed by the requirement of local coefficient of the first-order time derivative in the gauge generators ${\mathcal R}^\zI_\cza$.

Unfortunately, in the case of restricted gauge theories such a line of reasoning does not work directly because the reducible generators obtained by the projection procedure may become nonlocal in time. Moreover, the restriction of gauge theory generically leads to its canonical formalism with a much more complicated structure involving many generations of constraints. Therefore it is much harder to implement with the same level of generality the above scheme starting from the canonical quantization. For this reason we will choose a somewhat different approach in order to show that the equivalence to the canonical quantization fixes the basis of projected generators in restricted gauge theory by the requirement that the local parental theory generators $\hat{\mathcal R}^\zI_\cza$ are canonically induced, that is they satisfy a local normalization condition for their time derivative part.

Consider a parental gauge theory with the action $\hat S[\,\varphi\,]$ whose gauge generators $\hat{\mathcal{R}}^\zI_\cza$ form a closed algebra and are \emph{irreducible}, so that its generating functional is given by a standard Faddeev-Popov path integral. This integral can be represented on shell (with switched off sources) in three equivalent forms differing by the choice of gauge-fixing integration measure,
 \bea{three_measures}
  Z_{\rm parental}
  &\equiv& \int D\varphi\,\hat M^{\rm delta}_{(\chi)}\,e^{\,i\,\hat{S}[\,\varphi\,]}
  \,=\int D\varphi\,\hat M_{(\chi,\varkappa)}\,e^{\,i\,\hat{S}[\,\varphi\,]}\nonumber\\
  &=& \int D\varphi\,\hat M^{\rm delta}_{(\chi,\varkappa,\sigma)}\,e^{\,i\,\hat{S}[\,\varphi\,]}.
 \eea
The measure factors $\hat M^{\rm delta}_{(\chi)}$ and $\hat M_{(\chi,\varkappa)}$ respectively correspond to the delta-function-type and Gaussian-type gauge fixing with the full set of gauge conditions $\chi^\cza$, the latter involving the exponentiated gauge-breaking term with an invertible gauge-fixing matrix $\varkappa_{\cza\czb}$
  \bea{}
   \hat M^{\rm delta}_{(\chi)}
   &=&\det \hat Q^\cza_{\,\czb}
   \,\delta{( \chi^{\cza} )},       \label{BVMJ_irr_delta}\\
   \hat M_{(\chi,\varkappa)}
   &=&
   \det \hat Q^\cza_{\,\czb}\,
   (\det\varkappa_{\cza\czb})^{1/2}\,e^{-i\,\frac{1}2 \chi^\cza\varkappa_{\cza\czb}\chi^\czb}\!\!,\;\;\;    \label{BVMJ_irr_gauss}
   \eea
where $\hat Q^\cza_{\,\czb}$ is a standard Faddeev-Popov operator
 \beq{Q0}
  \hat Q^\cza_{\,\czb} \,=\, \chi^{\cza}_{,\zI} \hat{\mathcal R}^\zI_{\,\czb}.
 \eeq
Obviously, $\hat M^{\rm delta}_{(\chi)}=\hat M_{(\chi,\infty)}$ is a limit of $\hat M_{(\varkappa)}$ at $\varkappa_{\cza\czb}\to\infty$.

The third measure factor is less known and corresponds to the situation when a part of gauge conditions $\chi^p$ are of delta-function-type, whereas the rest of them are enforced via the gauge breaking term with the projector $\varPi_{\cza\czb}$
 \beq{combined}
  \hat M^{\rm delta}_{(\chi,\varkappa,\sigma)}
   \,=\,
   \frac{\det\hat Q^\cza_{\,\czb}\,\big(\!\det\varkappa_{\cza\czb}\big)^{1/2}}{\big(\!\det\kappa_{pq}\big)^{1/2}}
   \, e^{-\frac{i}2 \chi^\cza\varPi_{\cza\czb}\chi^\czb}
   \delta(\sigma^p_{\,\cza} \chi^\cza).
 \eeq
Here
 \beq{chi_p}
  \sigma^p_{\,\cza} \chi^\cza
  \equiv \chi^p
 \eeq
is a subset of gauge conditions obtained by projecting the full subset $\chi^\cza$ with the aid of a vielbein $\sigma^p_{\cza}=\kappa^{pq}\sigma^\czb_{\,q}\varkappa_{\czb\cza}$ which is dual to the set $\sigma^\cza_{\,p}$ introduced above in the formalism of reducible gauge generators,\footnote{\label{ftnt:combined}
 The proof of the last equality in (\ref{three_measures}) can be done by complementing the bases of $\sigma^\cza_{\,p}$ with the remaining vielbein vectors $\sigma^\cza_M$ which are orthogonal to $\sigma^\cza_{\,p}$ in the metric $\varkappa_{\cza\czb}$, $\sigma^\cza_{\,p}\varkappa_{\cza\czb}\sigma^\czb_M=0$ and, therefore, satisfy the determinant relation $\det\,([\begin{array}{cc}\!\sigma^\cza_{\,p}\!&\!\sigma^\cza_{\,M}\!\end{array}\!])
 (\det \varkappa_{\cza\czb})^{1/2}=(\det\kappa_{pq}\det\kappa_{MN})^{1/2}$, where $\kappa_{pq}=\sigma^\cza_{\,p}\varkappa_{\cza\czb}\sigma^\czb_{\,q}$ and $\kappa_{MN}=\sigma^\cza_{\,M}\varkappa_{\cza\czb}\sigma^\czb_{\,N}$ are respectively the metrics on subspaces spanned by $\sigma^\cza_{\,p}$ and $\sigma^\cza_{\,M}$. Then the full delta function of gauge conditions can be decomposed as $\delta(\chi^\cza)=\delta(\chi^p)\delta(\chi^M)
 (\det\varkappa_{\cza\czb})^{1/2}/(\det\kappa_{pq} \det\kappa_{MN})^{1/2}$ and the factor $\delta(\chi^M)/(\det\kappa_{MN})^{1/2}$ here, according to 't Hooft trick, being replaced by\, $\exp(-\frac{i}2\chi^M\kappa_{MN}\chi^N) =\exp(-\frac{i}2 \chi^\cza\varPi_{\cza\czb}\chi^\czb)$ --- the implementation of the second equality of (\ref{three_measures}) in the sector of $\chi^M$ gauges.
}
  \beq{dual_sigma}
    \sigma^p_{\,\cza} \sigma^\cza_{\,q}=\delta^p_{\,q},
    \,\quad\,
    \sigma^\cza_{\,p} \sigma^p_{\,\czb}=\delta^\cza_{\,\czb}-\varPi^\cza_\czb,
    \,\quad\,
    \varPi^\cza_\czb=\varkappa^{\cza\czc} \varPi_{\czc\czb}
  \eeq

On the other hand, inclusion of delta function of the subset of gauge conditions (or partial gauge fixing) can be interpreted as quantization of the {\em restricted} gauge theory with partial gauge-fixing conditions $\chi^p\HIDE{=0}$ playing the role of restriction constraints $\theta^p\HIDE{=0}$. According to our derivation above, this restricted gauge theory has reducible gauge generators $\mathcal{R}^\zI_\cza=\hat{\mathcal{R}}^\zI_\czb T^\czb_{\:\cza}(Q,k)$ built with the aid of the operator $Q^p_{\,\cza}=\chi^p_{,\zI}\hat{\mathcal{R}}^\zI_\cza$ and some set of vectors $k^\cza_{\,p}$, so that its path integral over {\em reduced} configuration space of $\phi^\zi$ should read
 \beq{partial}
  Z_{\rm partial\;gf}=
  \int D\phi\,\,M^{\rm red}_{(\chi,\varkappa,\sigma,\omega,k,\rho)} \,
  e^{\,i\,S^{\rm red}[\,\phi\,]},
 \eeq
where $S^{\rm red}[\,\phi\,]=\hat S[\,\varphi\,]\,|_{\,\chi^p=0}$ and the gauge-fixing integration measure according to the reducible gauge theory algorithm (\ref{Z_1loop_1st}) looks like (we distinguish the partial gauge fixing case from the restricted theory case by replacing the indices $a,b,...$ with the indices $p,q,...$ from the second part of Latin alphabet)\footnote{
 The reason for that which will be clarified below is that the set of restriction conditions $\theta^\ccza$ generically contains a gauge-fixing subset, which will be labeled by indices $p,q,...$ of range $m_1-m_2$, and the remaining conditions of range $m_2$ which restrict gauge invariants of the parental theory.
}
 \bea{}
   &&M^{\rm red}_{(\chi,\varkappa,\sigma,\omega,k,\rho)}=
   \frac{\det F^{\cza}_{\;\czb}\,\det \rho^p_{\,q} }{\det F^{p}_{\:q}}\nonumber\\
   &&\qquad\qquad\quad\quad
   \;\;\times\frac{\big(\!\det\varkappa_{\cza\czb}\big)^{1/2}}
   {\big(\!\det\kappa_{pq}\big)^{1/2}}\;
   e^{-\frac{i}2 \chi^\cza\varPi_{\cza\czb}\chi^\czb}, \;\;\;\\
   &&F^\cza_{\;\czb} \,=\, \hat Q^\cza_{\,\czc}\, T^\czc_{\:\czb}(Q,k)-\sigma^\cza_{\,p}\rho^{-1\,p}_{\;\;\;\;\;q}\omega^q_{\,\czb},\\
   &&F^p_{\:q}\,=\,(\omega \mathcal{Z})^p_q
   \,=\, (\omega k)^p_r\,(Qk)^{-1\,r}_{\;\;\;\;\;q},
   \label{FandF}\\
   &&\,Q \,\equiv\, Q^p_{\,\cza} \,=\, \sigma^p_{\,\czb}\,\hat Q^\czb_{\,\cza}.
 \eea
Here $Q^p_{\,\cza}$ is in fact a projected (from the left) Faddeev-Popov operator $\hat Q^\czb_{\,\cza}$ of the {\em parental} theory, $T^\czc_{\:\czb}(Q,k)$ --- the corresponding projector (\ref{tkQProj}) with kernels $Q^p_{\,\cza}$ and $k^\cza_{\,q}$, $F^\cza_{\;\czb}$ is the gauge-fixed ghost operator, $F^p_{\:q}$ is the ghosts-for-ghosts operator and first-stage reducibility generators $\mathcal{Z}^\cza_{\,p}$ are normalized according to (\ref{Zmu1}).

Remarkable property proven in Appendix \ref{ASect:Determinant_Relation} is the determinant relation which is valid for operators defined by Eqs.\,(\ref{Q0}) and (\ref{FandF}),
 \beq{det_rellation}
  \det\hat Q^\cza_{\,\czb}
  \,=\,
  \frac{\det F^{\cza}_{\;\czb}\,\det \rho^p_{\,q} }{\det F^p_{\:q}}.
 \eeq
It allows one to express the combined measure (\ref{combined}) in terms of the restricted theory measure
 \beq{measure_rel}
  \hat M^{\rm delta}_{(\chi,\varkappa,\sigma)} \,=\,
  M^{\rm red}_{(\chi,\varkappa,\sigma,\omega,k,\rho)}\,\delta(\sigma^p_{\,\cza}\chi^\cza),
 \eeq
whence one has
 \beq{}
  Z_{\rm partial\;gf}=
  \int D\varphi\,\,\hat M^{\rm delta}_{(\chi,\varkappa,\sigma)} \,
  e^{\,i\,\hat S[\,\varphi\,]}
  =\,Z_{\rm parental},
 \eeq
where the last equality follows from (\ref{three_measures}). This relation shows that the restricted theory with \emph{full-rank} operator $Q^p_{\,\cza}$ is nothing but a \emph{partial gauge fixing} concept --- the parental and restricted theories which are physically equivalent at the classical level remain equivalent at the quantum level.

The BFV canonical prescription of normalization for gauge generators in the Lagrangian BV approach for the parental theory with the irreducible gauge structure thus imply that correctly normalized reducible ``projected'' gauge generators $\mathcal{R}^\zI_\cza=\hat{\mathcal{R}}^\zI_\czb T^\czb_{\:\cza}(Q,k)$ of the restricted (here, partially gauge-fixed) formalism just should be constructed by projecting \emph{canonically normalized} parental gauge generators $\hat{\mathcal{R}}^\zI_\czb$. Note that this result also confirms the normalization (\ref{Zmu1}) of reducibility generators derived above from another (projector parameter independence) principle. 

Let us go over from the case of partial gauge fixing to generic restricted theory physically inequivalent to the parental one. Its path integral in reduced space representation is given by the analog of Eq.\,(\ref{partial}) with the set of restriction conditions $\theta^\ccza$ instead of $\chi^p$, whose operator $Q^\ccza_{\,\cza}$ is rank deficient, $\rank Q^\ccza_{\,\cza}=m_1-m_2< {\rm range}\: \ccza=m_1$,
 \bea{restricted}
  \hspace{-1cm}Z_{\rm restricted}&\;=\;&
  \int D\phi\,M^{\rm red}_{(\chi,\varkappa,\sigma,\omega,k,\rho)} \,
  e^{\,i\,S^{\rm red}[\,\phi\,]}\nonumber\\
  &=&\int D\varphi\,M^{\rm red}_{(\chi,\varkappa,\sigma,\omega,k,\rho)}\,\delta(\theta^\ccza)\,
  e^{\,i\,\hat S[\,\varphi\,]}.
 \eea
The transition to integration over the parental theory configuration space, $D\phi\,(\det G_{\zi\zj})^{1/2}=D\varphi\,(\det G_{\zI\zJ})^{1/2}\,(\det G^{\ccza\cczb})^{1/2}\,\delta(\theta^\ccza)$, is written here modulo local measure factors which we will disregard in what follows. The set of functions $\theta^\ccza$ according to the rank of $Q^\ccza_{\,\cza}$ can be split into the set of gauge-invariant functions $\theta^\zA$,\, $ {\rm range}\: \zA = m_2$, and the set $\theta^p$, $ {\rm range}\: p = m_1-m_2$, cf. Eqs.\,(\ref{rankQphys})--(\ref{invariant}), $\theta^\ccza\to(\theta^\zA,\theta^p)$ so that
 \beq{delta_factorization0}
  \delta(\theta^\ccza) \,=\, \delta(\theta^\zA)\,\delta(\theta^p)\,Y,
 \eeq
where $Y=\det\partial(\theta^\zA,\theta^p)/\partial\theta^\ccza$. The functions $\theta^p$ enumerated by letters from the second part of Latin alphabet play here the role conditions of partial gauge fixing, $\chi^p\equiv\theta^p\HIDE{=0}$, while the invariants $\theta^\zA$, which are forced to vanish in the path integral, are responsible for inequivalence of the restricted and parental theories.

Now, let us choose in the measure $M^{\rm red}_{(\chi,\varkappa,\sigma, \omega,k,\rho)}$ the full set of gauge conditions
 \beq{}
  \chi^\cza \,=\, \sigma^\cza_{\,p}\chi^p+\sigma^\cza_{\,M}\chi^{M}
 \eeq
with $\chi^p$ identified with $\theta^p$, $\chi^p=\theta^p$, $ {\rm range}\; p=m_1-m_2$ (cf. the footnote \ref{ftnt:combined} and Eq.\,(\ref{chi_p}) which applies here in view of orthogonality $\sigma^p_{\,\cza}\sigma^\cza_{\,M}=0$). Let the rest of gauge conditions $\chi^M$, $ {\rm range}\: M = m_0-m_1+m_2$, form any complementary set of gauges such that the total Faddeev-Popov operator $[\,\HIDE{\hat} Q^p_{\,\czb}\; \HIDE{\hat} Q^{M}_{\,\czb}\hspace{1pt}]$ is nondegenerate. Then, substituting the above relation (\ref{delta_factorization0}) in (\ref{restricted}) and using the relation (\ref{measure_rel}) between the measures one has
\bea{restricted1}
  Z_{\rm restricted}
  &=&
  \int D\varphi\, \hat M^{\rm delta}_{(\chi,\varkappa,\sigma)} Y\,\delta(\theta^\zA)\,
  e^{\,i\,\hat S[\,\varphi\,]}\nonumber\\
  &=&
  \int D\varphi\,
  \hat M^{\rm delta}_{(\chi)}Y\,\delta(\theta^\zA)\,
  e^{\,i\,\hat S[\,\varphi\,]}.
 \eea
Here, the last equality follows from the equality of measures Eq.\,(\ref{three_measures}) and the fact that
 \beq{delta_factorization}
  \det\!\big[\,\HIDE{\hat} Q^p_{\,\czb}\;\HIDE{\hat} Q^{M}_{\,\czb}\hspace{1pt}\big]
  \delta(\chi^p) \delta(\chi^{M})
  \,=\, \det\hat Q^\cza_{\,\czb} \delta(\chi^{\cza})
 \,\equiv\,  M^{\rm delta}_{(\chi)}
 \eeq
in view of invariance of the Faddeev-Popov delta-function-type measure with respect to linear transformations of the basis of gauge conditions.

Equation (\ref{restricted1}) allows us to make the needed statement: as long as the quantum measure $M^{\rm delta}_{(\chi)}$ can be derived from the canonical quantization with canonically normalized local generators $\hat{\mathcal R}^\zI_\cza$, then the same choice of parental theory generators should be used in the construction of restricted theory.

Note that the last expression for $Z_{\rm restricted}$ can be alternatively represented by expressing in terms of original special gauge conditions $\theta^p$ and $\chi^M$ and then using (\ref{delta_factorization0}). This allows us to get rid of a potentially nonlocal factor $Y$ and obtain the representation directly in terms of the restriction functions $\theta^\ccza$ and the complementary subset of gauge conditions,
 \beq{restricted2}
  \hspace{-0.3cm}Z_{\rm restricted}=\int D\varphi\,
  \det \big[\,\HIDE{\hat} Q^p_{\,\czb}\,\HIDE{\hat} Q^{M}_{\,\czb}\,\big]
  \delta(\chi^{M\!})\,\delta(\theta^\ccza)
  \,e^{\,i\,\hat S[\,\varphi\,]}.
 \eeq
It should be emphasized again that here the Faddeev-Popov operator $[\,\HIDE{\hat} Q^p_{\,\czb}\,\,\,\HIDE{\hat} Q^{M}_{\,\czb}\,]$ is built with respect to partial gauge fixing subset of $\theta^\ccza$ and any complementary to it set of gauge conditions $\chi^M$. This recipe may be less suitable in concrete applications, because explicit disentangling the subset $\theta^p$ from $\theta^\ccza$ may be involved, so the original form (\ref{restricted}) should be more useful, and it was explicitly used above for the derivation of the one-loop generating functional (\ref{Z_1loop_1st1}).

   \subsection{The difference between parental and restricted theories in the one-loop approximation}
    \label{SSect:1loop_EA_Difference}
For a class of theories with the Jacobian $Y$ independent of integration fields in (\ref{restricted1}) one can write down the representation for $Z_{\rm restricted}$ and its one-loop order (such irrelevant $Y$ will be omitted in the subsequent expressions in this section). In view of gauge invariance of $\theta^\zA$ Eq.\,(\ref{restricted1}) implies on shell the usual Faddeev-Popov integral with $\varkappa_{\alpha\beta}$-fixed gauge and with the \emph{gauge-invariant} insertion of $\delta(\theta^\zA)$
 \beq{restricted2}
  Z_{\rm restricted} \,=\,
  \int D\varphi\, \hat M_{(\chi,\varkappa)} \,\delta(\theta^\zA)\,
  e^{\,i\,\hat S[\,\varphi\,]},
 \eeq
which in the one-loop order by the mechanism of the identical transformation of Eq.\,(\ref{detF0}) goes over to a simple relation between the generating functionals of the restricted and parental theories
\beq{restricted10}
 Z_{\rm restricted}^{\rm 1-loop}
  \,=\, \hat Z^{\rm 1-loop}\,
  \big(\!\det\HIDE{\hat}\varTheta^{\zA\zB}\big)^{-1/2},
\eeq
where $\hat Z^{\rm 1-loop}$ is obviously the one-loop generating functional of the parental theory given in terms of its (hatted) gauge and ghost inverse propagators
  \bea{unrestricted}
   &&\hat Z^{\rm 1-loop} =
    \frac{\det\hat Q^{\cza}_{\;\czb}\,}{\big(\det\!\varkappa^{\cza\czb}\big)^{1/2}
    \big(\!\det\hat F_{\zI\zJ}\big)^{1/2}},\\
   &&\hat F_{\zI\zJ}
    \,=\, \hat S_{,\zI\zJ}  - \chi^\cza_{,\zI} \varkappa_{\cza\czb} \chi^\czb_{,\zJ}- \lambda_\zA \theta^\zA_{,\zI\zJ}
  \eea
and
  \beq{Theta_m2}
    \HIDE{\hat}\varTheta^{\zA\zB} \,\equiv\,
    \theta^\zA_{,\zI}\hat F^{-1\,\zI\zJ}\theta^\zB_{,\zJ}
  \eeq

The operator (\ref{Theta_m2}) is analogous to (\ref{Theta}), but it is acting in the space of indices $\zA$ which enumerate gauge-invariant functions $\theta^\zA$ disentangled from the full set of $\theta^\ccza$. Note that it is defined in terms of the Green's function of the gauge-fixed operator $\hat F_{\zI\zJ}$ of the parental theory with a source $\lambda_\zA$ (at gauge-invariant observable $\theta^\zA$). The presence of such source term in parental theory may be interpreted as going off shell and performing one-loop calculations on the family of backgrounds $\hat S_{,\zI} =\lambda_\zA \theta^\zA_{,\zI}$, which (together with $\theta^\zA(\varphi^\zI)=0$ conditions) specify saddle points of (\ref{restricted2}). This is in accordance with the fact that solutions of these background equations cover all possible backgrounds of the restricted theory. To compare $Z_{\rm restricted}^{\rm 1-loop}$ and $\hat Z^{\rm 1-loop}$ in (\ref{restricted10}) these objects of course should be calculated on the same backgrounds.\footnote{According to rather generic assumptions on restriction conditions $\theta^\za=0$ (see discussion in Sec.\,\ref{SSect:Restricted_and_Reduced_Reps}) the Lagrange multipliers $\lambda^\za$, and thus $\lambda^\zA$, are expressible in terms of the background fields. In direct analogy with (\ref{lambda_eq_lin}) here $\lambda_\zA(\varphi)=\hat S_{,\zI}(\varphi) \theta^\zI_\zA (\varphi)$ for $\theta^\zI_\zA (\varphi)$ being dual to $\theta^\zA_{,\zI}$.}

Another important observation confirming the consistency of the relation (\ref{restricted10}) is that the additional factor depending on the matrix $\HIDE{\hat}\varTheta^{\zA\zB}$ and this matrix itself are independent on shell of the choice of gauge, $\delta_{(\chi,\varkappa)}\HIDE{\hat}\varTheta^{\zA\zB}=0$, which can be checked by using the Ward identities for $\hat F^{-1\,\zI\zJ}$ derived above.

Finally, let us note on practical aspects regarding the structure of (\ref{Theta_m2}). Calculation of $\varTheta^{\zA\zB}$ based on nonlocal $\hat  F^{-1\,\zI\zJ}$ may be technically inconvenient. The transition to objects defined in terms of a local differential operator $\hat{F}_{\zI\zJ}$ significantly simplifies calculations. When the dual biorthogonal basis $(e^{\zI}_{\zI'},\theta^\zI_{\zA})$  satisfying $\theta^{\zA}_{,\zI}e^{\zI}_{\zI'}=0$ and $\theta^{\zA}_{,\zI}\theta^{\zI}_{\zB}=\delta^{\zA}_{\,\zB}$, is ``orthogonal'' with respect to the operator $\hat{F}_{\zI\zJ}$ in the sense that $e^{\zI}_{\zI'} \hat{F}_{\zI\zJ}\theta^\zJ_{\zA}=0$, then it is easy to calculate operator $\varTheta_{\zB\zA}$ inverse to $\varTheta^{\zA\zB}$:\; $\varTheta^{\zA\zB}\varTheta_{\zB\zC} = \delta^{\zA}_{\,\zC}$. Under the above assumptions the determinant $(\det{\varTheta^{\zB\zA}})^{-1/2}$ in (\ref{restricted10}) can be replaced by $(\det \varTheta_{\zA\zB})^{1/2}$ --- the inverse operator acquiring simple form in terms of local  $\hat{F}_{\zI\zJ}$
 \beq{invTheta}
  \varTheta_{\zA\zB} \,=\, \theta^{\zI}_{\zA} \hat F_{\zI\zJ} \theta^{\zJ}_{\zB}.
 \eeq

 \section{Unimodular gravity theory}
  \label{Sect:UMG_EA}
Application of the above formalism to unimodular gravity theory is straightforward. Its \emph{parental} theory is Einstein general relativity with the action --- the functional of the metric field
 \bea{}
  &&\varphi^\zI\;\mapsto\; g_{\mu\nu}(x),
  \,\quad\,
  \zI \;\mapsto\; (\mu\nu, x),\\
  &&\hat S[\,\varphi\,] \;\mapsto\; S_{\rm E}[\,g_{\mu\nu}]=\int d^4x\,g^{1/2}(x)\, R(g_{\mu\nu}(x)),
  \label{GR_theory}
  \qquad
 \eea
where $g(x)\equiv-\det g_{\mu\nu}(x)$ and $R(g_{\mu\nu}(x))$ is the scalar curvature of this metric (for brevity we work in units with $16\pi G=1$). The arrow signs ($\,\mapsto\,$) signify in what follows the realization of condensed notations of previous sections in this concrete field model.

Einstein theory is invariant under local gauge transformations $\var_\xi\varphi^\zI=\hat{\mathcal{R}}^{\zI}_\cza \xi^\cza$ --- metric diffeomorphisms generated by the vector field $\xi^\cza$, which read in explicit notations $\var_\xi g_{\mu\nu}=\nabla_\mu\xi_\nu+\nabla_\nu\xi_\mu$,\; $\xi^\cza \,\;\mapsto\;\, \xi^\cza(x)\equiv g^{\cza\mu}\xi_\mu(x)$, so that the gauge generators $\hat{\mathcal R}^\zI_\cza$ have the form
 \beq{gR_Diff_cov}
  \hat{\mathcal R}^\zI_\cza\;\mapsto\;
  2g_{\cza(\mu}\nabla_{\nu)}\delta(x,y),
  \,\quad\,
  I\;\mapsto\; (\mu\nu,x),
  \quad
  \cza\;\mapsto\; (\cza,y).
\eeq
By default, derivatives act to the right on the first spacetime-point argument of delta functions.

Unimodular restriction of the theory (\ref{GR_theory}) consists in the restriction to the subspace of metrics $\GG_{\mu\nu}(x)=g_{\mu\nu}(x)/g^{1/4}(x)$ with a unit determinant, $\GG(x)\equiv-\det\GG_{\mu\nu}(x)=1$. We will not introduce a special notation for nine independent variables per spacetime point playing the role of $\phi^\zi$ but just formulate the restriction constraints $\theta^\ccza(\varphi)$ as
 \bea{}
  \theta^\ccza &\mapsto& \theta^x\equiv\theta(x) = g^{1/2}(x) - 1,
  \quad \ccza\;\mapsto\;  x, \quad
   \label{UMG_rC_}\\
  \theta^\ccza_{,\zI} &\mapsto& \theta^{x,\:\mu\nu,y}= \tfrac12 g^{1/2}g^{\mu\nu}\delta(x,y), \nonumber\\
  &&\qquad \ccza \;\mapsto\;  x,
  \quad \zI\;\mapsto\; (\mu\nu,y),
    \label{UMG_drC_}
 \eea
and the gauge-restriction operator (\ref{def_rQ}) has the form
 \beq{Q_UMG}
  Q^\ccza_{\,\cza}\;\mapsto\;  Q^x_{\cza,y}
  =g^{1/2}\nabla_\cza\delta(x,y)
  =\partial_\cza \big( g^{1/2}(x)\delta(x,y)\big),
 \eeq
where $\nabla_\cza$ is a covariant derivative with Christoffel connection, here acting on a vector field.

Counterpart of restricted theory (\ref{S_fixed}) --- the Lagrange multiplier action of unimodular gravity, and its metric equations of motion read
 \bea{}
  &&\hspace{-0.4cm}S^\lambda_{\rm E}[\,g_{\mu\nu},\lambda\,]\!=\! \int d^4x\big(\,g^{1/2}R(g_{\mu\nu}\!) - \lambda(g^{1/2}\!-\!1)\big),\\
  &&\hspace{-0.4cm}\frac{\var S^\lambda_{\rm E}}{\var g_{\mu\nu}}
  =-g^{1/2}\big(R^{\mu\nu} - \tfrac12  g^{\mu\nu}R+ \tfrac12 \lambda g^{\mu\nu}\big)=0.
  \label{lambda_eom_grav}
 \eea
The latter resembles Einstein equations with the term mimicking cosmological constant: the Lagrange multiplier $\lambda(x)$ on shell is indeed \emph{constant} because according to (\ref{lambdaEoMQconsequence}) its nonzero part is a spacetime constant zero mode of the gauge-restriction operator (\ref{Q_UMG}),
  \beq{}
    \lambda_\ccza Q^\ccza_{\,\cza}
    =
    0 \; \mapsto \;
    \int d^4x\,\lambda(x)\,Q^x_{\cza,y}
    = -g^{1/2}\nabla_\cza\lambda(y)
    = 0 ,
  \eeq
where the covariant derivative is acting on scalar.
This result is obviously equivalent to contracting (\ref{lambda_eom_grav}) with covariant derivative and using contracted Bianchi identity for the Einstein tensor. On the other hand, tracing this equation one finds $\lambda=R/2$, and the set of ten metric equations becomes linearly dependent which corresponds to Eqs.\,(\ref{lambda_solution}) and (\ref{lambda_eq_lin}) of the general formalism of restricted gauge theories,
  \beq{EoM_UMG}
    R^{\mu\nu}- \tfrac14 g^{\mu\nu}R=0,
    \,\quad\,
    \nabla_\mu R=0.
  \eeq
The \emph{vacuum solution} of these equations is a generic Einstein space metric $\GG_{\mu\nu}$, $R_{\mu\nu}=\varLambda\,\GG_{\mu\nu}$, $\varLambda=\lambda/2={\rm const}$ with a unit determinant $\GG\equiv -\det\GG_{\mu\nu}=1$.

Thus, the left kernel of the operator $Q^\ccza_{\,\cza}$ of dimensionality $m_2=1$ spanned by the zero mode $Y^\zA_{\,\ccza} \,\;\mapsto\;\,  Y_x=1$,\,  $ {\rm range}\: \zA=m_2=1$, is what physically distinguishes unimodular gravity theory from Einstein (or Einstein-Hilbert theory with a cosmological constant term) because it allows one to prescribe any constant value of $\lambda$ from the initial conditions rather than postulate it as a fundamental constant in the Lagrangian of the theory. The relevant gauge-invariant physical degree of freedom constrained by the unimodular restriction according to Eqs.\,(\ref{invariant})--(\ref{invariant1}) above,
  \beq{theta_inv}
    \theta^A
    = Y^{\zA}_{\,\ccza}\theta^\ccza
    = 0
    \,\;\mapsto\;\,
    \bar{\theta}
    = \int d^4x\,(g^{1/2}(x)-1)
    = 0,
  \eeq
is the full spacetime volume $\int d^4x\,g^{1/2}\,|_{\,\theta=0}=\int d^4x$ which, of course, \emph{a priori} is not completely specified in Einstein gravity.

The construction of projectors (\ref{tkQProj}) with the aid of the matrix $k^\cza_{\,\cczb}$ satisfying the rank restriction conditions (\ref{g_rank}) suggests the following obvious choice which simultaneously with these conditions provides covariance with respect to spacetime diffeomorphisms,
  \bea{}
    k^\cza_{\,\cczb} &\mapsto& k^{\cza, x}_{\,y}=
    \nabla^\cza\delta(x,y) \HIDE{\,g^{-1/2}(y)}
    = g^{\cza\czb} \partial_\czb \delta(x,y)
    ,\nonumber\\
    &&\qquad \cza\;\mapsto\; (\cza,x),\quad  \cczb\;\mapsto\;  y,\label{k_UMG}\\
    (Qk)^\ccza_{\,\cczb} &\mapsto& (Qk)^x_{\,y}
    = g^{1/2} \Box \delta(x,y)
    = \partial_\mu \big( g^{1/2} g^{\mu\nu} \partial_\mu \delta(x,y)\big),\,
    \nonumber\\
    &&\qquad \ccza \;\mapsto\;  x, \quad
    \cczb\;\mapsto\;  y,                \label{Qk_UMG}
  \eea
which also provides covariance with respect to spacetime coordinate change. Here $\nabla^\cza$ is the covariant derivative with respect to the \emph{dynamical} metric $g_{\cza\czb}$ and $\Box$ is its covariant d'Alembertian,
  \bea{}
    \nabla^\cza
     = g^{\cza\czb}\nabla_\czb,
    \;\quad\;
    \Box
    = g^{\cza\czb}\nabla_\cza\nabla_\czb,
  \eea
which act here on the scalars \HIDE{(we define $\delta(x,y)$ as a scalar with respect to the first argument and as a density of a unit weight with respect to the second one)}(we use the standard definition of $\delta(x,y)$ as the symmetric kernel of the scalar identity operator), though in what follows we reserve for them the same notation when they will be acting on general tensors and tensor densities. With such choice projector (\ref{tkQProj}) reads
  \bea{}
    T^\cza_{\:\czb}(Q,k) &\mapsto&  T^{\cza,x}_{\:\czb,y} = T^\cza_{\:\czb}(\nabla)\delta(x,y),\nonumber\\
    &&
    \qquad T^\cza_{\:\czb}(\nabla)
    \,=\,
    \delta^\cza_{\,\czb}
    -\nabla^\cza \tfrac{1}{\Box} \nabla_\czb.
   \label{T_UMG}
  \eea
$T^\cza_{\:\czb}(\nabla)$ is a projector on the subspace of spacetime transverse vectors, and it is nonlocal because it is defined in terms of the Green's function of $\Box$ operator, which is understood in the Moore-Penrose sense associated with the rank deficiency of operators (\ref{Q_UMG}), (\ref{k_UMG}), and (\ref{Qk_UMG}) --- the number $m_2$ of their spacetime constant zero modes being just $1$.

   \subsection{Gauge fixing and propagators}
    \label{SSect:UMG_gf_Propagators}
Now we go over to the construction of auxiliary elements of gauge-fixing procedure $\chi^\cza$, $\varkappa^{\cza\czb}$, $\sigma^\cza_{\ccza}$, $\omega^\ccza_{\,\cza}$ and $\rho^\ccza_{\,\cczb}$. To preserve covariance of the formalism we will use background covariant gauge conditions which for simplicity will be linear in the dynamical (quantum) metric field $\varphi^\zI=g_{\mu\nu}$. The coefficient of $g_{\mu\nu}$ in $\chi^\cza$, that is $\chi^\cza_{,\zI}$, is $g_{\mu\nu}$-independent but explicitly depends on the \emph{background} metric $\GG_{\mu\nu}$ --- the one which on shell satisfies the equations of motion (\ref{EoM_UMG}) and is subject to unimodularity restriction $\GG=1$. Background covariance of such gauge conditions implies that the choice of this coefficient should be such that $\chi^\cza$ is covariant with respect to simultaneous diffeomorphisms of both $g_{\mu\nu}$ and $\GG_{\mu\nu}$. Usually as such a gauge one uses the linearized de Donder or DeWitt gauge which is linear in $h_{\mu\nu}(x)=g_{\mu\nu}(x)-\GG_{\mu\nu}(x)$ --- the quantum fluctuation of the metric on top of its background. The DeWitt gauge is $\chi^\cza \;\mapsto\;  \chi^\cza(x) = \GG^{1/2} \GG^{\cza\czb}(\nabla^\mu h_{\czb\mu}-\tfrac12\GG^{\mu\nu}\nabla_\czb h_{\mu\nu})$, where, as well as below, the covariant derivatives $\nabla_\czb$ and $\nabla^\mu\equiv\GG^{\mu\nu}\nabla_\nu$ are constructed in terms of the \emph{background} metric\HIDE{ $\GG_{\mu\nu}$}. In the unimodular case, however, the trace part $\GG^{\mu\nu}h_{\mu\nu}$ of the metric fluctuation is systematically projected out, so that it is worth using a simpler gauge
  \bea{UMG_chi_gf}
    \chi^\cza\;\mapsto\; \chi^\cza(x)
   \!&=&\!
    \GG^{1/2} \nabla^\mu h^\cza_\mu(x)
   \nonumber\\
   &\equiv& \!
    \GG^{1/2} \GG^{\mu\czb}\GG^{\cza\nu}\nabla_{\!\czb}\big(g_{\mu\nu}(x)
    -\GG_{\mu\nu}(x)\big) .
    \quad\quad\;\;
  \eea

For the same reasons of covariance the choice of gauge-fixing matrices $\varkappa^{\cza\czb}$, $\sigma^\cza_{\,\ccza}$, $\omega^\ccza_{\,\cza}$ and $\rho^\ccza_{\,\cczb}$ is also obvious. Just like $\chi^\cza_{,\zI}$ above we will construct them in terms of the background metric, part of them being directly related to the already introduced quantities
  \bea{}
    \varkappa^{\cza\czb}\! &\,\mapsto\,& \varkappa^{\cza,x\,\czb,y}\!
    =
    \GG^{1/2}\GG^{\cza\czb}\delta(x,y),               \label{varkappa_UMG}\\
    \sigma^\cza_{\,\ccza} \,&\,\mapsto\,& \sigma^{\cza,x}_{\,y}\!
    =
    - \GG^{1/2}\GG^{\cza\czb}\nabla_\czb\delta(x,y)
    =
    - \GG^{1/2}\GG^{\cza\czb}\partial_\czb\delta(x,y)
    \HIDE{\,\GG^{-1/2}(y)}
    \nonumber\\
    &&\qquad\; = - \GG^{1/2}(x)  k^{\cza,x}_{\,y}  |_{\,g_{\mu\nu}\to\,\GG_{\mu\nu}},
    \label{sigma_UMG}\\
    \omega^\ccza_{\,\cza}&\,\mapsto\,& \omega^x_{\cza,y}\!
    =
    \GG^{1/2}\nabla_\cza\delta(x,y)=
    \partial_\cza \big( \GG^{1/2}\delta(x,y)\big)\nonumber\\
    &&\qquad\;=
    Q^x_{\,\cza,y}|_{\,g_{\mu\nu}\to\,\GG_{\mu\nu}}, \label{omega_UMG}\\
    \rho^\ccza_{\,\cczb}
    &\,\mapsto\,&  \rho^{x}_{\,y}\!
    = \GG^{1/2} \delta(x,y),
   \label{rho_UMG}
  \eea
where in (\ref{sigma_UMG}) covariant derivative acts on scalar, whereas in (\ref{omega_UMG}) covariant derivative acts on vector (forming a covariant divergence).

As the on-shell results do not depend on the choice of these quantities, they could have been chosen in terms of the quantum field $g_{\mu\nu}$, but this would have lead to the origin of extra terms involving functional derivatives $\varkappa^{\cza\czb}_{\,,\zI}$, $\sigma^\cza_{\,\ccza,\zI}$, etc. Avoiding such terms essentially simplifies calculations and, particular, allows one to avoid explicit use of the extra ghost ${\cal C}'^\ccza$ entering (\ref{defX}), because in view of (\ref{X=chi})
 \bea{}
   X^\cza_{,\zI}=\chi^\cza_{,\zI} \,\;\mapsto\;\,  \GG^{1/2} \GG^{\cza(\mu}\nabla^{\nu)}\delta(x,y).
 \eea
It should be emphasized that, as long as we restrict ourselves with the one-loop approximation, after all needed functional derivatives have been taken everything gets computed at the background, so that the distinction between $g_{\mu\nu}$ and $\GG_{\mu\nu}$ disappears. For this reason we will basically write all the formalism below in terms of the metric $g_{\mu\nu}$ with the understanding that it should be restricted to the on-shell unimodular background $\GG_{\mu\nu}$ satisfying (\ref{EoM_UMG}).

Objects dual to those defined above, or raising and lowering condensed indices $\zI \;\mapsto\;  (\mu\nu,x)$, $\cza\;\mapsto\;(\cza,x)$, $\ccza \;\mapsto\; x$, can be attained by introducing a local configuration space metric $G_{\zI\zJ}$ and a similar metric in the space of gauge indices. In Einstein theory a natural choice is the DeWitt metric of the kinetic term of the action, $G_{\zI\zJ}\;\mapsto\; G^{\mu\nu,x \: \cza\czb,y}=\tfrac12g^{1/2}(g^{\mu(\cza}g^{\czb)\nu}-\tfrac12 g^{\mu\nu}g^{\cza\czb})\delta(x,y)$. In the unimodular context, again due to projecting out the trace part of metric fluctuations, it is more useful to choose
 \bea{}
  G_{\zI\zJ}
  \,\;\mapsto\;\, G^{\mu\nu,x \: \cza\czb,y}
  = \tfrac12  g^{1/2}g^{\mu(\cza}g^{\czb)\nu}\delta(x,y).
 \eea
As regards the sector of gauge and reducibility indices, the role of the relevant metrics can be played respectively by $\varkappa_{\cza\czb}$ and $\kappa_{\ccza\cczb}$ introduced in Sec.\,\ref{Sect:BV_EA_1stReducible}. (Let us remind that by ${\varkappa}_{\cza\czb}$ we denote the inverse to $\varkappa^{\czb\cza}$, and by $\kappa^{\cczb\ccza}$ --- the inverse to $\kappa_{\ccza\cczb}$.)

The gauge-fixing choice (\ref{varkappa_UMG}) and (\ref{sigma_UMG}) leads to operator (\ref{def_cSKSinvcSKS}) and dual operators (\ref{dual_sigma})
  \bea{}
   && \!\!\!\!\!
   \kappa_{\ccza\cczb}
   \equiv
    \sigma^{\cza}_{\,\ccza} {\varkappa}_{\cza\czb} \sigma^{\czb}_{\,\cczb}
     \,\;\mapsto\;\,
    \kappa_{xy}
    = -g^{1/2}\Box\delta(x,y),\;\;\;\;
   \\
   && \!\!\!\!\!
    \sigma^\ccza_{\czb}
   \equiv \kappa^{\ccza\cczb} \sigma^\cza_{\,\cczb} \varkappa_{\cza\czb}
    \,\;\mapsto\;\,
    \sigma^x_{\,\czb,y}
    = -\frac{1}{\Box} g^{-1/2} \nabla_\czb \delta(x,y) ,
        \qquad\;\;
    \label{Sigma_dual_UMG}
   \eea
which define the projector form (\ref{def_cProj})
   \bea{Pi_UMG}
   && \varPi_{\cza\czb}\equiv\varkappa_{\cza\czb}
    -  {\varkappa}_{\cza\czc} \sigma^{\czc}_{\,\ccza}\sigma^\ccza_{\,\czb}
    \,\;\mapsto\;\,\nonumber\\
   &&\quad \varPi_{\cza,x \: \czb,y}=
    \Big(g^{-1/2}g_{\cza\czb}-\nabla_\cza \frac{1}{\Box}g^{-1/2} \nabla_\czb\Big)
    \delta(x,y) .
    \qquad\;\;
  \eea
Partial derivatives acting on delta functions in (\ref{Sigma_dual_UMG}) and (\ref{Pi_UMG}) are covariant derivatives $\nabla_\czb$ since they finally act on vector densities (forming a covariant divergence).

With the above objects the construction of the metric and ghost fields inverse propagators (\ref{invPropagator2}) and (\ref{invgPropagator3}) is straightforward,
  \bea{}
   &&F_{\zI\zJ}
    \,\;\mapsto\;\, F^{\mu\nu\:\cza\czb}(\nabla)\,\delta(x,y), \nonumber\\
   &&\qquad F^{\mu\nu\:\cza\czb}(\nabla)
   =
    \tfrac12  g^{1/2} \Big( g^{\mu(\cza}g^{\czb)\nu}\Box+2R^{\mu(\cza\nu\czb)}
   \nonumber\\
   &&\qquad\qquad\qquad\qquad\qquad\quad
    -2\nabla^{(\mu}\nabla^{\nu)} \frac{1}{\Box} \nabla^{(\cza}\nabla^{\czb)}
   \nonumber\\
   &&\qquad\qquad\qquad\qquad\qquad\quad
     +  g^{\mu\nu} \nabla^{(\cza}\nabla^{\czb)}
     +  g^{\cza\czb} \nabla^{(\mu}\nabla^{\nu)}
   \nonumber\\
   &&\qquad\qquad\qquad\qquad\qquad\quad
     -  g^{\mu\nu}g^{\cza\czb} (\Box + \tfrac14 R)
    \Big),
    \label{tensor_operator}\\
   &&F^\cza_{\;\czb}
    \,\;\mapsto\;\,  F^\cza_{\;\czb}(\nabla)\,\delta(x,y),
   \nonumber\\
   &&\qquad F^\cza_{\;\czb}(\nabla)
    = g^{1/2} \Big( (\Box+ \tfrac14 R)\delta^\cza_{\,\czb} - \tfrac12 R\,\nabla^\cza \frac{1}{\Box} \nabla_\czb \Big),
    \nonumber\\
   \label{vector_operator}\\
   &&F^\ccza_{\;\cczb}
   =
    \delta^\ccza_{\,\cczb}
    \,\;\mapsto\;\,
    F^x_{\,y} = \delta(x,y).
   \label{gg_operator}
  \eea
  In the derivation of these expressions we used background equations of motion (\ref{EoM_UMG}), so these are essentially on-shell objects. In particular, applications beyond one-loop approximation would generate extra terms because of the necessity to distinguish background and quantum fields. The choice of the gauge-fixing matrix $\rho^\ccza_{\,\cczb}$ in (\ref{rho_UMG}) allows one to avoid extra term $\nabla^\cza\nabla_\czb$ in the local part of the vector operator $F^\cza_{\;\czb}(\nabla)$. Finally, simplification of the ghosts-for-ghosts operator $F^\ccza_{\;\cczb}$ is due to correct normalization of reducibility generators (\ref{Zmu1}) and a special choice (\ref{omega_UMG}) of $\omega^\ccza_{\,\cza} = Q^\ccza_{\,\cza}$.

Note that the operators acquired nonlocal parts containing inverse (scalar) d'Alemberti\-ans. They were generated due to nonlocal projectors in the gauge-breaking term and in the projected gauge generators. Moreover, even in the local part of the tensor operator covariant derivatives do not form overall d'Alembertian, and their indices get contracted with the indices of test functions on which the operator is acting. This situation is very different from the Einstein theory case for which DeWitt gauge conditions guarantee minimal nature of the operator of field disturbances which is an important property admitting direct use of the heat kernel method for the calculation of the effective action.

The local part of (\ref{tensor_operator}) can, however, be simplified. Its functional determinant enters the one-loop partition function via the combination (\ref{detF}), $\det F_{\zI\zJ}\, \det \varTheta^{\ccza\cczb}$, or (\ref{detFfin}), where $\varTheta^{\ccza\cczb}$ is defined in terms of $F_{\zI\zJ}$ by Eq.\,(\ref{Theta}). It is easy to check that this combination is invariant under the change of the operator $F_{\zI\zJ}$ of the form
\bea{}
\var F_{\zI\zJ}\,=\,\var \varOmega_{\ccza(\zI}\,\theta^\ccza_{,\zJ)},
\eea
because
\bea{}
&&\var \ln\big(\det F_{\zI\zJ}\,\det\varTheta^{\ccza\cczb}\big)
\,=\, \var \varOmega_{\ccza\zI}\,\theta^\ccza_{,\zJ}\,F^{-1\,\zJ\zI}\nonumber\\
&&\qquad\;-\,\theta^\ccza_{,\zI}F^{-1\,\zI\zJ} \var\varOmega_{\cczc\zJ}\,
\theta^\cczc_{,\zK}F^{-1\,\zK\zL}\theta^\cczb_{,\zL}\varTheta^{-1}_{\cczb\ccza}\,=\,0,
\quad\;\;\;\;
\eea
since $\theta^\cczc_{,\zK}F^{-1\,\zK\zL}\theta^\cczb_{,\zL}
=\varTheta^{\cczc\cczb}$. Therefore, we can omit in (\ref{tensor_operator}) the terms of the second line which are proportional to $\theta^\ccza_{,\zI}\sim g^{\mu\nu}$ and $\theta^\ccza_{,\zJ}\sim g^{\cza\czb}$ and replace this operator by
\bea{}
 \tilde F^{\mu\nu\,\cza\czb}(\nabla)
 &=&
 \tfrac12  g^{1/2}\Big(\, g^{\mu(\cza}g^{\czb)\nu}\Box+2R^{\mu(\cza\nu\czb)}\nonumber\\
 &&\qquad\quad
 -2\nabla^{(\mu}\nabla^{\nu)}
 \frac{1}{\Box} \nabla^{(\cza}\nabla^{\czb)}\Big),     \label{tilde_tensor_operator}
\eea
accordingly taking the operator $\tilde\varTheta^{\ccza\cczb}$ induced by $\tilde F^{-1\,\zI\zJ}=\tilde F^{-1}_{\mu\nu\,\cza\czb}(\nabla)\,\delta(x,y)$,
  \bea{}
    &&\!\!
    \tilde\varTheta^{\ccza\cczb}
    =
    \theta^\ccza_{,\zI}\tilde F^{-1\,\zI\zJ}\theta^\cczb_{,\zJ}
    \,\;\mapsto\;\,     \tilde\varTheta^{x\,y} = \HIDE{g^{1/2}}\tilde\varTheta(\nabla)\delta(x,y),
    \qquad\\
    &&\!\!
    \tilde\varTheta(\nabla)\,=\,
     \tfrac14 g^{1/2} g^{\mu\nu}\tilde F^{-1}_{\mu\nu\,\cza\czb}(\nabla)
    g^{\cza\czb} g^{1/2}.           \label{tilde_Theta}
  \eea

Still, all the vector (\ref{vector_operator}), tensor (\ref{tilde_tensor_operator}) and scalar (\ref{tilde_Theta}) operators remain nonlocal, and their remains a problem of reducing their determinants to some calculable form.

   \subsection{Reduction of functional determinants}
    \label{SSect:UMG_1loop_Det}
Reduction of the above determinants can be done by the decomposition of the space of tensor and vector fields into irreducible transverse and traceless components. However, this decomposition in concrete applications can be useful only when the underlying metric background is homogeneous and the bases of irreducible scalar, transverse vector and tensor harmonics with their explicit spectra are known. If one wants to work on generic backgrounds and use such general methods as heat kernel method or Schwinger-DeWitt technique of curvature expansion \cite{Schwinger-DeWitt}, then the above operators should be transformed to the form of differential or pseudo\~differential operators with simple principal symbols, preferably local and minimal ones  that are constructed of covariant derivatives which form powers of covariant d'Alembertians. This is especially important for the operator (\ref{tilde_Theta}) having essentially nonlocal structure with the Green's function of another nonlocal operator.

In order to reduce the calculation of the determinant of $\tilde F^{\mu\nu\,\cza\czb}(\nabla)$ to that of the minimal operator let us include it into the one-parameter family
\bea{}
\tilde F^{\mu\nu\,\cza\czb}(a|\nabla) &=& \tfrac12  g^{1/2}\Big( g^{\mu(\cza}g^{\czb)\nu}\Box+2R^{\mu(\cza\nu\czb)}\nonumber\\
&&\qquad\quad -2a\nabla^{(\mu}\nabla^{\nu)}
 \frac{1}{\Box} \nabla^{(\cza}\nabla^{\czb)}\Big),
 \quad
\eea
interpolating between $\tilde F^{\mu\nu\,\cza\czb}(1|\nabla)=\tilde F^{\mu\nu\,\cza\czb}(\nabla)$ and the local minimal operator $\tilde F^{\mu\nu\,\cza\czb}(0|\nabla)=\Delta^{\mu\nu\,\cza\czb}(\nabla)$,
\bea{}
 &&\!\!\!\Delta_{\zI\zJ}\!=G_{\zI\zK}\Delta^{\!\zK}_{\:\zJ}\!\;\mapsto\; \Delta^{\mu\nu\;\cza\czb}(\nabla)\!=\!
  \tfrac12  g^{1/2}g^{\mu\czc}g^{\nu\czd} \Delta_{\czc\czd}^{\;\;\cza\czb}(\nabla),  \nonumber\\
 &&\qquad \Delta_{\mu\nu}^{\;\;\cza\czb}(\nabla)\,=\,
 \Box\,\delta_{\mu\nu}^{\;\;\cza\czb}
 +2R_{\mu\;\;\nu}^{\,\,(\cza\;\;\czb)}.
 \eea

Differentiation of the family of their determinants gives
\bea{!0}
&&\hspace{-0.5cm}\frac{d}{da}\Tr\ln\tilde F^{\mu\nu\,\cza\czb}(a|\nabla)\nonumber\\
&&\hspace{-0.2cm}=-\!\Tr \Big(\,g^{1/2}\nabla^{(\cza}\nabla^{\czb)} \frac{1}{\Box} \nabla^{(\mu}\nabla^{\nu)}
\tilde F^{-1}_{\mu\nu\;\czc\delta}(a|\nabla)
\,\Big),\quad \label{derivative_det}
\eea
where the double derivative of the inverse operator $\tilde F^{-1}_{\mu\nu\;\czc\delta}(a|\nabla)$ can be obtained by the following sequence of transformations. When applied to $\tilde F^{\mu\nu\;\cza\czb}(a|\nabla)$ this double derivative reads as a local expression
\bea{}
  &&\!\!\!\!
  \nabla_\mu\nabla_\nu\tilde F^{\mu\nu\,\cza\czb}(a|\nabla)
 \nonumber\\
  &&\!\qquad=\,
  \tfrac14  g^{1/2}\big(\,2(1{-}2a)\Box
  +(1{-}a)R\,\big) \nabla^{(\cza}\nabla^{\czb)},
 \qquad
\eea
where we used the fact that $\nabla_\mu R^{\mu\cza\nu\czb}\!=0$ on Einstein background.
Functionally contracting this relation with the inverse of $\tilde F^{\mu\nu\;\cza\czb}(a|\nabla)$ on the right and with the inverse of the scalar operator $g^{1/2} \big( 2(1-2a)\Box+(1-a)R \big)$ on the left, we obtain
\bea{}
&&\nabla^\cza\nabla^\czb
\tilde F^{-1}_{\cza\czb\,\mu\nu}(a|\nabla)\nonumber\\
&&\qquad=\,
\frac2{\big(\Box+\frac{R}2\big)
-2a\big(\Box+\frac{R}4\big)}g^{-1/2}\nabla_{(\mu}\nabla_{\nu)}, \qquad
\label{nablaF^-1}
\eea
whence
 \bea{!}
  &&\hspace{-1cm}\nabla^\cza\nabla^\czb \tilde F^{-1}_{\cza\czb\,\mu\nu}(a|\nabla)\nabla^\mu\nabla^\nu\nonumber\\
  &&=\,
  2\frac{\Box\big(\Box+\frac{R}4\big)}{\big(\Box+\frac{R}2\big)
  -2a\big(\Box+\frac{R}4\big)}g^{-1/2},
 \eea
where we took into account that on Einstein manifold $\nabla_\mu\nabla_\nu\nabla^\mu\nabla^\nu=\Box\big(\Box+\frac{R}4\big)$ when this operator is acting on a scalar. Substituting this result in (\ref{derivative_det}), using cyclic permutation under the trace and integrating over $a$ from 0 to 1 we have
 \bea{!!}
  \Tr\ln\tilde F^{\mu\nu\,\cza\czb}(1|\nabla)
   &=&
  \Tr\ln\tilde F^{\mu\nu\,\cza\czb}(0|\nabla)\nonumber\\
   &&+\,{\Tr}'\ln\frac\Box{\Box+\frac{R}2}.
 \eea
Here ${\Tr}'$ implies taking the functional trace of the operator over the space of eigenmodes of the scalar operator $\Box$ {\em excluding} its constant zero mode. The explanation of this important fact follows from the observation that the action of the operator (\ref{!}) on a covariantly constant mode obviously gives zero in view of the positive power of $\Box$ in the numerator. Therefore, even the multiplication by $1/\Box$ in (\ref{!0}) does not make the contribution of this mode in the functional trace nonzero; this is obviously consistent with the Penrose-Moore prescription for the inverse of $\Box$ discussed above.

With the inclusion of the local measure factor (\ref{!!}) then becomes
\bea{det_tilde_F}
&& \det \tilde F^\zI_{\:\zJ} \,\;\mapsto\;\, \nonumber\\
&& \qquad \det\tilde F_{\mu\nu}^{\;\;\cza\czb}(\nabla)=
\det\Delta_{\mu\nu}^{\;\;\cza\czb}(\nabla)\,
{\det}'\frac{\Box}{\Box+\frac{R}2},\quad\;\;
\eea
where the prime in ${\det}'$ obviously implies the same rule --- omission of the zero eigenvalue of $\Box$ in the definition of the functional determinant, and this of course refers to both the numerator and denominator of the operator valued fraction under the sign of ${\det}'$.

Similar steps for the vector operator (\ref{vector_operator}) on the Einstein metric background result in
\bea{det_ghosts}
\hspace{-1cm}\det\tilde F^\cza_{\;\czb}(\nabla)=
\det\big(\Box\,\delta^\cza_{\,\czb}+ \tfrac14  R\,\delta^\cza_{\,\czb}\big)\,
{\det}'\frac{\Box}{\Box+\frac{R}2}.
\eea

The reduction of the determinant of the scalar operator $\tilde\varTheta(\nabla)$, which is defined by Eq.\,(\ref{tilde_Theta}) and does not at all have a local part, can also be done via the transformations of the above type. First of all consider the contraction which localizes the operator $\tilde F^{\mu\nu\,\cza\czb}(\nabla)$,
 \beq{}
  g_{\mu\nu}\tilde F^{\mu\nu\,\cza\czb}(\nabla)
  = \tfrac12  g^{1/2}\Big(\,\big(\Box+ \tfrac12 R\big)\,g^{\cza\czb}
  -2\nabla^\cza\nabla^\czb\,\Big).
 \eeq
Contracting this relation with $\tilde F^{-1}_{\cza\czb\,\mu\nu}(\nabla)$ and using Eq.\,(\ref{nablaF^-1}) with the parameter $a=1$ we get
 \beq{}
  g^{\cza\czb}\tilde F^{-1}_{\cza\czb\,\mu\nu}(\nabla)
  = \frac2{\Box+\frac{R}2}\,\Big(g_{\mu\nu}-\frac2\Box\nabla_\mu\nabla_\nu\Big)g^{-1/2},
 \eeq
so that the operator (\ref{tilde_Theta}) and its determinant take the form
\beq{det_tilde_Theta}
\tilde\varTheta(\nabla)=\frac1{\Box+\tfrac{R}2},
\,\quad\,
\det\tilde\varTheta(\nabla)=\frac1{\det\big(\Box+\tfrac{R}2\big)}.
\eeq

Assembling together in the tilde version of (\ref{Z_1loop_1st1}) the results  (\ref{det_tilde_F}), (\ref{det_ghosts}), (\ref{det_tilde_Theta}), trivial contribution of $\det\rho=1$ and noting that $\det\kappa_{\ccza\cczb}={\det}'\,\Box$\, we finally get
\beq{final_result}
Z^{\rm 1-loop}_{\rm UMG}
\!=
\frac{\det\big(\Box\,\delta^\cza_{\,\czb}+ \tfrac{R}4\delta^\cza_{\,\czb}\big)}
{\!\big[
\det(\Box\,\delta_{\mu\nu}^{\;\;\cza\czb}
\!+2R_{\,\mu\,\;\nu}^{\,\,(\cza\;\czb)}\!)\big]^{1/2}\!}
\!\left[\,\frac{\det\big(\Box+\tfrac{R}2\big)}
{{\det}'\big(\Box+\tfrac{R}2\big)}\right]^{1/2}
\!\!\!\!.
\eeq
Note the origin of the nontrivial factor
\bea{!!!}
\left[\,\frac{\det\big(\Box+\tfrac{R}2\big)}
{{\det}'\big(\Box+\tfrac{R}2\big)}\,\right]^{1/2}
\!\!=\, (2\varLambda)^{1/2},
\eea
which looks purely numerical, but in fact it is a function of the dynamical global degree of freedom $\varLambda$ belonging in UMG to the full configuration space of the theory.

Modulo this extra factor, the result (\ref{final_result}) exactly coincides with the one-loop contribution of gravitons in Einstein theory with the action
\bea{S_Lambda}
S_\varLambda[\,g_{\mu\nu}]\,=\int d^4x\,g^{1/2}(R-2\varLambda)
\eea
and the on-shell value of the cosmological constant $\varLambda=R/4$,
\bea{Einstein_result}
\hat Z^{\rm 1-loop}_{\rm E}(\varLambda)
=\frac{\det\big(\Box\,\delta^\cza_{\,\czb}+ \tfrac{R}4\delta^\cza_{\,\czb}\big)}{\big[
\det(\Box\,\delta_{\mu\nu}^{\;\;\cza\czb}
+2R_{\mu\;\;\nu}^{\,\,(\cza\;\;\czb)})\,\big]^{1/2}}\,
\Bigg|_{\,R_{\mu\nu}=\varLambda g_{\mu\nu}}.
\eea
The on-shell inverse propagator of this model in the DeWitt gauge,
 \bea{}
  \Delta_{\mu\nu}^{\;\;\cza\czb}(\varLambda\,|\,\nabla)
  &=&
  \Big(
  \Box\,\delta_{\mu\nu}^{\;\;\cza\czb}+2R_{\mu\;\;\nu}^{\,\,(\cza\;\;\czb)}
  +\delta_\mu^{(\cza}R^{\czb)}_\nu+\delta_\nu^{(\cza}R^{\czb)}_\mu\nonumber\\
  && \quad -
  R_{\mu\nu}g^{\cza\czb}-g_{\mu\nu}R^{\cza\czb}+(2\varLambda-R)\,
  \delta_{\mu\nu}^{\;\;\cza\czb}
  \nonumber\\
  && \quad +
  \tfrac12  R \,g_{\mu\nu}g^{\cza\czb}\Big)\Big|_{R_{\mu\nu}=\varLambda g_{\mu\nu}}
  \nonumber\\
  &=& \Box\,\delta_{\mu\nu}^{\;\;\cza\czb}
  +2R_{\mu\;\;\nu}^{\,\,(\cza\;\;\czb)},
 \eea
equals the above minimal operator $\Delta_{\mu\nu}^{\;\;\cza\czb}(\nabla)$ and the Faddeev-Popov ghost operator $Q^\cza_{\,\czb}=\Box\delta^\cza_{\,\czb} +R^\cza_\czb$ also coincides with the vector operator $(\Box+\tfrac14R)\delta^\cza_{\,\czb}$. So if we consider Einstein gravity with the cosmological constant as a parental theory of unimodular gravity then the relation (\ref{final_result}) can be interpreted as Eq.\,(\ref{restricted10}) relating the generating functionals of the restricted theory and the parental one with $\hat Z^{\rm 1-loop} \,\mapsto\, \hat Z^{\rm 1-loop}_{\rm E}(\varLambda)$, provided we can prove equality of factors (\ref{!!!}) and $(\det\varTheta^{AB})^{-1/2}$. This proof is straightforward.

The only invariant $\theta^\zA$ that can be built out of the restriction function (\ref{UMG_rC_})-(\ref{UMG_drC_}) by integrating it with the constant zero mode of the gauge-restriction operator is the following global quantity which reads along with its $\theta^\zA_{,\zI}$ as
 \bea{}
  \theta^\zA
  &\;\mapsto\;&
  \bar{\theta}\equiv\int d^4x\; \theta(x)
  =\int d^4x\; (g^{1/2}(x)-1),   \quad\;\;
            \label{UMG_rC_COPY}\\
  \theta^\zA_{,\zI}
  &\;\mapsto\;&
  \bar\theta^{\mu\nu}(y)=
  \tfrac12 g^{1/2}g^{\mu\nu}(y).
            \label{UMG_drC_COPY}
 \eea
In order to find its $\varTheta^{AB}$,
\beq{bartheta}
\bar\varTheta^{AB}
\,\;\mapsto\;\,
 \bar\varTheta\equiv\int d^4x\,d^4y\,
\bar\theta^{\mu\nu}(x)\hat F^{-1}_{\mu\nu\,\alpha\beta}(\nabla)
\delta(x,y)\bar\theta^{\alpha\beta}(y),
\eeq
we need the gauge field inverse propagator of the parental theory --- Einstein gravity with a cosmological constant, which reads on shell as
\bea{}
\hat F_{\zI\zJ} &\mapsto&
\hat F^{\mu\nu\,\alpha\beta}(\nabla)\,\big|_{\,R_{\mu\nu}=\varLambda g_{\mu\nu}}\nonumber\\
&&\;= \tfrac14\,g^{1/2}\big(2g^{\mu(\cza}g^{\czb)\nu}
-g^{\mu\nu}g^{\alpha\beta}\big)\Box\nonumber\\
&&\;\;\;\;\;+
g^{1/2}\big(R^{\mu(\alpha\nu\beta)}
-\tfrac18\,R\,g^{\mu\nu}g^{\alpha\beta}\big).
\eea
This operator satisfies an obvious\HIDE{on-shell} relation
  \beq{}
    g_{\mu\nu}\hat F^{\mu\nu\,\alpha\beta}(\nabla) \HIDE{\delta(x,y)}
   \,=\,
    -\tfrac12\,g^{1/2}\big(\Box+\tfrac12R\big)g^{\alpha\beta} \HIDE{\delta(x,y)},
  \eeq
which allows one to find the following contraction of the Green's function kernel with metric tensors
\beq{}
g^{\mu\nu}\hat F^{-1}_{\mu\nu\,\alpha\beta}(\nabla)\delta(x,y)g^{\alpha\beta}(y) \,= \frac{-8}{\Box+\tfrac12R}\delta(x,y)g^{-1/2}(y),
\eeq
whence the needed factor (\ref{bartheta}) equals
\beq{barvarTheta}
\bar\varTheta=-2\int d^4x\,g^{1/2}(x)\frac1{\Box+\tfrac12R} {\cdot} 1 \,=
-\frac1{\varLambda}\int d^4x\,g^{1/2}(x),
\eeq
where the last equality follows from the uniformity of $R=4\varLambda$ in spacetime. Of course on shell, $g_{\mu\nu}=\GG_{\mu\nu}$, we have $g^{1/2}=1$, so the square root of the metric determinant is retained entirely for the sake of manifest covariance, and in this way it represents invariant volume of spacetime. Thus,
\beq{}
(\det\varTheta^{AB})^{-1/2}
\,\;\mapsto\;\,
\bar\varTheta^{-1/2}=
\frac{\varLambda^{1/2}}{\big(\int d^4x\,g^{1/2}(x)\big)^{1/2}},
\eeq
which up to a constant factor coincides with (\ref{!!!}).\footnote{
 When calculating determinants we omit irrelevant overall numerical factors. Here in the determinant of a zero-dimensional matrix we neglect $-1$ factor.
}

For completeness we present here the same answer rewritten in the basis of irreducible subspaces of tensor and vector fields, which are defined via disentangling from the full tensor field its tranverse-traceless $h^{\rm TT}=h_{\mu\nu}^{\rm TT}$, transversal vector $h_{\rm T}=h_\mu^{\rm T}$ and two scalar parts \cite{York:1974psa},
 \bea{UMG_HtrlYorkDecomp}
  &&h_{\Zm\Zn} =  h^{\rm TT}_{\Zm\Zn}+ 2\nabla_{(\Zm} h^{\rm T}_{\Zn)} + \big(\nabla_{\Zm}\nabla_{\Zn} - \tfrac{1}4 g_{\Zm\Zn}\Box\big) s+ \tfrac14 g_{\mu\nu} h,\nonumber\\
  &&\nabla^\mu h^{\rm TT}_{\Zm\Zn}=g^{\mu\nu}h^{\rm TT}_{\Zm\Zn}=0,
  \,\quad\,\,
  \nabla^\mu h^{\rm T}_\mu=0.
 \eea
Similarly for a vector such a decomposition reads
\bea{York_vec}
v_\mu=v_\mu^{\rm T}+\nabla_\mu v,
\,\quad\,\,
\nabla^\mu v_\mu^{\rm T}=0.
\eea

Jacobians of transition $h_{\mu\nu}\to(h^{\rm TT},h^{\rm T},s,h)$, $v_\mu\to(v_\mu^{\rm T},v)$ in the functional integration over $h_{\mu\nu}$ and $v_\mu$ equal
\bea{Jacobians}
&&\frac{D(h_{\mu\nu})}{D(h^{\rm TT},h^{\rm T},s,h)}\nonumber\\
&&=\Big( {\det}'\Box\:{\det}'\big(\Box+\tfrac{R}3\big)\det\!_{\rm T}\big(\Box\delta^\mu_{\,\nu}+\tfrac{R}4\delta^\mu_{\,\nu}\big) \Big)^{1/2}\!\!\!\!,
\qquad\\
&&\frac{D(v_\mu)}{D(v^{\rm T},v)} \,=\, \big({\det}'\,\Box\,\big)^{1/2}.
\eea
Here ${\det}'$ is a functional determinant of a scalar operator with omitted zero mode of the $\Box$-operator (note that this is the omission of the (constant) zero mode of $\Box$, but {\em not} the zero mode of the operator whose determinant is being taken).\footnote{
 The omission of these spacetime constant zero modes of $\Box$ takes place because they do not contribute to the left-hand sides of Eqs.\,(\ref{UMG_HtrlYorkDecomp})--(\ref{York_vec}).
}
Similarly $\det\!_{\rm T}$ denotes the determinant of the vector operator taken on the space of transverse vector functions. Quadratic forms with tensor field $\Delta^{\mu\nu\;\cza\czb}(\nabla)$ and vector field $(\Box+R/4)g_{\mu\nu}$ kernels correspondingly read
\bea{}
&&\int d^4x\,g^{1/2}h_{\mu\nu}\Delta^{\mu\nu\;\cza\czb}(\nabla) h_{\cza\czb}\nonumber\\
&&\;\;=\int d^4x\,g^{1/2}\Big(\,h^{\rm TT}_{\mu\nu}\Delta^{\mu\nu\;\cza\czb}(\nabla)h^{\rm TT}_{\cza\czb}-h_\mu^{\rm T}\big(\Box+\tfrac{R}4\big)^2h^\mu_{\rm T}\nonumber\\
&&\;\;\;\;\;\;\;\;
+\frac34 s\Box(\Box+\tfrac{R}2)(\Box+\tfrac{R}3)s
+ \tfrac14 h(\Box+\tfrac{R}2)h\,\Big),\\
&&\int d^4x\,g^{1/2}v_\mu\big(\Box+\tfrac{R}4\big)v^\mu
\nonumber\\
&&\;\;=\int d^4x\,g^{1/2}\Big(\,v_\mu^{\rm T}\big(\Box
+\tfrac{R}4\big)v^\mu_{\rm T}
-v\Box\big(\Box
+\tfrac{R}2\big)v\,\Big).\quad
\eea
Taking the Gaussian integrals with these quadratic forms and using the above Jacobians one can find the representation for the determinants of Eq.\,(\ref{final_result}) in terms of their irreducible counterpart --- transverse-traceless tensor $\det\!_{\rm TT}$ and transverse vector $\det\!_{\rm T}$ ones,
\bea{}
&&\det\big(\Box\,\delta_{\mu\nu}^{\;\;\cza\czb}
+2R_{\mu\;\;\nu}^{\,\,(\cza\;\;\czb)}\big)
\nonumber\\
&&\qquad\;\; =\det\!_{\rm TT}\big(\Box\,\delta_{\mu\nu}^{\;\;\cza\czb}
+2R_{\mu\;\;\nu}^{\,\,(\cza\;\;\czb)}\big)
\det\!_{\rm T} \big(\Box\delta^\mu_{\,\nu} +\tfrac{R}4\delta^\mu_{\,\nu}\big)
\nonumber\\
&&\qquad\quad\;\; \times
\det\big(\Box+\tfrac{R}2\big)\,{\det}'\big(\Box+\tfrac{R}2\big),\\
&&\det\big(\Box\delta^\mu_{\,\nu}\!+\tfrac{R}4\delta^\mu_{\,\nu}\!\big)
\nonumber\\
&&\qquad\;\;=
\det\!_{\rm T} \big(\Box\delta^\mu_{\,\nu} +\tfrac{R}4\delta^\mu_{\,\nu}\big)
\,{\det}'\big(\Box+\tfrac{R}2\big),
\eea
whence the partition function in Einstein theory with the cosmological constant (\ref{Einstein_result}) reads
\bea{irreducible_result}
&&\hat Z^{\rm 1-loop}_{\rm E}=\left[\,\frac{\det\!_{\rm T} \big(\Box\,\delta^\mu_{\,\nu} +\tfrac{R}4\delta^\mu_{\,\nu}\big)}{\det\!_{\rm TT}\big(\Box\,\delta_{\mu\nu}^{\;\;\cza\czb}
+2R_{\mu\;\;\nu}^{\,\,(\cza\;\;\czb)}\big)}\,\right]^{1/2}\nonumber\\
&&\qquad\qquad\;\;\times\left[\,\frac{{\det}'\big(\Box+\tfrac{R}2\big)}
{{\det}\,\big(\Box+\tfrac{R}2\big)}\,\right]^{1/2}.
\eea

In terms of these determinants on irreducible subspaces of transverse-traceless modes it differs by extra factor from the usually claimed form \cite{Tseytlin}. This factor originates on account of a constant zero mode of a scalar d'Alembertian. It should be emphasized that on homogeneous de Sitter or anti-de Sitter background other vector and tensor operators also have zero modes associated with Killing symmetries of these backgrounds (see e.g. \cite{Tseytlin}). Here we disregard them because we consider generic inhomogeneous Einstein metric spacetimes for all of which this zero mode of $\Box$ always exists.

Thus, in terms of determinants on constrained (irreducible) fields the one-loop result for Einstein theory with the cosmological constant differs from a conventional expression by the contribution of one constant eigenmode of the operator $\Box+\tfrac{R}2$. Curiously, for unimodular gravity this contribution in the same representation completely cancels by additional factor in (\ref{final_result}), and we have
\bea{UMG_irreducible}
\hspace{-0.8cm}Z^{\rm 1-loop}_{\rm UMG}=\left[\,\frac{\det\!_{\rm T} \big(\Box\,\delta^\mu_{\,\nu}
+\tfrac{R}4\delta^\mu_{\,\nu}\big)}{\det\!_{\rm TT}\big(\Box\,\delta_{\mu\nu}^{\;\;\cza\czb}
+2R_{\mu\;\;\nu}^{\,\,(\cza\;\;\czb)}\big)}\,\right]^{1/2}\!\!.
\eea
This result coincides with the one claimed in \cite{deLeonArdon:2017qzg,Percacci:2017fsy}. It manifestly exhibits the counting of local physical degrees of freedom --- 5 traceless-tensor modes minus 3 transverse vector modes.

 \section{Conclusion}
  \label{Sect:Conclusion}
To summarize our results, we worked out a full set of gauge-fixing elements in generic gauge theory of the first-stage reducibility and constructed a workable algorithm for its one-loop effective action. We also derived the set of tree-level Ward identities for gauge field, ghost and ghosts-for-ghosts propagators, which allow one to prove on-shell gauge independence of the effective action from the choice of auxiliary elements of gauge-fixing procedure. We showed that Lagrangian quantization of a restricted theory originating from its parental gauge theory can be performed within the BV formalism for models with linearly-dependent gauge generators of the first-stage reducibility. It turns out that new physics contained in the restricted theory as compared to its parental theory model is associated with the rank deficiency of a special gauge-restriction operator reflecting the gauge transformation properties of the restriction constraints functions. The choice of first-stage reducibility generators, or zero vectors of the projected gauge generators induced from the parental theory, has a certain freedom limited only by a special rank restriction condition, but the on-shell independence of physical results from this choice is provided by a special normalization of these vectors.

These general results are applied to the quantization of unimodular gravity theory. Its one-loop effective action, initially obtained in terms of complicated nonlocal pseudodifferential operators, is transformed to functional determinants of minimal second-order differential operators calculable on generic backgrounds by Schwinger-DeWitt technique of local curvature expansion. This also confirms the known representation of one-loop contribution in unimodular gravity theory in terms of functional determinants on irreducible transverse and transverse traceless subspaces of tensor, vector, and scalar modes. The one-loop order in unimodular gravity turns out to be equivalent to that of Einstein gravity theory with a cosmological term only up to a special contribution of the global degree of freedom associated with the variable value of the cosmological constant.

From the viewpoint of local phenomenology, the new physics in unimodular gravity turns out to be of a somewhat borderline nature. Classically it is manifested in the fact that the cosmological constant in UMG becomes a part of initial conditions rather than a fundamental constant in the Lagrangian of the Einstein theory. At the one-loop level the contribution of this extra global and spacetime constant degree of freedom is very peculiar, and the way it shows up depends on the representation of the theory. In fact we have two somewhat complementary representations for both theories: in terms of functional determinants of differential operators on full field spaces or on spaces constrained by irreducible representations. The one-loop Einstein gravity in the full space representation (\ref{Einstein_result}) does not reveal the contribution of this mode, whereas the constrained determinants representation (\ref{irreducible_result}) makes it manifest. With two representations for unimodular gravity (\ref{final_result}) and (\ref{UMG_irreducible}) this situation is reversed.

The manifestation of this contribution is the power-like dependence on (\ref{barvarTheta}) in the partition function $\sim\int d^4x\,g^{1/2}/\varLambda$ which becomes in the effective action a logarithmic essentially nonlocal contribution. Off shell, that is in transition to gradient expansion for nonconstant curvature scalar, $R\to R(x)$, it may go over into the structures like $\ln\int d^4x\,g^{1/2}(x)\big(1/R(x)+O(\nabla R)\big)$. These structures might be important in Euclidean quantum gravity responsible for tunneling phenomena and gravitational thermodynamics. In fact, thermodynamics reveals the duality relation between Einstein theory and unimodular gravity as the analogy of the Laplace transform relating the statistical ensemble with fixed volume vs the fixed pressure ensemble. Qualitatively, this can be shown as follows.

By identifying in (\ref{theta_inv}) the coordinate 4-volume $\int d^4x=V$ as the fixed argument of the generating functional (\ref{restricted2}) we have the expression for this functional in UMG theory (in Euclidean picture this is a partition function at fixed volume $V$)
\bea{}
Z_{\rm UMG}(V)&\;=\;&\int Dg_{\mu\nu} \hat M\,e^{-S_E[\,g_{\mu\nu}]}\nonumber\\
&&\qquad\times\,\delta\big({\textstyle\int} d^4x\,g^{1/2}-V\big),
\eea
where the measure $\hat M$ incorporates, just like in (\ref{restricted2}), the full local gauge fixing of all four dimensional diffeomorphisms. Then, it is obvious that the Laplace transform with respect to the volume variable converts the UMG partition function $Z_{\rm UMG}(V)$ to that of the Einstein theory $Z_{\rm E}(\varLambda)$ with a fixed value of $\varLambda$ --- the cosmological constant dual to $V$,
\bea{}
&&\int_0^\infty dV\,e^{-\frac{\varLambda V}{8\pi G}}Z_{\rm UMG}(V)\nonumber\\
&&\;\;\;\;=\int Dg_{\mu\nu}\, \hat M\,\exp\big(-S_E[\,g_{\mu\nu}]-\tfrac\varLambda{8\pi G}{\textstyle\int} d^4x\,g^{1/2}\big)\nonumber\\
&&\;\;\;\;\equiv\, Z_{\rm E}(\varLambda),
\eea
(here we reinsert the gravitational constant factor $1/16\pi G$ and note the opposite sign of the Euclidean version of the action (\ref{S_Lambda})). Of course, this derivation should be regulated by specifying the boundary conditions which fully determine a finite value of $V$ (or its infinite limit) and a finite value of the action, achieved by a subtraction of proper surface terms.  This can be done along the lines of \cite{Fiol:2008vk}, but in the present form it already conveys the essence of duality between Einstein theory and UMG gravity.

It should be emphasized that throughout our derivations we used Moore-Penrose concept of inverting operators with zero modes, which by and large corresponds to the omission of the zero-mode subspace. This makes all derivations, as discussed in Appendix \ref{ASect:Projectors_Variation}, consistent, but apparently leaves room for nontrivial effects of the above extra contributions based on a careful treatment of boundary conditions. Despite the fact that the functional determinants of the scalar d'Alembertian in Eqs.\,(\ref{Jacobians})--(\ref{irreducible_result}), which are vulnerable to zero-mode treatment, completely cancel out, there still may be a subtlety in their treatment and this might amount to the extension of the BV method beyond first-stage reducibility. Note that $m_2$-dimensional zero-mode subspace of gauge-restriction operator (\ref{def_rQ}) is exactly the playground for second-stage reducibility in the general formalism. The smallness of the phase-space sector of this mode in UMG, $m_2=1$, does not make it less important and might be at the core of cosmological constant problem. All this, however, goes beyond the scope of this paper and remains a subject of further research to be reported elsewhere.

Another direction of further research might be the generalized unimodular gravity (GUMG) of \cite{Barvinsky:2017pmm,Barvinsky:2019agh}, which is interesting in view of its dark energy and inflation theory implications. This model is more complicated than UMG, it has more complicated canonical formalism encumbered by the presence of second-class Dirac constraints and it strongly breaks diffeomorphism and Lorentz symmetry because of replacement of the UMG restriction condition $\det g_{\mu\nu}=-1$ by the Lorentz noninvariant relation between the lapse function and spatial metric. New physics in this model is associated with the origin of the dark perfect fluid which might serve as a source of dark energy or play the role of inflaton, i.e. scalar graviton degree of freedom \cite{Barvinsky:2019qzx}. Covariant quantization of this model along the lines of the BV method applied to parental Einstein gravity is also a good nontrivial playground for our technique.

 \section*{Acknowledgements}
   \label{Sect:Acknowledgements}
We wish to thank N.Kolganov and S.Lyakhovich for stimulating discussions. Also we are grateful to the anonymous referee for drawing our attention to the duality relation between Einstein and unimodular gravity theories. The work was supported by the Russian Science Foundation grant No. \href{https://rscf.ru/en/project/23-12-00051}{23-12-00051}.

\appendix

\renewcommand{\theequation}{\thesection\arabic{equation}}

 \section{Moore-Penrose inverse and variation of projectors}
  \label{ASect:Projectors_Variation}
\newcommand{\Qk}{(Q k)}
\newcommand{\invQk}{{(Q k)^{-1}} \vphantom{|}}
\newcommand{\invM}{{M^{-1\!}}\vphantom{|}}
Gauge theory restriction implies introduction of the transversal projector (\ref{tkQProj}) which is defined by its left kernel $Q^\ccza_{\,\cza}$ --- the gauge-restriction operator $Q^\ccza_{\,\cza}=\theta^\ccza_{,\zI}\hat{\mathcal{R}}^\zI_{\cza}$.\footnote{
Remember that we consider the restriction conditions $\theta^\ccza=0$ irreducible and independent of equations of motion of the parental theory $\hat{S}_{,\zI}=0$.} Peculiar feature of this operator is its \emph{rank deficiency}. While the range of index $\ccza$ labelling the  restriction condition functions $\theta^\ccza(\varphi)$ is $m_1$, the rank of $Q^\ccza_{\,\czb}$ is $m_1-m_2$. As discussed in Sec.\,\ref{Sect:Restricting_and_Reducibility} it is physically important, for if $Q^\ccza_{\,\czb}$ were a full-rank matrix ($m_2=0$) then this would be just a partial gauge fixing and the parental and restricted theories would be physically equivalent. New physics comes when $m_2>0$, which means that $m_2$ out of $m_1$ restriction conditions $\theta^\ccza(\varphi)=0$ annihilate several gauge-invariant functions --- physical observables.

Right kernel $k^{\cza}_{\,\cczb}$ of the projector (\ref{tkQProj}) is a parameter of the particular projector family, which must satisfy the rank condition $    \rank{(Qk)^\ccza_{\,\cczb}} = \rank{ Q^{\cza}_{\,\ccza}}= \rank{ k^{\czb}_{\,\cczb}} = m_1-m_2$     (\ref{g_rank}).
Thus, critically important feature of the projector $T^{\cza}_{\:\czb}(Q,k)$ (\ref{tkQProj}),
 \beq{tkQProj_COPY}
  T^{\cza}_{\:\czb}
  \,=\, \delta^{\cza}_{\:\czb} -
  k^{\cza}_{\,\ccza} {\invQk}^\ccza_{\,\cczb}  Q^{\cczb}_{\,\czb},
  \nonumber
 \eeq
is that it includes the \emph{inverse} of the \emph{degenerate} operator $\Qk^\ccza_{\,\cczb}$.  This inverse can be uniquely defined as Moore-Penrose inverse \cite{MoorePenrose}, and a brief reminder on this construction for the generalized inverse of generic matrices is in order here.

All objects below are of matrix nature so for the sake of readability we omit indices in most of equations implying ordinary matrix multiplication. When introducing matrices we explicitly correlate them with their counterparts in index notation using symbol $\leftrightarrow$.

For generic $m_1 \times m_1$ matrix $M\,\leftrightarrow\, M^{\ccza}_{\,\cczb}$ which is rank deficient with $\rank M = m_1 - m_2$ the following properties unambiguously define Moore-Penrose inverse  $\invM \,\leftrightarrow\, \invM^{\ccza}_{\,\cczb}$ \cite{MoorePenrose}
 \bea{MoorePenroseInverse}
  \begin{array}{cl}
    1)& M \invM M = M \\
    2)& \invM M \invM = \invM \\
    3)& P_1\equiv M \invM  = P_1^* \\
    4)& P_2\equiv \invM M  = P_2^* \\
  \end{array}
 \eea
where $P_1\,\leftrightarrow\,{P_1}^\ccza_{\,\cczb}$ and $P_2\,\leftrightarrow\,{P_2}^\ccza_{\,\cczb}$ are Hermitian projectors with idempotent properties $P_1^2=P_1$, $P_2^2=P_2$, and the star denotes Hermitian conjugation of matrices.\footnote{For real matrices the Hermitian conjugation is just a transposition, and Hermitian projectors become orthonormal projectors.} These properties imply the equality of ranks, $\rank \invM = \rank M = \rank P_1 = \rank P_2$.

Introduce for convenience the complementary projectors $L_1 = I - P_1$, $L_2 = I - P_2$ where $I$ is the identity $m_1 \times m_1$ matrix. The complete set of projector properties, which could serve as their definitions, is
\bea{MP_projector_properties}
P_1M=MP_2=M, \,\quad\, \invM P_1=P_2\invM=\invM.
\eea
Similarly the action of complimentary projectors respectively from the left and from the right on $M$ and $\invM$ annihilate them,
 \beq{ML_projector_properties}
  L_1 M = M L_2=0, \,\quad\, \invM L_1=L_2\invM=0.
 \eeq
Therefore, the rows of $L_1$ (reducibly) span left kernel of $M$ and its columns span right kernel of $\invM$, while $L_2$ (reducibly) spans the right kernel of $M$ and the left kernel of $\invM$.

With this definition of the inverse matrix for $M^\ccza_{\,\cczb}=\Qk^\ccza_{\,\cczb} \equiv Q^\ccza_{\,\cza} k^\cza_{\,\cczb} \,\leftrightarrow\, M = (Qk)$ the restricted theory $m_0\times m_0$ projector $T^\cza_{\:\czb} \,\leftrightarrow\, T$ (\ref{tkQProj}) satisfies all needed properties. In particular, its left kernel is spanned by the gauge-restriction operator $Q \,\leftrightarrow\, Q^\ccza_{\,\cza}$ because $QT=Q-(Qk)(Qk)^{-1}Q=L_1Q=0$. The latter equality is true since matrices $Q$ and $M=(Qk)$ have the same ranks and the same coranks with respect to the left index with range $m_1$ and thus matrices $Q$ and $M$ share the same left kernel so that $L_1 Q = 0$ in view of (\ref{ML_projector_properties}). The right kernel of $T$ is spanned by $k \,\leftrightarrow\, k^\cza_{\,\cczb}$ provided the rank restriction condition (\ref{g_rank}) is satisfied. This follows from the relation $Tk=k-k(Qk)^{-1}(Qk)=kL_2=0$ where $k$ is annihilated by $L_2$ from the right for the analogous reason as for the left kernel.

Such a definition is fully consistent until one has to vary inverse matrices. When the variation of $M$ changes its rank the inverse matrix and projectors become discontinuous and fail to be differentiable. As was shown in \cite{StewartIzumino} the rank preserving matrix variations $\var M$ which guarantee continuity of the Moore-Penrose inverse matrix and its projectors should satisfy the property
\beq{var_rank_preserving} 
  L_1 \var M L_2 = 0.
 \eeq
The variation\HIDE{ even without Hermiticity conditions} of equation 2) in (\ref{MoorePenroseInverse}) with respect to such $\var M$ gives
 \bea{var_invM} 
  \!\!\!\var \big(\invM\big)
  = - \invM \var M \invM
       \!\hspace{-0.5pt}+ \var \big(\invM\big)  L_1
       \!\hspace{-0.5pt}+ L_2 \var \big(\invM\big)
       \hspace{-0.5pt} .
       \;\;\;\;
 \eea
The last two \emph{anomalous} terms here are responsible for the deviation from the usual variational equation for the inverse of nondegenerate matrices. The first standard term is in fact transverse because in view of (\ref{MP_projector_properties}) $\invM \var M \invM = P_2 \invM \var M \invM P_1$.

It is important that the anomaly terms in (\ref{var_invM}) cancel when the variation of the inverse matrix is performed inside the contraction from the left and from the right respectively with structures annihilated by projectors $L_2$ and $L_1$, so that $A =A P_2$ and $B=P_1 B$,
 \bea{var_invM_projected}
     A\,\var \big(\invM\big) B\,
     &=& - A P_2 \invM \var M \invM P_1 B\nonumber\\
     &=& - A \invM \var M \invM B.
 \eea

In concrete applications generically one can not restrict the variation of the rank-deficient matrix $M$ to preserve its rank. However, when the matrix $M$ has a special internal structure it may be a part of the solution circumventing this difficulty. This is exactly the case of  $m_1 \times m_1$ rank-deficient operator $\Qk^\ccza_{\,\cczb} \equiv Q^\ccza_{\,\cza} k^\cza_{\,\cczb}$. The rank restriction conditions (\ref{g_rank}), $\rank{(Q k)} = \rank{Q} = \rank{k}$, provide the rank preservation (\ref{var_rank_preserving}) under arbitrary variations of $k$ and $Q$,
 \beq{var_rank_preserving_Qk} 
  {L_1} \var \Qk {L_2}
  = {L_1} \var Q \, k {L_2}
  + {L_1} Q \,\var k\, {L_2} =0,
 \eeq
where ${L_1}$, ${L_2}$ are the projectors onto left and right kernels of the operator $(Q k)$.

Since an arbitrary variation of the matrix $\Qk^\ccza_{\,\cczb}$  satisfies the rank preserving condition \ref{var_rank_preserving} one can use Eq.\,(\ref{var_invM}) for the variation of its inverse. Moreover in the
projector $T$ (\ref{tkQProj}) the variation of $\invQk$ stands inside the structure $ k\, \var {\invQk} Q \,\leftrightarrow\, k^{\czb}_{\,\ccza} \var {\invQk}^\ccza_{\,\cczb} \, Q^{\cczb}_{\,\czb}$
which, due to ${L_1} Q =0$ and
$k\, {L_2}=0$, guarantees the cancelation of the anomalous terms in (\ref{var_invM}) and  implies (\ref{var_invM_projected})
 \beq{k_varQk_Q}
  k\, \var {\invQk}  Q^{\cczb}_{\,\czb}
  \,=\,
  - k {\invQk} \var {\Qk} {\invQk} Q.
 \eeq
This variational equation underlies the variation formula for the projector $T$ (\ref{tkQProj}), which finally reads
 \bea{tkQProj_var_deriv} 
  \var T
  \,=\, - T \,\var k \, {\invQk} Q -
  k \,\invQk \var Q \, T.
 \eea
It should be emphasized again that this relation crucially relies on rank restriction conditions (\ref{g_rank}).

 \section{Gauge algebra of the restricted theory from a parental theory}
  \label{ASect:Gauge_Struct_Restriction}
The algebra (\ref{Jacobi_Id_Lagrange_projected_COPY}) follows directly by substituting the projected generators $\mathcal{R}^{\zI}_{\czb}= \hat{\mathcal{R}}^{\zI}_{\cza} T^{\cza}_{\:\czb}$ (\ref{gRprnt_projected}) into their Lie bracket,
  \bea{Jacobi_Id_projected_CALC}
   &&\!\!\!\mathcal{R}^\zI_{\cza,\zJ} \mathcal{R}^\zJ_{\czb} -  \mathcal{R}^\zI_{\czb,\zJ} \mathcal{R}^\zJ_{\cza}
  \,=\,
   (\hat{\mathcal{R}}^\zI_{\czc,\zJ}\hat{\mathcal{R}}^\zJ_{\czd}
   -
   \hat{\mathcal{R}}^\zI_{\czd,\zJ} \hat{\mathcal{R}}^\zJ_{\czc})
    T^{\czc}_{\:\cza}  T^{\czd}_{\:\czb}\nonumber\\
   &&\qquad\qquad+
   (\hat{\mathcal{R}}^\zI_{\czc} T^{\czc}_{\,\cza,\zJ} \hat{\mathcal{R}}^\zJ_{\czd} T^{\czd}_{\:\czb}
   -
   \hat{\mathcal{R}}^\zI_{\czd} T^{\czd}_{\,\czb,\zJ} \hat{\mathcal{R}}^\zJ_{\czc} T^{\czc}_{\:\cza}),
  \eea
and noting that in view of the open algebra relation for $\hat{\mathcal{R}}^{\zI}_{\cza}$ the first group of terms here reads
 \bea{_contrib1}
  &&\hspace{-1cm}\big( \hat{\mathcal{R}}^\zI_{\czc,\zJ}\hat{\mathcal{R}}^\zJ_{\czd}
   -
  \hat{\mathcal{R}}^\zI_{\czd,\zJ} \hat{\mathcal{R}}^\zJ_{\czc} \big)
    T^{\czc}_{\:\cza}  T^{\czd}_{\:\czb}\nonumber\\
  &&\qquad\qquad=\,
  \hat{\mathcal{R}}^\zI_{\cze} \hat{C}^{\cze}_{\czc\czd}  T^{\czc}_{\:\cza}  T^{\czd}_{\:\czb}
  +
  \hat{E}^{\zI\zJ}_{\czc\czd}\hat{S}_{,\zJ}  T^{\czc}_{\:\cza}  T^{\czd}_{\:\czb}.
 \eea

The second group of terms in the right-hand side of (\ref{Jacobi_Id_projected_CALC}) requires differentiation of the projector $T^{\czc}_{\:\cza}$ (\ref{tkQProj})\HIDE{ contained in the reduced theory generators}. This projector involves the procedure of inverting the matrix $(Qk)^\ccza_{\,\cczb}$ whose rank is lower than the range of its indices and, therefore, requires the Moore-Penrose construction of the generalized matrix inversion \cite{MoorePenrose}. This in turn leads to subtleties of variational procedure for $(Qk)^{-1\,a}_{\,\,\;\;\;\;b}$ discussed in Appendix \ref{ASect:Projectors_Variation}. As shown there, the variational property of the projector is {\em effectively} equivalent to the naive use of the variational rule $\var (Qk)^{-1}=-(Qk)^{-1}\var(Qk)\,(Qk)^{-1}$, provided the rank restriction condition (\ref{g_rank}) holds, and it reads as (\ref{tkQProj_var_deriv}). This can be directly applied to\HIDE{ obtain the expected representation for} the second group of terms in the right-hand side of (\ref{Jacobi_Id_projected_CALC}) on using the symmetry $\theta^{\cczb}_{,\zI\zJ}=\theta^{\cczb}_{,\zJ\zI}$, the algebra of parental theory generators, and their corollary
  $Q^{\cczb}_{\czc,\zJ}
   T^{\czc}_{\:\cza}
   \hat{\mathcal{R}}^\zJ_{\czd} T^{\czd}_{\:\czb}
   -
  (\cza\leftrightarrow\czb)
  =
  \theta^{\cczb}_{,\zI}(\hat{\mathcal{R}}^\zI_{\cze} \hat{C}^{\cze}_{\czc\czd}
   T^{\czc}_{\:\cza}
   T^{\czd}_{\:\czb}
  +
  \hat{E}^{\zI\zJ}_{\czc\czd}\hat{S}_{,\zJ}
   T^{\czc}_{\:\cza}
   T^{\czd}_{\:\czb})$.
As a result
\bea{_contrib2} 
   &&\big(\hat{\mathcal{R}}^\zI_{\czc} T^{\czc}_{\,\cza,\zJ} \hat{\mathcal{R}}^\zJ_{\czd} T^{\czd}_{\:\czb}
   -
   \hat{\mathcal{R}}^\zI_{\czd} T^{\czd}_{\,\czb,\zJ} \hat{\mathcal{R}}^\zJ_{\czc} T^{\czc}_{\:\cza} \big)
   \nonumber\\
   &&\quad=
   -\big(
   \hat{\mathcal{R}}^\zI_{\czc}
   T^{\czc}_{\:\cze}
   k^{\cze}_{\,\ccza,\zJ}
   {(Qk)^{-1}} \vphantom{|} ^\ccza_{\,\cczb} Q^{\cczb}_{\,\cza}
    \hat{\mathcal{R}}^\zJ_{\czd} T^{\czd}_{\:\czb}
   -
  (\cza\leftrightarrow\czb) \big)
   \nonumber\\
   &&\quad\quad
   - \big(
   \hat{\mathcal{R}}^\zI_{\czc}
   k^{\czc}_{\,\ccza} {(Qk)^{-1}} \vphantom{|} ^\ccza_{\,\cczb}
   Q^{\cczb}_{\cze,\zJ}
   T^{\cze}_{\:\cza}
   \hat{\mathcal{R}}^\zJ_{\czd} T^{\czd}_{\:\czb}
   -
  (\cza\leftrightarrow\czb) \big)
   \nonumber\\
   &&\quad=
   -\hat{\mathcal{R}}^\zI_{\czc}
   L^{\czc}_{\,\cze}
   \hat{C}^{\cze}_{\czc\czd}
   T^{\czc}_{\:\cza}
   T^{\czd}_{\:\czb}-
    \mathcal{R}^\zI_{\czc}
    N^{\czc}_{\,\cza\czb}
   +
    \mathcal{R}^\zI_{\czc}N^{\czc}_{\,\czb\cza}\nonumber\\
  &&\quad\quad-
  \hat{\mathcal{R}}^\zI_{\cze}
   k^{\cze}_{\,\ccza} {(Qk)^{-1}} \vphantom{|} ^\ccza_{\,\cczb}
   \theta^{\cczb}_{,\zK}\hat{E}^{\zK\zJ}_{\czc\czd}\hat{S}_{,\zJ}
   T^{\czc}_{\:\cza}
   T^{\czd}_{\:\czb}.
 \eea
Here $N^{\czc}_{\,\cza\czb}$ is defined by Eq.\,(\ref{N}) and $L^{\cza}_{\,\czb}$ is the \emph{longitudinal} projector complementary to \emph{transverse} projector $T^{\cza}_{\:\czb}$,
 \beq{lkQProj} 
  L^{\cza}_{\,\czb}
   \,\equiv\,
  \delta^{\cza}_{\,\czb} - T^{\cza}_{\:\czb}
  \,=\,
  k^{\cza}_{\,\ccza} {(Qk)^{-1}} \vphantom{|} ^\ccza_{\,\cczb} Q^{\cczb}_{\,\czb}.
 \eeq

Therefore, summing the contributions (\ref{_contrib1}) and (\ref{_contrib2}) one observes that their first terms form out of $\hat{\mathcal{R}}^\zI_{\cze}$ the projected generator $\mathcal{R}^\zI_{\cze}$, so that we get
 \bea{Jacobi_Id_projected_RES}
  \mathcal{R}^\zI_{\cza,\zJ} \mathcal{R}^\zJ_{\czb} -  \mathcal{R}^\zI_{\czb,\zJ} \mathcal{R}^\zJ_{\cza}
 \! &=&\!
  \mathcal{R}^\zI_{\cze} \hat{C}^{\cze}_{\czc\czd}  T^{\czc}_{\:\cza}  T^{\czd}_{\:\czb}
    -  \mathcal{R}^\zI_{\czc}
    N^{\czc}_{\,\cza\czb}\!
    +
    \mathcal{R}^\zI_{\czc}N^{\czc}_{\,\czb\cza}
   \nonumber\\
   &&
  +D^{\zI}_{\,\zK}
   \hat{E}^{\zK\zJ}_{\czc\czd}\hat{S}_{,\zJ}
   T^{\czc}_{\:\cza}
   T^{\czd}_{\:\czb}
 \eea
where $D^{\zI}_{\,\zK}$ is defined by Eq.\,(\ref{D}),
\bea{}
    D^{\zI}_{\,\zK}
   \equiv\,
   \delta^{\zI}_{\,\zK}
  -\hat{\mathcal{R}}^\zI_{\cze}k^{\cze}_{\,\ccza}
  {(Qk)^{-1}} \vphantom{|} ^\ccza_{\,\cczb}
   \theta^{\cczb}_{,\zK},
\eea
and we used the fact that $\mathcal{R}^\zI_{\cze}=\mathcal{R}^\zI_{\czb}T^{\czb}_{\:\cze}$, $k^{\cza}_{\,\ccza}=L^{\cza}_{\,\czb} k^{\czb}_{\,\ccza}$ and $Q^{\cczb}_{\,\czc} = Q^{\cczb}_{\,\czb} L^{\czb}_{\:\czc}$. Here the last term is not explicitly antisymmetric in indices $\zI$ and $\zJ$. However,
due to the property $\hat{S}_{,\zJ}=\hat{S}_{,\zI} D^{\zI}_{\,\zJ}$ of the projector $D^{\zI}_{\,\zJ}$ it can equivalently be rewritten in the antisymmetric form with
${E}^{\zI\zJ}_{\cza\czb} =
  D^{\zI}_{\,\zK}
  D^{\zJ}_{\,\zL}
   \hat{E}^{\zK\zL}_{\czc\czd} \,
   T^{\czc}_{\:\cza}
   T^{\czd}_{\:\czb}$.
This finally leads to the algebra (\ref{Jacobi_Id_Lagrange_projected_COPY})  with structure functions (\ref{cT_restricted})-(\ref{D}).

The oblique projector $D^{\zI}_{\,\zJ}$ defined by (\ref{D}) enters the formalism when the parental algebra is open. Its left kernel is spanned by ${(Q k)^{-1}} \vphantom{|} ^\ccza_{\,\cczb}\, \theta^{\cczb}_{,\zI}$ and the right kernel is spanned by the set of longitudinal gauge vectors $\hat{\mathcal{R}}^\zJ_{\czb} k^{\czb}_{\,\ccza}$
 \bea{DProj_Properties}
    &&D^{\zI}_{\,\zJ} D^{\zJ}_{\,\zK} \,=\, D^{\zI}_{\,\zK},\,
    {(Q k)^{-1}} \vphantom{|} ^\ccza_{\,\cczb}\, \theta^{\cczb}_{,\zI}  D^{\zI}_{\,\zJ}
    \,=\, 0,\quad \nonumber\\
    &&D^{\zI}_{\,\zJ} \hat{\mathcal{R}}^\zJ_{\czb} k^{\czb}_{\,\ccza}
    \,=\, 0.
 \eea
Moreover, this projector in the $\varphi^\zI$-space converts the parental generator into the projected one in the space of gauge indices, $D^{\zI}_{\,\zJ} \hat{\mathcal{R}}^\zJ_{\czb}=\hat{\mathcal{R}}^\zI_{\cza}
T^{\cza}_{\:\czb} \equiv {\mathcal{R}}^\zI_{\czb}$, so that $\theta^{\cczb}_{,\zI}  D^{\zI}_{\,\zJ} \hat{\mathcal{R}}^\zJ_{\czb} = 0$.

When $\rank \theta^\ccza_{,\zI}=\rank Q^\ccza_{\,\cza} = m_1$ then the second of relations (\ref{DProj_Properties}) can be simplified to $\theta^{\cczb}_{,\zI}  D^{\zI}_{\,\zJ} =0$. For rank-deficient $Q^\ccza_{\,\cza}$, when $\rank \theta^\ccza_{,\zI}= m_1$ and $\rank Q^\ccza_{\,\cza} = m_1-m_2$, the correct property is $\theta^{\cczb}_{,\zI}  D^{\zI}_{\,\zJ} =\big(1-(Qk)(Qk)^{-1}\big)\vphantom{|}^\cczb_{\,\ccza}\theta^\ccza_{,\zI}\equiv {L_1}^\cczb_{\,\ccza} \,\theta^\ccza_{,\zI}$, where ${L_1}^\cczb_{\,\ccza}$ is a projector on left zero vectors of $Q^\ccza_{\,\cza}$. For this generic (physically interesting case) the rank deficiency of $D^{\zI}_{\,\zJ}$ equals the rank of gauge-restriction operator (\ref{def_rQ}), $\corank D^{\zI}_{\,\zJ} = \rank Q^\ccza_{\,\czb} = \rank k^{\cza}_{\,\cczb}$.

Note that the resulting open algebra of the restricted theory (\ref{Jacobi_Id_Lagrange_projected_COPY}) closes on shell of the parental theory $\hat{S}_{,\zI}=0$ rather than on its own shell $\hat{S}_{,\zI} - \lambda_\ccza \theta^\ccza_{,\zI}=0$. In view of the above relations one has
$\hat{S}_{,\zI} D^{\zI}_{\,\zJ}
= (\hat{S}_{,\zI} - \lambda_\ccza \theta^\ccza_{,\zI}) D^{\zI}_{\,\zJ}
+\lambda_\ccza {L_1}^\ccza_{\,\cczb}\theta^\cczb_{,\zJ}$, so that the algebra (\ref{Jacobi_Id_Lagrange_projected_COPY}) can be rewritten in the form
  \bea{Lagrange_projected_openalgebra}
   &&\hspace{-0.5cm}\mathcal{R}^\zI_{\cza,\zJ} \mathcal{R}^\zJ_{\czb} -  \mathcal{R}^\zI_{\czb,\zJ} \mathcal{R}^\zJ_{\cza}
   \,=\,  \mathcal{R}^\zI_{\czc} C^{\czc}_{\cza\czb}\nonumber\\
   &&\quad\quad+ {E}^{\zI\zJ}_{\cza\czb} \big(\hat{S}_{,\zJ} - \lambda_\ccza \theta^\ccza_{,\zJ}\big)
   + {E}^{\zI\zJ}_{\cza\czb} \lambda_\ccza {L_1}^\ccza_{\,\cczb}\theta^\cczb_{,\zJ},
  \eea
where the last term, unless it is zero, breaks its closure on shell of the restricted theory. Note that $\lambda_\ccza {L_1}^\ccza_{\,\cczb}$ is the part of the full set of Lagrange multipliers $\lambda_\ccza$ which stay unrestricted by the equation of motion (\ref{lambdaEoM}) for Lagrange multipliers.

 \section{Determinant relation}
  \label{ASect:Determinant_Relation}

\newcommand{\A}{a}

In this section we prove the determinant relation (\ref{det_rellation}). All involved quantities are matrix (two-index) structures which allow us to omit indices implying standard matrix multiplication. There are indices of two types --- lowercase Greek indices of some range $m_0$ and lowercase Roman indices of the lower range $m_1<m_0$\footnote{Regarding the rank of quantities with indices $\ccza, \cczb, ..$ see a brief discussion in the end of this section.} Quadratic  $m_0\times m_0$  matrices are denoted by capital Roman letters (e.g. $\fatm{T}$). The matrices mapping from $m_0$-dimensional space to a lower $m_1$-dimensional space and back will be either underlined or will respectively carry a line over them. Thus we have
 \bea{}
  &&\fatm{Q} \,\leftrightarrow\, \hat{Q}^\cza_{\,\czb}, \quad
  \fatm{I} \,\leftrightarrow\, \delta^\cza_{\,\czb}, \,
  \nonumber\\
  &&
  \underline \A \,\leftrightarrow\, \A^\ccza_\cza, \quad
  \underline\sigma \,\leftrightarrow\, \sigma^\ccza_{\,\cza}, \quad
  \overline\sigma \,\leftrightarrow\, \sigma^\cza_{\,\ccza}, \quad
  \overline k \,\leftrightarrow\, k^\cza_{\,\ccza},\nonumber\\
  &&\overline{\sigma}\,\underline{\A}
  \,\leftrightarrow\, (\overline{\sigma}\,\underline{\A})^\cza_\czb
  \equiv \sigma^\cza_{\,\ccza} \A^\ccza_{\,\czb}, \quad
  \underline\sigma\,\fatm{Q}\,\overline k
  \,\leftrightarrow\, (\underline\sigma\fatm{Q}\overline k)^\ccza_{\,\cczb}
  \equiv \sigma^\ccza_{\,\cza} \hat{Q}^\cza_{\,\czb} k^\czb_{\,\cczb}, \nonumber\\
  &&\underline{\A}\,\overline{k}
  \,\leftrightarrow\, (\underline{\A}\,\overline{k})^\ccza_{\,\cczb}
  \equiv {\A}^\ccza_{\,\cza}{k}^\cza_{\,\cczb}.
 \eea
Note that the matrix $\fatm{Q}$ here is a quadratic $m_0\times m_0$ matrix and in Sec.\,\ref{SSect:Canon_Measure_Normalization} it stands for the complete Faddeev-Popov operator for the rank-$m_0$ parental gauge symmetry. We also assume that $\underline\sigma$ and $\overline\sigma$ form dual vectors in the sense that
\bea{duality_sigma}
\underline\sigma\,\overline\sigma = I
\,\;\leftrightarrow\;\,
\underline\sigma^\ccza_{\,\cza}\,
\overline\sigma^\cza_{\,\cczb}=\delta^\ccza_{\,\cczb}.
\eea

Then one can prove the following relation between the determinants of these matrices
 \bea{relation}
  &&\fatm{\det}\,\fatm{Q} = \frac{\det(\underline\sigma\fatm{Q}\overline k)}{\det (\underline{\A}\overline{k})}\nonumber\\
  &&\qquad\quad\times\;\fatm{\det}\,
  \Big(\underbrace{\fatm{Q}\,\big(\fatm{I}-\overline{k}\, (\underline{\sigma}\fatm{Q}\overline{k})^{-1}\underline{\sigma}\,\fatm{Q}\big)
  + \overline{\sigma}\,\underline{\A}}_{\fatm{F}}\Big)\,
  .\quad
 \eea
It is also assumed that all determinants in (\ref{relation}) are nonzero.

Note that the $m_0\times m_0$ matrix in the first determinant in the right-hand side of (\ref{relation}) can be written as
\bea{fatF}
\fatm{F} = \fatm{Q}\, \fatm{T} + \overline{\sigma}\,\underline{\A},
\,\quad\,
\fatm{T} = \fatm{I} - \overline{k}\,(\underline{\sigma}\fatm{Q}\overline{k})^{-1}
\underline{\sigma}\,\fatm{Q},
\eea
where $\fatm{T} \,\leftrightarrow\, T^\cza_{\:\czb}(\underline\sigma\fatm{Q},k)$ is in fact the oblique projector (\ref{tkQProj}) with the left kernel $\underline\sigma\,\fatm{Q} \,\leftrightarrow\, \sigma^\ccza_\czb \hat{Q}^\czb_{\,\cza} =  Q^\ccza_{\,\cza}$ which can be identified with the gauge-restriction operator. The first term of the matrix $\fatm{F}$ is therefore degenerate and has as right and left zero vectors $\overline k$ and $\underline\sigma$ respectively, $\fatm{Q}\,\fatm{T}\,\overline{k}=\underline{\sigma}\, \fatm{Q}\, \fatm{T}=0$. Therefore it can be interpreted as the analog of the Hessian of a gauge-invariant action, whereas the second term in $\fatm{F}$ plays the role of a gauge-fixing term providing invertibility of this matrix.

The proof of the relation (\ref{relation}) can be done by using the basis in which the matrix $\fatm{F}$ acquires a block-triangular form, but it is easier to use the analog of Ward identities in order to prove that the right-hand side is actually independent of the choice of arbitrary elements $\underline\sigma$, $\underline{\A}$ and $\overline k$ and then check that this relation indeed holds under a special choice of these elements. Such a choice is obvious and reads as $\overline\sigma=\fatm{Q}\,\overline k$ and $\underline{\A}=\underline\sigma\,\fatm{Q}$ (when $\underline{\sigma}\,\fatm{Q}\,\overline{k} = \underline{\A}\,\overline k=I$), so that it remains to check that the right-hand side is indeed $(\underline\sigma,\underline{\A},\overline k)$-independent.

Multiplying the matrix (\ref{fatF}) from the left and from the right by zero vectors of its first term $\fatm{Q}\,\fatm{T}$ one finds two Ward identities and their corollary,
 \beq{}
  \fatm{F}^{-1}\overline\sigma
  =\overline k\, (\underline{\A}\overline k)^{-1} \!,
  \;\;\,
  \underline{\A}\,\fatm{F}^{-1}\!
  =\underline\sigma,
  \;\;\,
  \fatm{F}^{-1}\fatm{Q}\,\fatm{T} = \fatm{I}-\overline k\,(\underline{\A}\overline k)^{-1}\underline{\A}.
 \eeq
Then, direct variation of the right-hand side of the relation (\ref{relation}) with respect to $\underline{\A}$, $\overline k$ and $\underline\sigma$ shows that it is indeed independent of these quantities on account of the above identities and the variational version of duality relation (\ref{duality_sigma}), $\underline\sigma\,\var\overline\sigma +\var\underline\sigma\,\overline\sigma=0$.

We finish this discussion with the note on applicability of the above relation to the case when the objects labeled by indices $\ccza$ are rank-deficient (with the rank $m_1-m_2 <m_1$). In Sec.\,\ref{SSect:Canon_Measure_Normalization}, where this determinant relation was used, instead of indices $\ccza$ we explicitly used the indices $p$ belonging to the range $m_1-m_2$ which symbolized the maximal rank (irreducible) representation. In such a representation all matrices are nondegenerate and quantities with mixed
indices are of the maximal rank. However the maximal rank
representation is not necessary and this proof runs equally well in
the reducible representation of rank-deficient objects $(\underline\sigma,\overline\sigma,\underline{\A},\overline k)$ of rank $m_1-m_2$, provided the composite quantities $\overline{\sigma}\,\underline{\A}$, $(\underline{\A}\overline k)$, $(\underline{\sigma}\fatm{Q}\overline{k})$ have the same rank $m_1-m_2$\footnote{This is the analog of the rank condition (\ref{g_rank}).} and the inverse degenerate matrices $(\underline{\sigma}\fatm{Q}\overline{k})^{-1}$, $(\underline{\A}\overline k)^{-1}$ are treated in the Moore-Penrose sense (as was discussed in Appendix \ref{ASect:Projectors_Variation}).
Determinants of degenerate operators of rank $m_1-m_2$ should be understood in an appropriate regularized sense, for example by omitting the contribution of zero eigenvalues in the eigenvector bases. If one goes into details of the variational proof, then the variations of regularized determinants turn out to be $\var \det{(\underline{\sigma}\fatm{Q}\overline{k})} = \mathrm{tr}\, \big[ \var (\underline{\sigma}\fatm{Q}\overline{k}) (\underline{\sigma}\fatm{Q}\overline{k})^{-1} \big]$ with the Moore-Penrose definition of the inverse matrix. The uniqueness of $\fatm{T}$ projector variation was shown in Appendix \ref{ASect:Projectors_Variation}. The uniqueness of the regularized determinant variation has a similar mechanism.

This line of reasoning justifies the validity of one-loop contributions to the effective action of the restricted theory of Secs.\,\ref{Sect:Restricting_and_Reducibility} and \ref{SSect:1loop_EA_Restricted} (even though in Sec.\,\ref{Sect:BV_EA_1stReducible} these contributions were formally based on full-rank quantities with reducibility indices $\ccza$).







\newcommand{\zzz}{} 
{}

\end{document}